\documentclass[journal,twoside,web,dvipsnames,table]{ieeecolor}

\usepackage{generic}
\usepackage{cite}
\usepackage{amsmath,amssymb,amsfonts}
\usepackage{algorithmic}
\usepackage{graphicx}
\usepackage{textcomp}
\usepackage{tikz}
\usepackage{adjustbox}
\usepackage{pgfplots}
\usetikzlibrary{arrows.meta, calc, arrows, patterns, patterns.meta, spy}
\usepackage{xcolor}
\usepackage{colortbl}
\usepackage{pgfplotstable}
\usepackage{multirow}
\usepackage{scalerel}
\usepackage{gensymb}
\usepackage{subcaption}
\usepackage{caption}
\usepackage{url}
\usepackage{rotating}
\usepackage{afterpage}

\pgfplotsset{compat=1.9}

\DeclareMathOperator*{\argmin}{\arg\min} 

\newcommand{\norm}[1]{\left\lVert#1\right\rVert}

\newcolumntype{C}[1]{>{\centering\arraybackslash}m{#1}}

\pgfplotstableset{
    /color cells/min/.initial=0,
    /color cells/max/.initial=1000,
    /color cells/textcolor/.initial=,
    color cells/.code={%
        \pgfqkeys{/color cells}{#1}%
        \pgfkeysalso{%
            postproc cell content/.code={%
                \begingroup
                \pgfkeysgetvalue{/pgfplots/table/@preprocessed cell content}\value
\ifx\value\empty
\endgroup
\else
                \pgfmathfloatparsenumber{\value}%
                \pgfmathfloattofixed{\pgfmathresult}%
                \let\value=\pgfmathresult
                \pgfplotscolormapaccess
                    [\pgfkeysvalueof{/color cells/min}:\pgfkeysvalueof{/color cells/max}]%
                    {\value}%
                    {\pgfkeysvalueof{/pgfplots/colormap name}}%
                \pgfkeysgetvalue{/pgfplots/table/@cell content}\typesetvalue
                \pgfkeysgetvalue{/color cells/textcolor}\textcolorvalue
                \toks0=\expandafter{\typesetvalue}%
                \xdef\temp{%
                    \noexpand\pgfkeysalso{%
                        @cell content={%
                            \noexpand\cellcolor[rgb]{\pgfmathresult}%
                            \noexpand\definecolor{mapped color}{rgb}{\pgfmathresult}%
                            \ifx\textcolorvalue\empty
                            \else
                                \noexpand\color{\textcolorvalue}%
                            \fi
                            \the\toks0 %
                        }%
                    }%
                }%
                \endgroup
                \temp
\fi
            }%
        }%
    }
}

\begin{document}

\title{Anatomically Parameterized Statistical Shape Model: Explaining Morphometry through Statistical Learning}

\author{Arnaud Boutillon, Asma Salhi, Valérie Burdin, and Bhushan Borotikar
\thanks{This work was funded by IMT, Fondation Mines-Télécom and Institut Carnot TSN through the Futur \& Ruptures program, and supported by region of Brittany funds (email: arnaud.boutillon@imt-atlantique.fr)}
\thanks{A. Boutillon, A. Salhi and V. Burdin are with IMT Atlantique, Brest, France and LaTIM UMR 1101, Inserm, Brest, France.}
\thanks{B. Borotikar was with LaTIM UMR 1101, Inserm, Brest, France and CHRU de Brest, Brest, France. He is now with SCMIA, SIU, Pune, India.}
\thanks{\copyright 2022 IEEE.  Personal use of this material is permitted. Permission from IEEE must be obtained for all other uses, in any current or future media, including reprinting/republishing this material for advertising or promotional purposes, creating new collective works, for resale or redistribution to servers or lists, or reuse of any copyrighted component of this work in other works.}}

\maketitle

\begin{abstract}
\textit{Objective:} Statistical shape models (SSMs) are a popular tool to conduct morphological analysis of anatomical structures which is a crucial step in clinical practices. However, shape representations through SSMs are based on shape coefficients and lack an explicit one-to-one relationship with anatomical measures of clinical relevance. While a shape coefficient embeds a combination of anatomical measures, a formalized approach to find the relationship between them remains elusive in the literature. This limits the use of SSMs to subjective evaluations in clinical practices. We propose a novel SSM controlled by anatomical parameters derived from morphometric analysis. \textit{Methods:} The proposed anatomically parameterized SSM (ANAT$_{\text{SSM}}$) is based on learning a linear mapping between shape coefficients (latent space) and selected anatomical parameters (anatomical space). This mapping is learned from a synthetic population generated by the standard SSM. Determining the pseudo-inverse of the mapping allows us to build the ANAT$_{\text{SSM}}$. We further impose orthogonality constraints to the anatomical parameterization (OC-ANAT$_{\text{SSM}}$) to obtain independent shape variation patterns. The proposed contribution was evaluated on two skeletal databases of femoral and scapular bone shapes using clinically relevant anatomical parameters within each (five for femoral and six for scapular bone). \textit{Results:} Anatomical measures of the synthetically generated shapes exhibited realistic statistics. The learned matrices corroborated well with the obtained statistical relationship, while the two SSMs achieved moderate to excellent performance in predicting anatomical parameters on unseen shapes. \textit{Conclusion:} This study demonstrates the use of anatomical representation for creating anatomically parameterized SSMs and as a result, removes the limited clinical interpretability of standard SSMs. \textit{Significance:} The proposed models could help analyze differences in relevant bone morphometry between populations, and be integrated in patient-specific pre-surgery planning or in-surgery assessment. 

\end{abstract}


\section{Introduction}
\label{sec:introduction}

Automatic interpretation and analysis of three-dimensional (3D) anatomical structures is key in medical applications. To this end, statistical shape modeling (SSM) is a popular tool that provides a compact representation of a family of objects as a normal distribution of their shape variations \cite{luthi_gaussian_2018}. These models adopt an analysis-by-synthesis approach in which to explain and interpret a 3D object, one needs to be able to synthesize it. Popularity of these models lies in their ability to model biological shapes that naturally have a high variability and complexity. Furthermore, being linear representations, these models are mathematically convenient \cite{mutsvangwa_automated_2015}. Literature reports that SSMs have been integrated into medical workflows \cite{ambellan_statistical_2019} to help clinicians diagnose pathologies \cite{von_tycowicz_efficient_2018, cerveri_predicting_2020}, design implants \cite{vanden_berghe_virtual_2017, burton_assessment_2019}, reconstruct 3D anatomy from 2D radiographs \cite{chen_automatic_2013} or plan patient-specific intervention \cite{rajamani_statistical_2007, salhi_statistical_2020}. Consequently, SSMs of bony structures have been developed in the literature which include but are not limited to the femur \cite{von_tycowicz_efficient_2018, cerveri_representative_2019, lynch_statistical_2019, cerveri_predicting_2020}, humerus \cite{mutsvangwa_automated_2015, sintini_investigating_2018}, pelvis \cite{vanden_berghe_virtual_2017}, scapula \cite{mutsvangwa_automated_2015, plessers_virtual_2018, salhi_statistical_2020}, tibia \cite{lynch_statistical_2019, cerveri_predicting_2020}, vertebrae \cite{hollenbeck_statistical_2018}, and wrist \cite{chen_automatic_2013}.

For clinical use, the key property of the SSM lies in the dense correspondence established during the registration process, which identifies the points sharing the same anatomical characteristics \cite{rasoulian_group-wise_2012, mutsvangwa_automated_2015, borotikar_augmented_2017}. This feature has been effectively used to embed bony SSM with landmark-based anatomical information such as muscle insertions \cite{salhi_subject-specific_2017} or identify cortical bone thickness \cite{sintini_investigating_2018}. Moreover, the generative capabilities of these models enable the exploration of the shape coefficient representation within the valid anatomical shape variation \cite{mutsvangwa_automated_2015, cerveri_predicting_2020}. Shapes generated in such manner allow the user to understand the anatomical feature having higher variability by virtue of varying individual principal components. However, the shape variation patterns observed through individual change of principal components do not explicitly correspond to a unique anatomical parameter used in morphometric analysis by the clinicians. Thus, clinicians are often left with a visual guesswork upon how changing a single anatomical parameter can affect the remaining parameters. 

Morphometry refers to the quantitative analysis of the shape in terms of lengths, widths, angles, and masses \cite{von_schroeder_osseous_2001, polguj_correlation_2011}. In the context of anatomical structure analysis, this information is used to quantify the morphological development over time or due to a disorder, detect changes or abnormalities in the shape, understand variation in population (healthy or impaired), and deduce functional relationship (normal or affected) \cite{ambellan_statistical_2019}. Hence, clinicians have developed indexes that are correlated with the instability of bony structures and musculoskeletal joints. For example, for the upper limb, extreme critical shoulder angle, defined on the scapular bone, is an indicator of degenerative rotator cuff tears and glenohumeral osteoarthritis \cite{cherchi_critical_2016}. Similarly, the kinematic instability of the knee joint is characterized by the tibiofemoral alignment index, which integrates five angle measurements \cite{laxafoss_alignment_2013}. Therefore, anatomical parameters are used as predictive tools to assess biomechanical disorders and instabilities and to make an informed treatment decision. However, the computation of these indexes relies on manual landmarking on medical images or on 3D reconstructed surfaces. Both these approaches are either time consuming or suffer from intra- and inter-observer variability \cite{subburaj_computer-aided_2010, ghafurian_computerized_2016}. Hence, in order to improve the robustness and accuracy of quantifying anatomical parameters, automatic geometrical methods have been proposed in the literature \cite{subburaj_computer-aided_2010, ghafurian_computerized_2016}.

Recent studies have proposed to use point-to-point correspondence, established during the SSM building process, to automatically compute anatomical measurements \cite{vanden_berghe_virtual_2017, hollenbeck_statistical_2018, plessers_virtual_2018, cerveri_representative_2019}, while others have suggested classification methodologies based on shape coefficients \cite{taghizadeh_statistical_2017, lynch_statistical_2019, cerveri_predicting_2020}. These automatic approaches provide accurate measurements of the anatomical parameters and reliable assessment of the instability of knee and shoulder joints. Furthermore, certain studies have investigated the relationship between shape coefficients and anatomical parameters \cite{hollenbeck_statistical_2018, sintini_investigating_2018, cerveri_representative_2019} in which shape coefficients were found to encode anatomical parameters. However, to the best of our knowledge, none of these studies have presented a mathematical framework in which each mode of variation of the SSM represented an individual anatomical parameter regarded as relevant in clinical practice.

In this paper, we propose a novel technique to build SSMs which integrate the relationship between the shape coefficients (a shape representation arising from principal component analysis) and the anatomical parameters (a shape representation derived from relevant morphometric analysis). To that end, we define a mapping between both these representations, which is learned through linear regression on a synthetic population generated by the standard SSM. With this approach, we develop two types of SSMs with anatomical parameters as their modes of variation. The first one is an anatomically parameterized SSM which allows us to explore the relationship between the entire shape and the selected anatomical parameters. The second one is a constrained version of the first model with an orthogonal anatomical parameterization to obtain independent shape variation patterns. We apply and evaluate the proposed methodology to the femoral and scapular bone shapes using clinically relevant anatomical parameters. The contributions of this study are three-fold:
\begin{enumerate}
    \item Generation of synthetic populations with automatic anatomical measurements performed using dense point-to-point correspondence and landmark tracking.
    \item Development of two anatomically parameterized SSMs controlled by anatomical parameters derived from morphometric analysis, thanks to a statistical learning of the mapping between shape coefficient and anatomical parameters representations.
    \item Assessment of the prediction performance of the developed models and exploration of the novel representation arising from anatomical parameters.
\end{enumerate}

The remainder of this paper is structured as follows. Section \ref{sec:methods} presents the anatomically parameterized statistical shape model (Section \ref{sec:anat_ssm}), the constrained model (Section \ref{sec:ind_anat_ssm}) and a measure of shape variability (Section \ref{sec:shape_variability_anatomical_parameters}). The experiments are explained in Section \ref{sec:experiments} which mainly encompasses the automatic anatomical measurements derivation (Section \ref{sec:automatic_derivation_anatomical_parameters}), the synthetic populations generation (Section \ref{sec:synthetic_populations_generation}) and the assessment of the two models (Section \ref{sec:assessment_anat_ssms}). The results reported in Section \ref{sec:results} validate the automatic derivation approach (Section \ref{sec:validation_automatic_computation}), the synthetic populations (Section \ref{sec:synthetic_population_characteristics}) and the developed statistical models (Section \ref{sec:validation_anat_ssm}). Section \ref{sec:discussion} follows with the discussion of results (Section \ref{sec:comparison_synthetic_real}), clinical benefits (Section \ref{sec:benefits_clinical_practice}) and limitations of the proposed methodology (Section \ref{sec:limitations}). Finally, Section \ref{sec:conclusion} presents the conclusion and perspectives of the study.

\section{Methods}
\label{sec:methods}

In this section, we first provide an overview of the mathematical formalism of standard SSMs (BASE$_{\text{SSM}}$) with principal component analysis used for dimensionality reduction. We then present the formalism of the proposed novel generative model (ANAT$_{\text{SSM}}$) based on anatomical parameterization derived from least squares regression. We further propose a constrained form of ANAT$_{\text{SSM}}$ leading to an orthogonal anatomical parameterization model (OC-ANAT$_{\text{SSM}}$) which is formalized as an orthogonal Procrustes problem. Finally, similar to BASE$_{\text{SSM}}$, we define the shape variability induced by the two proposed models. 

\subsection{Overview of statistical shape modeling}
\label{sec:overview_ssm}

Statistical shape models assume that the space of all possible shape deformations can be learned from a set of example shapes. Let $\Gamma_1,...,\Gamma_n$ be a set of $n$ shapes in which each instance $\Gamma_i$ is represented by a mesh that is a discrete set of $N$ landmark points, $\Gamma_i = \{\gamma^{k}_{i} \ | \ \gamma^{k}_{i} \in \mathbb{R}^3, k=1,...,N \}$. The points among the shapes are assumed to be in correspondence, which means that for two shapes $\Gamma_i$ and $\Gamma_j$ the $k$-th landmark points, $\gamma^{k}_{i}$ and $\gamma^{k}_{j}$, represent the same anatomical mesh point. In the case of dense set of points, correspondence is usually established automatically using a registration algorithm \cite{mutsvangwa_automated_2015}. In order to build the model, each shape $\Gamma_i$ is described as a vector shape $s_i \in \mathbb{R}^{3N}$, where the $x,y,z$- coordinates of each point $\gamma^{k}_{i} = (x^{k}_i, y^{k}_i, z^{k}_i)$ are concatenated as follows: $s_i = (x^{1}_{i}, y^{1}_{i}, z^{1}_{i}, ..., x^{N}_{i}, y^{N}_{i}, z^{N}_{i})^{T}$.

Based on the vectorial representation, it is possible to construct a probability distribution over shapes by applying multivariate statistics \cite{albrecht_posterior_2013, luthi_gaussian_2018}. We assume that the shape variations can be modeled using a multivariate normal distribution $s \sim \mathcal{N}(\mu, \Sigma)$, in which the mean $\mu$ and covariance matrix $\Sigma$ are estimated from the example vector shapes, $s_1, ..., s_n$. The covariance matrix $\Sigma$ is usually intractable as the number of points $N$ is large ($N \gg n$). However, since the rank of the covariance matrix is at most $n-1$, performing principal component analysis (PCA) on $\Sigma$ generates a new representation defined by $n-1$ basis vectors which leads to the following model:
\begin{align}
    \textit{BASE}_{\textit{SSM}} &= \mu + \sum_{i=1}^{n-1} \alpha_i \sqrt{\lambda_i} \phi_i
\label{eq:SSM_equation}
\end{align}
\noindent where $(\lambda_i, \phi_i)_{1 \leq i \leq n-1}$ are the eigenvalues and eigenvectors of the covariance matrix $\Sigma$, while $\alpha_i$ are the shape coefficients. The eigenvectors $\phi_i$ are orthogonal and form the principal components of the model. Each of these principal components represents an independent shape variation pattern, while the corresponding eigenvalue $\lambda_i$ quantifies their variance. The eigenvalues and their corresponding eigenvectors are typically arranged from largest to smallest, and the first principal components correspond to the main modes of variation \cite{luthi_gaussian_2018}.

If we adopt the notation $P = (\phi_1, ..., \phi_{n-1}) \in \mathbb{R}^{3N,n-1}$, $D = \text{diag}(\sqrt{\lambda_1}, ...,\sqrt{\lambda_{n-1}}) \in \mathbb{R}^{n-1,n-1}$ and $\alpha = (\alpha_1, ..., \alpha_{n-1})^{T} \in \mathbb{R}^{n-1}$, (\ref{eq:SSM_equation}) is reformulated as follows:
\begin{align}
    \textit{BASE}_{\textit{SSM}} = \mu + P D \alpha
\end{align}
Assuming that the shape coefficients $\alpha$ are distributed according to $\mathcal{N}(0,I_{n-1})$, then the shapes follow a normal multivariate distribution $\mathcal{N}(\mu,\Sigma)$. Thus, the BASE$_{\text{SSM}}$ is a model of deformation added to the mean shape, and the shape distribution is efficiently parameterized by the shape coefficients \cite{luthi_gaussian_2018, salhi_statistical_2020}. 

However, in the context of morphometric analysis, the shape variation patterns of the principal components of the BASE$_{\text{SSM}}$ do not explicitly represent a one-to-one relationship between them and the anatomical parameters that are typically employed to characterize these structures. Hence, we developed an SSM parameterized by a novel representation arising from the anatomical parameters. This novel representation is derived from a learned mapping between shape coefficients and anatomical parameters, and its formalism is explained in the following sections. 

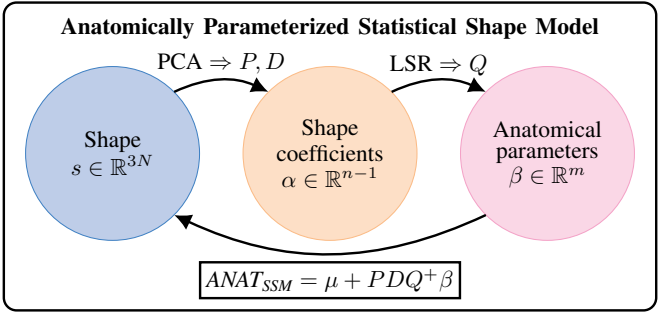
\begin{figure}[t]
\centering
\begin{adjustbox}{width=.48\textwidth}
\begin{tikzpicture}

\node[anchor=north] at (9, 17) {\scalebox{2.75}{\textbf{Anatomically Parameterized Statistical Shape Model}}};
\filldraw[color=NavyBlue!80, fill=NavyBlue!25, very thick](0,11.25) circle (3.6);
\filldraw[color=Apricot!95, fill=Apricot!40, very thick](9,11.25) circle (3.6);
\filldraw[color=Lavender!95, fill=Lavender!40, very thick](18,11.25) circle (3.6);

\draw[line width=1mm, -{Latex[length=20pt, width=20pt]}] (2.55,13.8) to[out=25,in=155] (6.45,13.8);
\draw[line width=1mm, -{Latex[length=20pt, width=20pt]}] (11.55,13.8) to[out=25,in=155] (15.45,13.8);
\draw[line width=1mm, -{Latex[length=20pt, width=20pt]}] (15.45,8.7) to[out=-155,in=-25] (2.55,8.7);

\node at (0, 11.75) {\scalebox{2.75}{Shape}};
\node at (0, 10.75) {\scalebox{2.75}{$s \in \mathbb{R}^{3N}$}};

\node at (9, 12.25) {\scalebox{2.75}{Shape}};
\node at (9, 11.25) {\scalebox{2.75}{coefficients}};
\node at (9, 10.25) {\scalebox{2.75}{$\alpha \in \mathbb{R}^{n-1}$}};

\node at (18, 12.25) {\scalebox{2.75}{Anatomical}};
\node at (18, 11.25) {\scalebox{2.75}{parameters}};
\node at (18, 10.25) {\scalebox{2.75}{$\beta \in \mathbb{R}^{m}$}};

\node at (4.5, 15) {\scalebox{2.75}{$\text{PCA} \Rightarrow P, D$}};
\node at (13.5, 14.9) {\scalebox{2.75}{$\text{LSR} \Rightarrow Q$}};
\node[draw, line width=1mm, inner sep=7.5] at (9, 6) {\scalebox{2.75}{$\textit{ANAT}_{\textit{SSM}} = \mu + PDQ^{+}\beta$}};

\draw[line width=1mm, color=black, rounded corners=15] (22.5, 17.5) -- (-4.5,17.5) -- (-4.5,4.75) -- (22.5,4.75) -- cycle;

\end{tikzpicture}
\end{adjustbox}
  \caption{Proposed anatomically parameterized statistical shape model (ANAT$_{\text{SSM}}$) which encompasses a deformation model added to the mean shape. The linear mapping between the shape space and the representation arising from shape coefficients is based on principal component analysis (PCA), while least squares regression (LSR) is used to estimate the mapping from shape coefficients to anatomical parameters.}
  \label{fig:diagram}
\end{figure}

\subsection{Anatomically parameterized statistical shape model (ANAT$_{\text{SSM}}$)}
\label{sec:anat_ssm}

In the morphometric analysis of an anatomical structure, its shape can be characterized by a small set of $m \leq n-1$ clinically relevant anatomical parameters $\beta = (\beta_{c_1}, ..., \beta_{c_m})^{T} \in \mathbb{R}^{m}$ \cite{von_schroeder_osseous_2001, polguj_correlation_2011}. Here, $c_1, ..., c_m$ are the labels of the anatomical parameters. Hence, similar to the shape coefficients $\alpha$, the anatomical parameters based representation of shape $\beta$ enables another way of compactly describing the shape distribution. Thus, each shape instance can be represented either by using the shape coefficients or using the anatomical parameters. Therefore, we proposed that there exists a mapping between the two representations which can be effectively determined. Specifically, we proposed that the anatomical parameters based representation can be derived from the shape coefficients based representation (Fig. \ref{fig:diagram}).

In this proposed method, we presume that the mapping between the shape coefficients $\alpha$ and the anatomical parameters $\beta$ is linear, and there exist a matrix $Q \in \mathbb{R}^{m, n-1}$ such that:
\begin{align}
    \beta = Q \alpha
\end{align}
\noindent The matrix $Q$ is learned by employing least squares regression on a set of shapes with their corresponding shape coefficients and anatomical parameters. This set of shapes can be obtained using the BASE$_{\text{SSM}}$ as a generative model parameterized by the coefficients $\alpha$. Then, the parameters $\beta$ are automatically estimated from the synthetic shapes. Additionally, if the shape coefficients follow a multivariate normal distribution $\mathcal{N}(0,I_{n-1})$, then the anatomical parameters $\beta$ would be distributed according to $\mathcal{N}(0,QQ^T)$. Hence, $QQ^T$ corresponds to the covariance matrix associated with the distribution of the anatomical parameters.

Next, we inverse the learned linear mapping to obtain a generative model controlled by the anatomical parameters, referred to as anatomically parameterized statistical shape model (ANAT$_{\text{SSM}}$). The matrix $Q$ is reversed by computing its Moore-Penrose pseudo-inverse, $Q^{+} = Q^{T}(QQ^{T})^{-1}$, assuming that $Q$ is of full rank $m$. This inverse mapping leads to the following generative model:  
\begin{align}
    \textit{ANAT}_{\textit{SSM}} &= \mu + P D Q^{+} \beta \\
    &= \mu + \sum_{j=1}^{m} \beta_{c_j} \sum_{i=1}^{n-1} Q_{i,j}^{+} \sqrt{\lambda_i} \phi_i
\end{align}
\noindent where the shape deformation distribution added to the mean shape is efficiently parameterized by the anatomical parameters $\beta$. As the anatomical parameters are typically correlated (i.e. $QQ^T \ne I_m$), the shape deformation vectors $(\sum_{i=1}^{n-1} Q^{+}_{i,j} \sqrt{\lambda_i} \phi_i )_{1 \leq j \leq m}$ that best-fitted anatomical parameters are not orthogonal, and thus the shape variation patterns arising from these vectors are not independent.

However, it would be relevant for clinical practice to build a model with an independent parameterization in order to understand the effect of modifying one anatomical parameter at a time during pre-surgery planning. To this end, we orthogonally constrained our covariance matrix to enforce independent shape deformation, as explained in the next section.

\subsection{Orthogonally constrained and anatomically parameterized statistical shape model (OC-ANAT$_{\text{SSM}}$)}
\label{sec:ind_anat_ssm}

To obtain the closest independent anatomical parameters, as the distribution (of parameters) is assumed to be Gaussian, we enforce the covariance to be the identity matrix by employing the nearest matrix $K \in \mathbb{R}^{m,n-1}$ to $Q$ subject to $KK^{T} = I_m$:
\begin{align}
    K = \argmin_{KK^{T} = I_m} \norm{Q - K}_{F}^{2}
\label{eq:K_optimization}
\end{align}
\noindent where $\norm{.}_{F}$ is the Frobenius norm. This optimization problem is known as the orthogonal Procrustes problem for which the closed-form solution is known \cite{schonemann_generalized_1966}. The problem is solved in two steps. First, we compute the reduced singular value decomposition of $Q$, $Q = U \Delta V^{T}$, with $U \in \mathbb{R}^{m \times m}$, $\Delta \in \mathbb{R}^{m \times m}$, $V \in \mathbb{R}^{n-1 \times m}$. Second, the solution to (\ref{eq:K_optimization}) optimization problem is given by $K = U V^{T}$.

Therefore, we obtain a new linear mapping between the shape coefficients $\alpha$ and the uncorrelated anatomical parameters $\Tilde{\beta}$ defined as follows: 
\begin{align}
\Tilde{\beta} = K \alpha
\end{align}
\noindent where $\Tilde{\beta}$ follows a multivariate normal distribution with identity covariance matrix $\mathcal{N}(0,KK^T) = \mathcal{N}(0,I_m)$.

To build the generative model parameterized by the principal anatomical components, we can reverse the linear system by employing the matrix $K^T$. Hence, the orthogonally constrained and anatomically parameterized statistical shape model (OC-ANAT$_{\text{SSM}}$) can be defined as follows:
\begin{align}
    \textit{OC-ANAT}_{\textit{SSM}} &= \mu + P D K^{T} \Tilde{\beta} \\
    &= \mu + \sum_{j=1}^{m} \Tilde{\beta}_{c_j} \sum_{i=1}^{n-1} K^{T}_{i,j} \sqrt{\lambda_i} \phi_i
\end{align}
\noindent in which the deformation vectors $(\sum_{i=1}^{n-1} K^{T}_{i,j} \sqrt{\lambda_i} \phi_i )_{1 \leq j \leq m}$ are independent. These vectors are closest to the shape deformations vectors of the ANAT$_{\text{SSM}}$, while subjected to the orthogonality constraints. Hence, the shape variation patterns arising from these vectors are independent.

\subsection{Shape variability induced by anatomical parameters}
\label{sec:shape_variability_anatomical_parameters}

In BASE$_{\text{SSM}}$, the shape variation corresponding to the $i$-th eigenvector is quantified by the eigenvalue $\lambda_i$, and the principal components are ordered from largest to smallest eigenvalues as explained in Section \ref{sec:overview_ssm}. Similarly, we define the shape variation induced by the $j$-th anatomical parameters of ANAT$_{\text{SSM}}$ and OC-ANAT$_{\text{SSM}}$ as follows:
\begin{align}
    \kappa_{c_j} &= \sum_{i=1}^{n-1} Q^{+^{2}}_{i,j} \lambda_i \\
    \Tilde{\kappa}_{c_j} &= \sum_{i=1}^{n-1} K^{T^{2}}_{i,j} \lambda_i
\end{align}
\noindent The shape variance $\kappa_{c_j}$ and $\Tilde{\kappa}_{c_j}$ of ANAT$_{\text{SSM}}$ and OC-ANAT$_{\text{SSM}}$ can be determined by the matrices $Q^{+}$ and $K^{T}$, and the eigenvalues $\lambda_i$. Therefore, for each anatomical parameter, a scalar value represents the shape variability induced by this parameter. Finally, we sorted the anatomical parameters in decreasing order (from largest to smallest) based on their corresponding variance, with the first anatomical parameter producing the largest shape variation.

\section{Experiments}
\label{sec:experiments}

In this section, we present the experiments conducted to assess and validate the proposed methods. First, we provide an overview of the femoral and scapular bone datasets employed and BASE$_{\text{SSM}}$ building process. We then summarize the frameworks to automatically derive  the five femoral and six scapular anatomical measurements using the BASE$_{\text{SSM}}$ and landmark tracking. The BASE$_{\text{SSM}}$ are subsequently employed to generate synthetic populations from which the anatomical measurements are automatically extracted. The generated synthetic populations enable the learning of the matrices $Q$ and $K$ and the creation of anatomically parameterized SSMs, ANAT$_{\text{SSM}}$ and OC-ANAT$_{\text{SSM}}$. Finally, we assess the characteristics of the obtained matrices and SSMs.

\subsection{Femoral and scapular statistical shape models}
\label{sec:ssm_femoral_scapular_bones}

Experiments were conducted on two human skeleton datasets of femoral and scapular bone shapes.

\noindent \textbf{Femoral dataset.} The femoral bone images were extracted from an open whole body CT scan dataset maintained by Sicas Medical Image Repository (SMIR) \cite{kistler_virtual_2013} and publicly available at (\url{https://www.smir.ch/}). This dataset consisted of $n = 50$ whole body CT scans acquired using SIEMENS SOMATOM Force Dual Source scanner (Siemens Healthcare, Germany) with a resolution of $0.99\times0.99\times0.50$ mm$^{3}$. The images were manually segmented by a medically trained annotator (years of experience = 12) to extract 3D surface models of femur bones (right only) with $N = 17000$ points.

\noindent \textbf{Scapular dataset.} The scapular dataset contained $n = 76$ samples previously acquired from the Department of Anatomy at the regional University Hospital (CHRU de Brest, France). CT scan images were acquired using the SIEMENS SOMATOM Definition AS scanner (Siemens Healthcare, Germany) with a resolution of $0.96\times0.96\times0.60$ mm$^{3}$. The images were evaluated for anatomical integrity and manually segmented by two radiologists (years of experience: R1 = 19 years, R2 = 12 years). Surface 3D mesh models with $N = 15000$ points were obtained using Amira software (Amira, FEI, Hillsboro, V5.4).

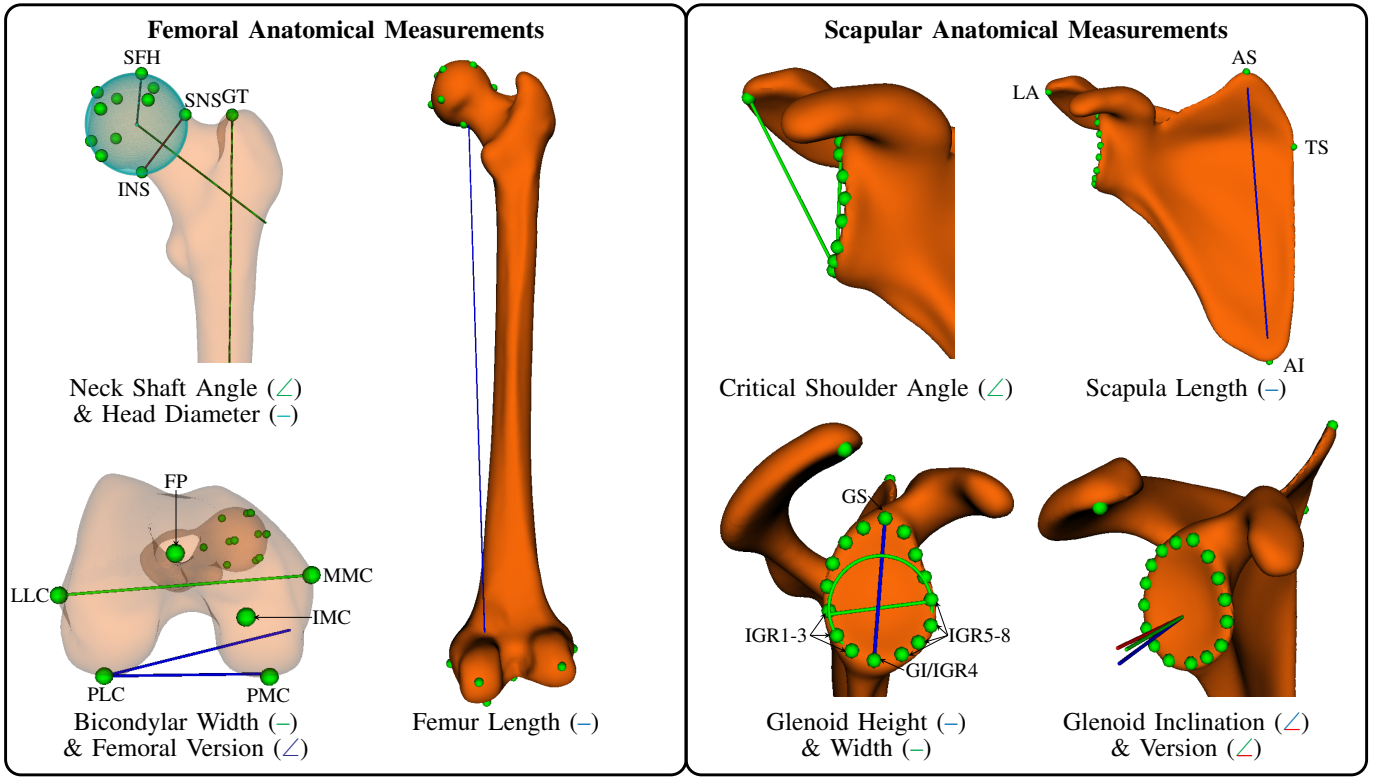
\begin{figure*}[t]
\centering
\begin{adjustbox}{width=\textwidth}
\begin{tikzpicture}

\node[inner sep=0pt] at (-.35,-.25)
    {\includegraphics[width=.205\textwidth]{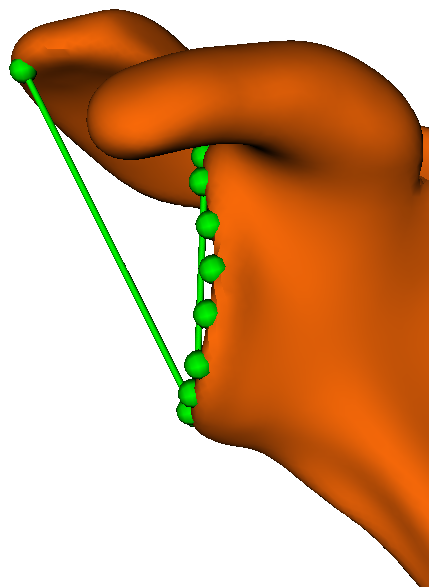}};
\node[inner sep=0pt, anchor=south] at (5.5,-8.6)
    {\includegraphics[width=.3\textwidth]{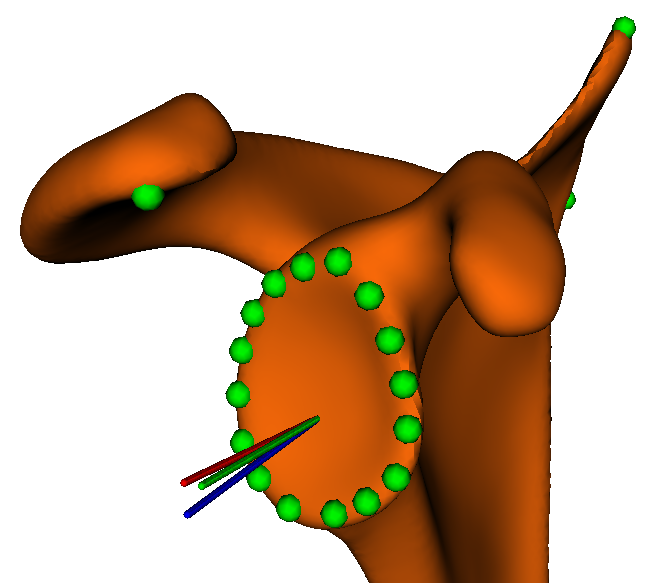}};
\node[inner sep=0pt,anchor=south] at (0,-8.6)
    {\includegraphics[width=.29\textwidth]{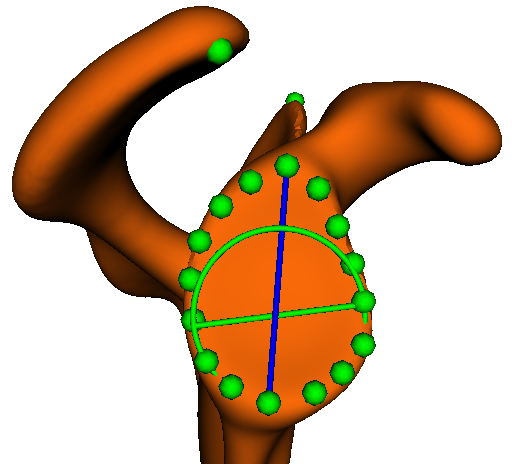}};
\node[inner sep=0pt] at (5.25,-.35)
    {\includegraphics[width=.25\textwidth]{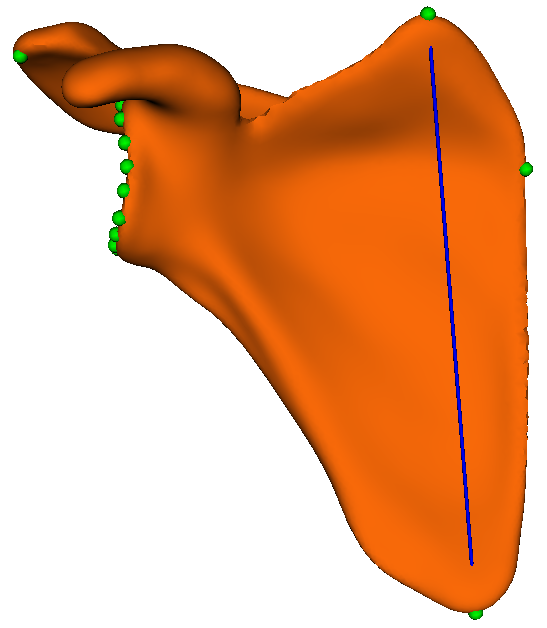}};
    
\node[inner sep=0pt] at (-11.7,-.25)
    {\includegraphics[width=.215\textwidth]{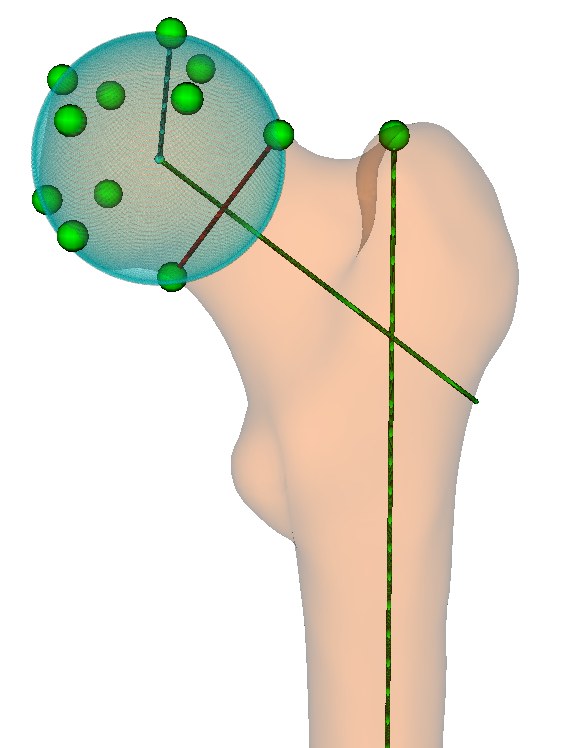}};
\node[inner sep=0pt] at (-6.2,-3.2)
    {\includegraphics[width=.16\textwidth]{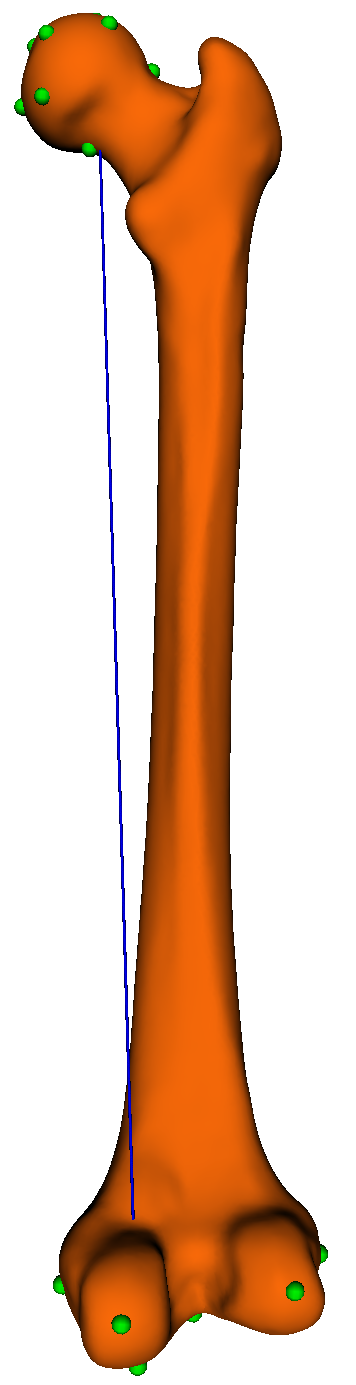}};
\node[inner sep=0pt] at (-11.7,-6.6)
    {\includegraphics[width=.275\textwidth]{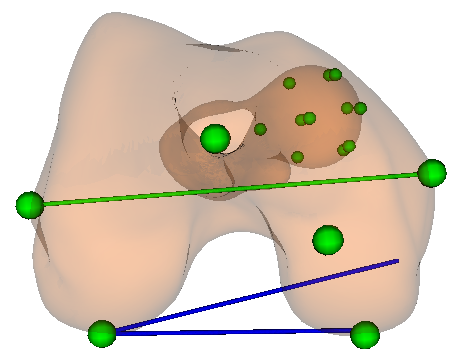}};

\node[anchor=north] at (2.75, 3.15) {\scalebox{1.25}{\textbf{Scapular Anatomical Measurements}}};
\node[anchor=north] at (0, -3) {\scalebox{1.25}{Critical Shoulder Angle ($\textcolor{Green}{\angle}$)}};
\node[anchor=north] at (5.5, -8.75) {\scalebox{1.25}{Glenoid Inclination ($\textcolor{NavyBlue}{\angle}$)}};
\node[anchor=north] at (5.5, -9.2) {\scalebox{1.25}{\& Version ($\textcolor{Green}{\angle}$)}};
\node[anchor=north] at (0, -8.75) {\scalebox{1.25}{Glenoid Height (\textcolor{NavyBlue}{\---})}};
\node[anchor=north] at (0, -9.2) {\scalebox{1.25}{\& Width (\textcolor{Green}{\---})}};
\node[anchor=north] at (5.5, -3) {\scalebox{1.25}{Scapula Length (\textcolor{NavyBlue}{\---})}};

\node at (7.75, .85) {\scalebox{1}{TS}};
\node at (6.5, 2.4) {\scalebox{1}{AS}};
\node at (2.75, 1.8) {\scalebox{1}{LA}};
\node at (7.35, -2.9) {\scalebox{1}{AI}};
\node at (1.3, -8.15) {\scalebox{1}{GI/IGR4}};
\node at (-.2, -5.15) {\scalebox{1}{GS}};
\node at (1.95, -7.55) {\scalebox{1}{IGR5-8}};
\node at (-1.55, -7.55) {\scalebox{1}{IGR1-3}};

\node at (-12.45, 2.4) {\scalebox{1}{SFH}};
\node at (-10.85, 1.72) {\scalebox{1}{GT}};
\node at (-11.4, 1.66) {\scalebox{1}{SNS}};
\node at (-12.59, 0.13) {\scalebox{1}{INS}};
\node at (-8.91, -6.52) {\scalebox{1}{MMC}};
\node at (-14.38, -6.85) {\scalebox{1}{LLC}};
\node at (-13.07, -8.56) {\scalebox{1}{PLC}};
\node at (-10.28, -8.58) {\scalebox{1}{PMC}};
\node at (-9.19, -7.24) {\scalebox{1}{IMC}};
\node at (-11.87, -4.9) {\scalebox{1}{FP}};

\draw[line width=0.1mm, stealth-] (.28, -5.4) -- (0.03, -5.15);
\draw[line width=0.1mm, stealth-] (.23, -7.96) -- (.65, -8.15);
\draw[line width=0.1mm, stealth-] (-11.87, -6.02) -- (-11.87, -5.05);
\draw[line width=0.1mm, stealth-] (-10.535, -7.24) -- (-9.53, -7.24);
\draw[line width=0.1mm, stealth-] (.73, -7.9) -- (1.4, -7.55);
\draw[line width=0.1mm, stealth-] (.99, -7.67) -- (1.4, -7.55);
\draw[line width=0.1mm, stealth-] (1.2, -7.37) -- (1.4, -7.55);
\draw[line width=0.1mm, stealth-] (1.2, -6.97) -- (1.4, -7.55);
\draw[line width=0.1mm, stealth-] (-.35, -7.83) -- (-.97, -7.55);
\draw[line width=0.1mm, stealth-] (-.62, -7.55) -- (-.97, -7.55);
\draw[line width=0.1mm, stealth-] (-.7, -7.17) -- (-.97, -7.55);
\draw[line width=0.2mm, red] (7.19,-9.17) -- (7.45,-9.17);
\draw[line width=0.2mm, red] (6.37,-9.62) -- (6.63,-9.62);

\node[anchor=north] at (-8.95, 3.15) {\scalebox{1.25}{\textbf{Femoral Anatomical Measurements}}};
\node[anchor=north] at (-11.7, -3) {\scalebox{1.25}{Neck Shaft Angle ($\textcolor{Green}{\angle}$)}};
\node[anchor=north] at (-11.7, -3.45) {\scalebox{1.25}{\& Head Diameter (\textcolor{BlueGreen}{\---})}};
\node[anchor=north] at (-6.2, -8.75) {\scalebox{1.25}{Femur Length (\textcolor{NavyBlue}{\---})}};
\node[anchor=north] at (-11.7, -8.75) {\scalebox{1.25}{Bicondylar Width (\textcolor{Green}{\---})}};
\node[anchor=north] at (-11.7, -9.2) {\scalebox{1.25}{\& Femoral Version ($\textcolor{Blue}{\angle}$)}};

\draw[line width=.5mm, color=black, rounded corners=7.5] (-3.1, 3.3) -- (8.6,3.3) -- (8.6,-9.95) -- (-3.1,-9.95) -- cycle;
\draw[line width=.5mm, color=black, rounded corners=7.5] (-3.1, 3.3) -- (-14.8,3.3) -- (-14.8,-9.95) -- (-3.1,-9.95) -- cycle;

\end{tikzpicture}
\end{adjustbox}
  \caption{Automatic derivation of femoral and scapular anatomical measurements based on 18 femoral and 20 scapular landmarks ($\textcolor{Green}{\bullet}$) selected on the mean shape of the respective BASE$_{\text{SSM}}$. Femoral anatomical measurements include: neck shaft angle (NSA), femoral version (FV), bicondylar width (BW), head diameter (HD) and femur length (FL), while scapular anatomical measurements encompass: critical shoulder angle (CSA), glenoid inclination, version, height and width (GI, GV, GH and GW) as well as scapula length (SL). Please refer to the supplementary material for definitions of the landmarks and measurements.}
  \label{fig:automatic_computation_anatomical_parameters}
\end{figure*}

For each anatomical structure, we used the IMCP-GMM (iterative median closest point-Gaussian mixture model) pipeline \cite{mutsvangwa_automated_2015} to create the BASE$_{\text{SSM}}$ which comprised three steps: 1) Rigid alignment of shapes using iterative median closest point (IMCP) algorithm and creating a virtual manifold \cite{jacq_performing_2008}, 2) non-rigid alignment of datasets to establish dense correspondence using a coherence point drift (CPD) algorithm  \cite{myronenko_point_2010} and 3) BASE$_{\text{SSM}}$ creation using GPMM as reported in \cite{salhi_statistical_2020}. After establishing point-to-point correspondence across all shapes in step 2), experiments were performed following a leave-one-out strategy, in which one shape was retained for evaluation and the remaining ones were employed to build the models. The BASE$_{\text{SSM}}$ were implemented using the open-source toolbox for scalable image analysis and shape modeling (\url{https://scalismo.org/}). Please refer to the supplementary material for an assessment of the robustness of the BASE$_{\text{SSM}}$.

\subsection{Automatic derivation of anatomical measurements from landmarks}
\label{sec:automatic_derivation_anatomical_parameters}

The femoral and scapular morphologies were characterized by sets of anatomical measurements selected based on their relevance to hip, knee and shoulder joint replacement procedures \cite{plessers_virtual_2018, salhi_statistical_2020, casciaro_towards_2014, verma_morphometry_2017}. For each anatomical structure, these anatomical measurements were automatically computed using a set of anatomical landmarks selected on the mean shape of the BASE$_{\text{SSM}}$ (Fig. \ref{fig:automatic_computation_anatomical_parameters}).

\noindent \textbf{Femoral anatomical measurements.} We employed $m = 5$ femoral anatomical measurements: neck shaft angle (NSA), femoral version (FV), bicondylar width (BW), head diameter (HD), and femur length (FL), which were automatically computed using a set of 18 anatomical landmarks \cite{verma_morphometry_2017, hartel_determination_2016, terzidis_gender_2012, casciaro_towards_2014, unnanuntana_evaluation_2010, wei_approach_2020}. Angulation (NSA), torsion (FV) and dimension (HD) of the proximal femur are essential for surgical planning of hip joint replacements \cite{verma_morphometry_2017, hartel_determination_2016, casciaro_towards_2014, unnanuntana_evaluation_2010, wei_approach_2020}, whereas BW is an important measurement for knee joint replacements \cite{terzidis_gender_2012} and FL provides an evaluation of the size of the femur \cite{verma_morphometry_2017}.

\noindent \textbf{Scapular anatomical measurements.} The scapular shape was characterized by a set of $m = 6$ anatomical measurements: critical shoulder angle (CSA), glenoid inclination (GI), glenoid version (GV), glenoid height (GH), glenoid width (GW), and scapula length (SL) using 20 landmarks \cite{von_schroeder_osseous_2001, cherchi_critical_2016, plessers_virtual_2018, verhaegen_can_2018, burton_assessment_2019}. Orientation (GI, GV) and size (GH, GW) of the glenoid are crucial for shoulder joint replacements \cite{plessers_virtual_2018, verhaegen_can_2018, burton_assessment_2019}, while CSA is a robust indicator of glenohumeral osteoarthritis \cite{cherchi_critical_2016} and SL assesses the global dimension of the scapula \cite{von_schroeder_osseous_2001}.

During leave-one-out evaluation, the selected anatomical landmarks on the mean shape were transferred to the retained shape by using the established point-to-point correspondence between the BASE$_{\text{SSM}}$ and the retained shape. The tracked landmarks were then used to automatically compute the anatomical measures for the retained shape. To determine the accuracy of the automatic computation of anatomical measures, we evaluated the absolute error between anatomical measures derived from manual (expert) landmarking and those derived from automatically transferred landmarks.

\subsection{Synthetic population generation}
\label{sec:synthetic_populations_generation}

To learn the matrices $Q$ and $K$ of each anatomical structure, we generated synthetic datasets of shapes with their corresponding shape coefficients and anatomical measures. These synthetic populations were obtained by using the femoral and scapular BASE$_{\text{SSM}}$ (developed in Section \ref{sec:ssm_femoral_scapular_bones}) as generative models parameterized by the shape coefficients $\alpha \sim \mathcal{N}(0, I_{n-1})$. Then, as a point-to-point correspondence between the BASE$_{\text{SSM}}$ and the synthetic shapes was already established, we could track the selected landmarks in each generated shape. Using the automatic methods described earlier (Section \ref{sec:automatic_derivation_anatomical_parameters}) we computed the anatomical measures in each generated shape. In total, we generated $1000$ synthetic shapes for each bone to learn the matrices $Q$ and $K$. The anatomical measures were then used as anatomical parameters for characterizing the shapes, mapping their relationship with the respective shape coefficients, and ultimately to build the proposed models. 

To determine if the distribution of the anatomical parameters among the synthetic shapes was represented by a normal distribution, we employed the Shapiro-Wilk normality test. This test computed the \textit{p}-value for the significance of normality of each marginal distribution (individual parameter distributions), which was compared against the typical significance level of $0.01$. We then computed the histogram and estimated the mean and variance of each marginal distribution. A normal distribution was fitted to each histogram based on the estimated mean and variance. For the rest of this study, the marginal distributions were normalized to have zero mean and unit variance. Furthermore, to assess the statistical relationship between the anatomical parameters, we computed the Pearson correlation coefficients between each pair of parameters. We also evaluated the correlation between shape coefficients and anatomical parameters using Pearson correlation coefficients. 

\subsection{Assessment of the learned matrices}
\label{sec:assessment_matrices}

To assess the learned matrices $Q$ and $K$, which represented the linear mapping from the shape coefficients to the anatomical parameters, we computed the mean absolute difference between the learned matrices weights and the Pearson correlation coefficients computed between the shape coefficients and anatomical parameters. Similarly, to evaluate the matrix $QQ^T$, which represented the learned covariance between anatomical parameter, we computed the mean absolute difference between the matrix $QQ^T$ weights and the Pearson correlation coefficients computed between anatomical parameters pairs. Finally, we assessed the orthogonality constraints by verifying that $KK^T$ was the identity matrix.

The generative models ANAT$_{\text{SSM}}$ and OC-ANAT$_{\text{SSM}}$ were then built using the matrices $Q$ and $K$ as formalized in Sections \ref{sec:anat_ssm} and \ref{sec:ind_anat_ssm}.

\subsection{Assessment of ANAT$_{\text{SSM}}$ and OC-ANAT$_{\text{SSM}}$}
\label{sec:assessment_anat_ssms}

The predictive performance of ANAT$_{\text{SSM}}$ and OC-ANAT$_{\text{SSM}}$ were assessed at each iteration of the leave-one-out evaluation by computing the absolute error between anatomical measures derived from manual landmarking and those obtained by both models on the retained shape. We used the correspondence between ANAT$_{\text{SSM}}$ and OC-ANAT$_{\text{SSM}}$ and the retained shape to automatically extract the anatomical parameters.

Furthermore, to assess and compare the shape variation patterns induced by the two models in each anatomical structure, we generated shapes by changing the value of the $j$-th anatomical parameter between $\pm3\sigma_{c_j}$. For each anatomical parameter, we performed visual comparison of the shape variation patterns of ANAT$_{\text{SSM}}$ and OC-ANAT$_{\text{SSM}}$. We also computed the shape variability ($\kappa_{c_j}$ and $\Tilde{\kappa}_{c_j}$) induced by the $j$-th anatomical parameters in the two models, and their corresponding sub-models. Sub-models were derived from the anatomical models by retaining the anatomical parameter with largest anatomical variability sequentially. More specifically, at each step of the ablation study, the sub-models ANAT$_{\setminus \beta_{c_j}}$ and OC-ANAT$_{\setminus \Tilde{\beta}_{c_j}}$ were obtained by removing the $j$-th column of the matrices $Q$ and $K$ associated with the $c_j$ anatomical measurements (i.e. sub-matrices). The obtained values were normalized by total BASE$_{\text{SSM}}$ shape variability ($\sum_{i=1}^{n-1}\lambda_i$). 

\section{Results}
\label{sec:results}

\begin{table*}[t]
\centering
\caption{Leave-one-out assessment of the absolute error between BASE$_{\text{SSM}}$, ANAT$_{\text{SSM}}$ and OC-ANAT$_{\text{SSM}}$ predictions and manually derived anatomical measurements. Mean, standard deviation (STD), maximum and minimum are reported.}
    \begin{tabular}{|C{.2cm}|C{1.4cm}||C{.8cm}|C{.8cm}|C{.8cm}|C{.8cm}||C{.8cm}|C{.8cm}|C{.8cm}|C{.8cm}||C{.8cm}|C{.8cm}|C{.8cm}|C{.8cm}|} 
    
    \hline
    \multicolumn{2}{|c||}{\multirow{2}{*}{Absolute Error}} & \multicolumn{4}{c||}{BASE$_{\text{SSM}}$} & \multicolumn{4}{c||}{ANAT$_{\text{SSM}}$} & \multicolumn{4}{c|}{OC-ANAT$_{\text{SSM}}$} \\ \cline{3-14}
    \multicolumn{2}{|c||}{} & mean & STD & min & max & mean & STD & min & max & mean & STD & min & max \\ 

    \hline \hline
    \multirow{5}{*}{\rotatebox[origin=c]{90}{Femur}} & NSA ($\degree$) & 1.7 & 1.3 & $<0.1$ & 5.3 & 2.1 & 1.5 & 0.1 & 6.7 & 2.0 & 1.6 & $<0.1$ & 6.3 \\\cline{2-14}
    & FV ($\degree$) & 1.7 & 1.4 & $<0.1$ & 6.7 & 2.7 & 2.2 & 0.1 & 9.9 & 3.1 & 2.6 & 0.1 & 10.1 \\\cline{2-14}
    & BW (mm) & 0.7 & 0.5 & $<0.1$ & 1.8 & 1.1 & 0.9 & $<0.1$ & 3.2 & 3.4 & 2.5 & 0.2 & 9.7 \\\cline{2-14}
    & HD (mm) & 0.5 & 0.3 & $<0.1$ & 1.4 & 1.0 & 0.8 & $<0.1$ & 3.9 & 2.2 & 1.6 & 0.1 & 7.7 \\\cline{2-14}
    & FL (cm) & 0.1 & 0.1 & $<0.1$ & 0.3 & 0.1 & 0.1 & $<0.1$ & 0.3 & 1.4 & 1.0 & $<0.1$ & 4.6  \\
    \hline\hline
    
    \multirow{6}{*}{\rotatebox[origin=c]{90}{Scapula}} & CSA ($\degree$) & 1.6 & 1.1 & $<0.1$ & 4.6 & 2.0 & 1.8 & $<0.1$ & 9.6 & 2.4 & 2.0 & 0.1 & 7.8 \\\cline{2-14}
    & GI ($\degree$) & 2.0 & 1.5 & 0.1 & 6.6 & 2.4 & 2.1 & 0.1 & 8.7 & 2.6 & 2.3 & 0.1 & 9.3 \\\cline{2-14}
    & GV ($\degree$) & 1.2 & 1.0 & $<0.1$ & 5.2 & 2.1 & 1.8 & $<0.1$ & 9.6 & 2.6 & 2.1 & $<0.1$ & 9.7 \\\cline{2-14}
    & GH (mm) & 1.4 & 1.2 & $<0.1$ & 5.7 & 1.5 & 1.2 & 0.1 & 6.2 & 2.4 & 1.7 & 0.1 & 8.6\\\cline{2-14}
    & GW (mm) & 1.0 & 1.1 & $<0.1$ & 7.0 & 1.3 & 1.4 & $<0.1$ & 8.5 & 1.8 & 1.6 & $<0.1$ & 8.6 \\\cline{2-14}
    & SL (mm) & 0.8 & 0.6 & $<0.1$ & 2.7 & 1.4 & 1.0 & 0.1 & 4.0 & 5.1 & 3.7 & 0.1 & 15.8 \\
    \hline

    \end{tabular}
\label{tab:automatic_computation_error}
\end{table*}

\begin{figure*}[t]
\begin{tikzpicture}

\begin{axis}[name=plot1,
        title={Critical Shoulder Angle ($\degree$)},
        title style={font=\footnotesize, yshift=-7.5},
        height=4cm,
        width=4cm, 
        xmin=15.5, 
        xmax=44.8, 
        ymin=0,
        ymax=0.11,
        xtick={20, 30, 40},
        xticklabels={20, 30, 40},
        ytick={0, 0.02, 0.04, 0.06, 0.08, 0.1},
        yticklabels={0, 0.02, 0.04, 0.06, 0.08, 0.1},
        ticklabel style = {font=\scriptsize, major tick length=2pt},
        yticklabel style = {xshift=2},
        xticklabel style = {yshift=2}]
\addplot[ybar interval, color=NavyBlue] 
table [x, y] {csa.dat};
\addplot[BrickRed, samples=100, domain=15.5:44.8] {1/(4.3*sqrt(2*pi)) * exp(-(x-29.3)^2/(2*4.3^2))};
\node[rotate=90, anchor=south, BrickRed, font=\scriptsize] at (axis cs: 41.8,0.055) {$\mu_{\scaleto{CSA}{3pt}}$=$29.3$};
\node[rotate=90, anchor=south, BrickRed, font=\scriptsize] at (axis cs: 44.8,0.055) {$\sigma_{\scaleto{CSA}{3pt}}$=$4.3$};
\node[rotate=90,anchor=north, NavyBlue, font=\scriptsize] at (axis cs: 15.5,0.055) {$p_{\scaleto{CSA}{3pt}}$=$0.17$};
\end{axis}

\begin{axis}[name=plot2,
        at={($(plot1.east)+(.60cm,0)$)},
        anchor=west,
        title={Glenoid Inclination ($\degree$)},
        title style={font=\footnotesize, yshift=-7.5},
        height=4cm,
        width=4cm, 
        xmin=-4.4, 
        xmax=21.4, 
        ymin=0,
        ymax=0.165,
        xtick={0, 10, 20},
        xticklabels={0, 10, 20},
        ytick={0, 0.03, 0.06, 0.09, 0.12, 0.15},
        yticklabels={0, 0.03, 0.06, 0.09, 0.12, 0.15},
        ticklabel style = {font=\scriptsize, major tick length=2pt},
        yticklabel style = {xshift=2},
        xticklabel style = {yshift=2}]
\addplot[ybar interval, color=NavyBlue] 
table [x, y] {inclination.dat};
\addplot[BrickRed, samples=100, domain=-4.4:21.4] {1/(4.1*sqrt(2*pi)) * exp(-(x-8.5)^2/(2*4.1^2))};
\node[rotate=90, anchor=south, BrickRed, font=\scriptsize] at (axis cs: 18.8,0.0825) {$\mu_{\scaleto{GI}{3pt}}$=$8.5$};
\node[rotate=90, anchor=south, BrickRed, font=\scriptsize] at (axis cs: 21.4,0.0825) {$\sigma_{\scaleto{GI}{3pt}}$=$4.1$};
\node[rotate=90,anchor=north, NavyBlue, font=\scriptsize] at (axis cs: -4.4,0.0825) {$p_{\scaleto{GI}{3pt}}$=$0.17$};
\end{axis}

\begin{axis}[name=plot3,
        at={($(plot2.east)+(.60cm,0)$)},
        anchor=west,
        title={Glenoid Version ($\degree$)},
        title style={font=\footnotesize, yshift=-7.5},
        height=4cm,
        width=4cm, 
        xmin=-27.3, 
        xmax=17.9,
        ymin=0,
        ymax=0.11,
        xtick={-20, -5, 10},
        xticklabels={-20, -5, 10},
        ytick={0, 0.02, 0.04, 0.06, 0.08, 0.1},
        yticklabels={0, 0.02, 0.04, 0.06, 0.08, 0.1},
        ticklabel style = {font=\scriptsize, major tick length=2pt},
        yticklabel style = {xshift=2},
        xticklabel style = {yshift=2}]
\addplot[ybar interval, color=NavyBlue] 
table [x, y] {version.dat};
\addplot[BrickRed, samples=100, domain=-27.3:17.9] {1/(6.8*sqrt(2*pi)) * exp(-(x+4.7)^2/(2*6.8^2))};
\node[rotate=90, anchor=south, BrickRed, font=\scriptsize] at (axis cs:13.3,0.055) {$\mu_{\scaleto{GV}{3pt}}$=$-4.7$};
\node[rotate=90, anchor=south, BrickRed, font=\scriptsize] at (axis cs: 17.9,0.055) {$\sigma_{\scaleto{GV}{3pt}}$=$6.8$};
\node[rotate=90,anchor=north,NavyBlue, font=\scriptsize] at (axis cs: -27.3,0.055) {$p_{\scaleto{GV}{3pt}}$=$0.53$};
\end{axis}

\begin{axis}[name=plot4,
        at={($(plot3.east)+(.60cm,0)$)},
        anchor=west,
        title={Glenoid Height (mm)},
        title style={font=\footnotesize, yshift=-7.5},
        height=4cm,
        width=4cm, 
        xmin=25.1, 
        xmax=48.3, 
        ymin=0,
        ymax=0.165,
        xtick={30, 35, 40, 45},
        xticklabels={30, 35, 40, 45},
        ytick={0, 0.03, 0.06, 0.09, 0.12, 0.15},
        yticklabels={0, 0.03, 0.06, 0.09, 0.12, 0.15},
        ticklabel style = {font=\scriptsize, major tick length=2pt},
        yticklabel style = {xshift=2},
        xticklabel style = {yshift=2}]
\addplot[ybar interval, color=NavyBlue] 
table [x, y] {glenoid_height.dat};
\addplot[BrickRed, samples=100, domain=25.1:48.3] {1/(3.4*sqrt(2*pi)) * exp(-(x-36.7)^2/(2*3.4^2))};
\node[rotate=90, anchor=south, BrickRed, font=\scriptsize] at (axis cs: 45.9,0.0825) {$\mu_{\scaleto{GH}{3pt}}$=$36.7$};
\node[rotate=90, anchor=south, BrickRed, font=\scriptsize] at (axis cs: 48.3,0.0825) {$\sigma_{\scaleto{GH}{3pt}}$=$3.4$};
\node[rotate=90,anchor=north,NavyBlue, font=\scriptsize] at (axis cs: 25.1,0.0825) {$p_{\scaleto{GH}{3pt}}$=$0.12$};
\end{axis}

\begin{axis}[name=plot5,
        at={($(plot4.east)+(.60cm,0)$)},
        anchor=west,
        title={Glenoid Width (mm)},
        title style={font=\footnotesize, yshift=-7.5},
        height=4cm,
        width=4cm, 
        xmin=19.1, 
        xmax=34.5, 
        ymin=0,
        ymax=0.22,
        xtick={20, 25, 30, 35},
        xticklabels={20, 25, 30, 35},
        ytick={0, 0.04, 0.08, 0.12, 0.16, 0.20},
        yticklabels={0, 0.04, 0.08, 0.12, 0.16, 0.20},
        ticklabel style = {font=\scriptsize, major tick length=2pt},
        yticklabel style = {xshift=2},
        xticklabel style = {yshift=2}]
\addplot[ybar interval, color=NavyBlue] 
table [x, y] {glenoid_width.dat};
\addplot[BrickRed, samples=100, domain=18.8:35] {1/(2.5*sqrt(2*pi)) * exp(-(x-26.8)^2/(2*2.5^2))};
\node[rotate=90, anchor=south, BrickRed, font=\scriptsize] at (axis cs: 32.9,0.11) {$\mu_{\scaleto{GW}{3pt}}$=$26.8$};
\node[rotate=90, anchor=south, BrickRed, font=\scriptsize] at (axis cs: 34.5,0.11) {$\sigma_{\scaleto{GW}{3pt}}$=$2.5$};
\node[rotate=90,anchor=north, NavyBlue, font=\scriptsize] at (axis cs: 19.1,0.11) {$p_{\scaleto{GW}{3pt}}$=$0.14$};
\end{axis}

\begin{axis}[name=plot6,
        at={($(plot5.east)+(.60cm,0)$)},
        anchor=west,
        title={Scapula Length (mm)},
        title style={font=\footnotesize, yshift=-7.5},
        height=4cm,
        width=4cm, 
        xmin=110.6, 
        xmax=195, 
        ymin=0,
        ymax=0.055,
        xtick={120, 150, 180},
        xticklabels={120, 150, 180},
        ytick={0, 0.01, 0.02, 0.03, 0.04, 0.05},
        yticklabels={0, 0.01, 0.02, 0.03, 0.04, 0.05},
        ticklabel style = {font=\scriptsize, major tick length=2pt},
        yticklabel style = {xshift=2},
        xticklabel style = {yshift=2},
        scaled y ticks = false]
\addplot[ybar interval, color=NavyBlue] 
table [x, y] {scapula_length.dat};
\addplot[BrickRed, samples=100, domain=110.6:195] {1/(12.8*sqrt(2*pi)) * exp(-(x-152.8)^2/(2*12.8^2))};
\node[rotate=90, anchor=south, BrickRed, font=\scriptsize] at (axis cs: 186.4,0.0275) {$\mu_{\scaleto{SL}{3pt}}$=$152.8$};
\node[rotate=90, anchor=south, BrickRed, font=\scriptsize] at (axis cs: 195,0.0275) {$\sigma_{\scaleto{SL}{3pt}}$=$12.8$};
\node[rotate=90,anchor=north, NavyBlue, font=\scriptsize] at (axis cs: 110.6,0.0275) {$p_{\scaleto{SL}{3pt}}$=$0.99$};
\end{axis}

\begin{axis}[name=plot7,
        at={($(plot1.east)+(.3cm,2.3cm)$)},
        anchor=south,
        title={Neck Shaft Angle ($\degree$)},
        title style={font=\footnotesize, yshift=-7.5},
        height=4cm,
        width=4cm, 
        xmin=118.1, 
        xmax=135.5, 
        ymin=0,
        ymax=0.22,
        xtick={120, 125, 130, 135},
        xticklabels={120, 125, 130, 135},
        ytick={0, 0.04, 0.08, 0.12, 0.16, 0.20},
        yticklabels={0, 0.04, 0.08, 0.12, 0.16, 0.20},
        ticklabel style = {font=\scriptsize, major tick length=2pt},
        yticklabel style = {xshift=2},
        xticklabel style = {yshift=2}]
\addplot[ybar interval, color=NavyBlue] 
table [x, y] {nsa.dat};
\addplot[BrickRed, samples=100, domain=118.1:135.5] {1/(2.4*sqrt(2*pi)) * exp(-(x-126.8)^2/(2*2.4^2))};
\node[rotate=90, anchor=south, BrickRed, font=\scriptsize] at (axis cs: 133.7,0.11) {$\mu_{\scaleto{NSA}{3pt}}$=$126.8$};
\node[rotate=90, anchor=south, BrickRed, font=\scriptsize] at (axis cs: 135.5,0.11) {$\sigma_{\scaleto{NSA}{3pt}}$=$2.4$};
\node[rotate=90,anchor=north, NavyBlue, font=\scriptsize] at (axis cs: 118.1,0.11) {$p_{\scaleto{NSA}{3pt}}$=$0.62$};
\end{axis}

\begin{axis}[name=plot8,
        at={($(plot7.east)+(.60cm,0)$)},
        anchor=west,
        title={Femoral Version ($\degree$)},
        title style={font=\footnotesize, yshift=-7.5},
        height=4cm,
        width=4cm, 
        xmin=-18.6, 
        xmax=45.4, 
        ymin=0,
        ymax=0.11,
        xtick={-10, 10, 30},
        xticklabels={-10, 10, 30},
        ytick={0, 0.02, 0.04, 0.06, 0.08, 0.10},
        yticklabels={0, 0.02, 0.04, 0.06, 0.08, 0.10},
        ticklabel style = {font=\scriptsize, major tick length=2pt},
        yticklabel style = {xshift=2},
        xticklabel style = {yshift=2},
        scaled y ticks = false]
\addplot[ybar interval, color=NavyBlue] 
table [x, y] {femur_version.dat};
\addplot[BrickRed, samples=100, domain=-18.6:45.4] {1/(8.7*sqrt(2*pi)) * exp(-(x-13.4)^2/(2*8.7^2))};
\node[rotate=90, anchor=south, BrickRed, font=\scriptsize] at (axis cs: 38.8,.055) {$\mu_{\scaleto{FV}{3pt}}$=$13.4$};
\node[rotate=90, anchor=south, BrickRed, font=\scriptsize] at (axis cs: 45.4,0.055) {$\sigma_{\scaleto{FV}{3pt}}$=$8.7$};
\node[rotate=90,anchor=north,NavyBlue, font=\scriptsize] at (axis cs: -18.6,0.055) {$p_{\scaleto{FV}{3pt}}$=$0.87$};
\end{axis}

\begin{axis}[name=plot9,
        at={($(plot8.east)+(.60cm,0)$)},
        anchor=west,
        title={Bicondylar Width (mm)},
        title style={font=\footnotesize, yshift=-7.5},
        height=4cm,
        width=4cm, 
        xmin=62.4, 
        xmax=102, 
        ymin=0,
        ymax=0.11,
        xtick={70, 80, 90, 100},
        xticklabels={70, 80, 90, 100},
        ytick={0, 0.02, 0.04, 0.06, 0.08, 0.10},
        yticklabels={0, 0.02, 0.04, 0.06, 0.08, 0.10},
        ticklabel style = {font=\scriptsize, major tick length=2pt},
        yticklabel style = {xshift=2},
        xticklabel style = {yshift=2}]
\addplot[ybar interval, color=NavyBlue] 
table [x, y] {bicondylar_width.dat};
\addplot[BrickRed, samples=100, domain=62.4:102] {1/(6.4*sqrt(2*pi)) * exp(-(x-82.2)^2/(2*6.4^2))};
\node[rotate=90, anchor=south, BrickRed, font=\scriptsize] at (axis cs: 97.9,0.055) {$\mu_{\scaleto{BW}{3pt}}$=$82.2$};
\node[rotate=90, anchor=south, BrickRed, font=\scriptsize] at (axis cs: 102,0.055) {$\sigma_{\scaleto{BW}{3pt}}$=$6.4$};
\node[rotate=90,anchor=north,NavyBlue, font=\scriptsize] at (axis cs: 62.4,0.055) {$p_{\scaleto{BW}{3pt}}$=$0.73$};
\end{axis}

\begin{axis}[name=plot10,
        at={($(plot9.east)+(.60cm,0)$)},
        anchor=west,
        title={Head Diameter (mm)},
        title style={font=\footnotesize, yshift=-7.5},
        height=4cm,
        width=4cm, 
        xmin=36.9, 
        xmax=58.7, 
        ymin=0,
        ymax=0.165,
        xtick={40, 45, 50, 55},
        xticklabels={40, 45, 50, 55},
        ytick={0, 0.03, 0.06, 0.09, 0.12, 0.15},
        yticklabels={0, 0.03, 0.06, 0.09, 0.12, 0.15},
        ticklabel style = {font=\scriptsize, major tick length=2pt},
        yticklabel style = {xshift=2},
        xticklabel style = {yshift=2}]
\addplot[ybar interval, color=NavyBlue] 
table [x, y] {head_diameter.dat};
\addplot[BrickRed, samples=100, domain=36.9:58.7] {1/(3.8*sqrt(2*pi)) * exp(-(x-47.8)^2/(2*3.8^2))};
\node[rotate=90, anchor=south, BrickRed, font=\scriptsize] at (axis cs: 56.5,0.0825) {$\mu_{\scaleto{HD}{3pt}}$=$47.8$};
\node[rotate=90, anchor=south, BrickRed, font=\scriptsize] at (axis cs: 58.7,0.0825) {$\sigma_{\scaleto{HD}{3pt}}$=$3.8$};
\node[rotate=90,anchor=north, NavyBlue, font=\scriptsize] at (axis cs: 36.9,0.0825) {$p_{\scaleto{HD}{3pt}}$=$0.38$};
\end{axis}

\begin{axis}[name=plot11,
        at={($(plot10.east)+(.60cm,0)$)},
        anchor=west,
        title={Femur Length (cm)},
        title style={font=\footnotesize, yshift=-7.5},
        height=4cm,
        width=4cm, 
        xmin=36, 
        xmax=54, 
        ymin=0,
        ymax=0.165,
        xtick={40, 45, 50},
        xticklabels={40, 45, 50},
        ytick={0, 0.03, 0.06, 0.09, 0.12, 0.15},
        yticklabels={0, 0.03, 0.06, 0.09, 0.12, 0.15},
        ticklabel style = {font=\scriptsize, major tick length=2pt},
        yticklabel style = {xshift=2},
        xticklabel style = {yshift=2},
        scaled y ticks = false]
\addplot[ybar interval, color=NavyBlue] 
table [x, y] {femur_length.dat};
\addplot[BrickRed, samples=100, domain=36:54] {1/(3.1*sqrt(2*pi)) * exp(-(x-45)^2/(2*3.1^2))};
\node[rotate=90, anchor=south, BrickRed, font=\scriptsize] at (axis cs: 52.2,0.0825) {$\mu_{\scaleto{FL}{3pt}}$=$45.0$};
\node[rotate=90, anchor=south, BrickRed, font=\scriptsize] at (axis cs: 54,0.0825) {$\sigma_{\scaleto{FL}{3pt}}$=$3.1$};
\node[rotate=90,anchor=north, NavyBlue, font=\scriptsize] at (axis cs: 36,0.0825) {$p_{\scaleto{FL}{3pt}}$=$0.29$};
\end{axis}

\node at (8.75,6.55) {\scalebox{1}{Femur}};
\node at (8.75,3) {\scalebox{1}{Scapula}};

\end{tikzpicture}
\caption{Histograms of femoral and scapular marginal (individual) anatomical parameters (\textcolor{NavyBlue}{\---}) derived from the synthetically generated populations. The \textit{p}-values ($p_{NSA}, ..., p_{SL}$) assessed the normality of the distributions while normal distributions (\textcolor{BrickRed}{\---}) were fitted based on estimated mean ($\mu_{NSA}, ..., \mu_{SL}$) and variance ($\sigma_{NSA}, ..., \sigma_{SL}$).}
\label{fig:anatomical_parameters_histograms}
\end{figure*}
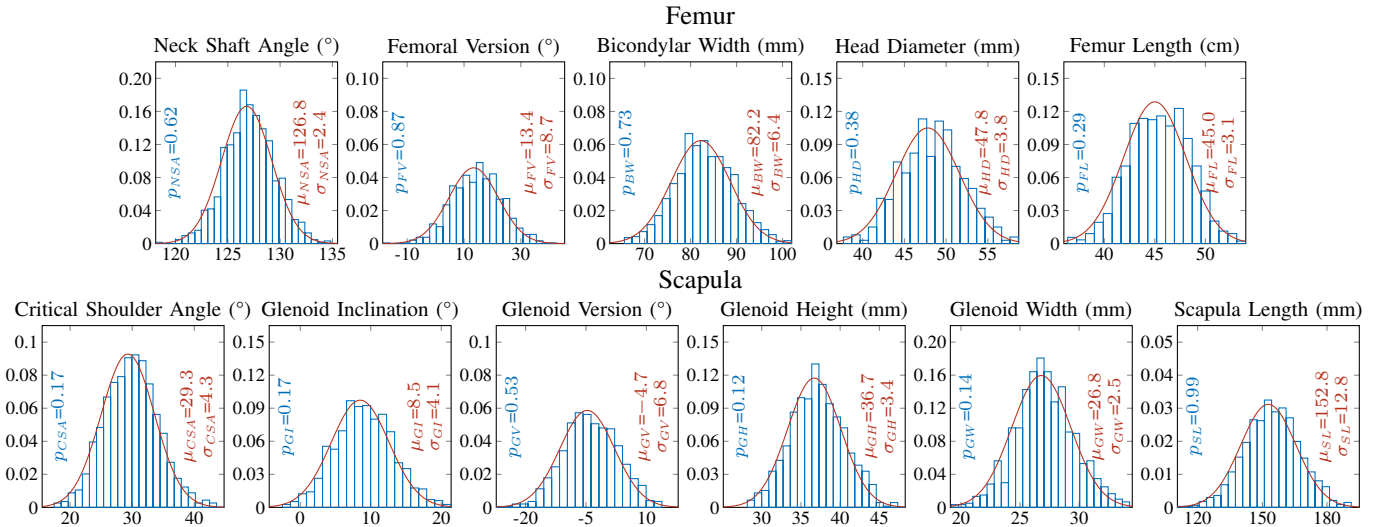{}

\subsection{Accuracy of automatic derivation of anatomical measurements}
\label{sec:validation_automatic_computation}

Automatic determination of anatomical measurements achieved moderate to excellent accuracy in comparison with manual measurements (Table \ref{tab:automatic_computation_error}). In both anatomical structures, angle measures were harder to predict as compared to size measurements. For angle measurements, the mean absolute error of the NSA, FV, CSA, GI and GV measures were comparable ($1.7\degree$, $1.7\degree$, $1.6\degree$, $2.0\degree$ and $1.2\degree$ respectively). Automatic computation of femoral size measures (BW, HD and FL) was found excellent with mean absolute errors below $1.0$ mm ($0.7$ mm, $0.5$ mm and $1.0$ mm respectively). For scapular size measurements, the mean absolute error of the GH measure was highest ($1.4$ mm) while GW and SL measures performed only marginally lower ($1.0$ mm and $0.8$ mm respectively). Prediction of FV and GI measures were the least robust with a standard deviation of $1.4\degree$ ($1.5\degree$ respectively) and a maximum absolute error of $6.7\degree$ ($6.6\degree$ respectively).

\begin{table}[t]
\centering
\caption{Pearson correlation coefficients between anatomical parameters in synthetic populations.}
\begin{subtable}{.5\textwidth}
\centering
\caption{Correlation in femoral synthetic population.}
\pgfplotstabletypeset[%
    color cells={min=-1,max=1,textcolor=black},
    /pgfplots/colormap={orangewhiteorange}{rgb255=(255,170,0) color=(white) rgb255=(255,170,0)},
    /pgf/number format/fixed,
    /pgf/number format/precision=3,
    col sep=comma,
    columns/Corr./.style={reset styles,string type, column type=|C{.75cm}},
    before row=\hline,every last row/.style={after row=\hline},
    columns/NSA/.style={column type=|C{.75cm}, column name=$\beta_{NSA}$},
    columns/FV/.style={column type=|C{.75cm}, column name=$\beta_{FV}$},
    columns/BW/.style={column type=|C{.75cm}, column name=$\beta_{BW}$},
    columns/HD/.style={column type=|C{.75cm}, column name=$\beta_{HD}$},
    columns/FL/.style={column type=|C{.75cm}|, column name=$\beta_{FL}$}
]{
Corr., NSA, FV, BW, HD, FL 
$\beta_{NSA}$, 1, 0.37, 0.07, 0.21, 0.26 
$\beta_{FV}$, 0.37, 1, -0.06, 0.03, 0.07
$\beta_{BW}$, 0.07, -0.06, 1, 0.89, 0.79
$\beta_{HD}$, 0.21, 0.03, 0.89, 1, 0.85
$\beta_{FL}$, 0.26, 0.07, 0.79, 0.85, 1
}
\label{tab:correlation_anatomical_parameters_femur}
\end{subtable}

\bigskip
\begin{subtable}{.5\textwidth}
\centering
\caption{Correlation in scapular synthetic population.}
\pgfplotstabletypeset[%
    color cells={min=-1,max=1,textcolor=black},
    /pgfplots/colormap={orangewhiteorange}{rgb255=(255,170,0) color=(white) rgb255=(255,170,0)},
    /pgf/number format/fixed,
    /pgf/number format/precision=3,
    col sep=comma,
    columns/Corr./.style={reset styles,string type, column type=|C{.75cm}},
    before row=\hline,every last row/.style={after row=\hline},
    columns/CSA/.style={column type=|C{.75cm}, column name=$\beta_{CSA}$},
    columns/GI/.style={column type=|C{.75cm}, column name=$\beta_{GI}$},
    columns/GV/.style={column type=|C{.75cm}, column name=$\beta_{GV}$},
    columns/GH/.style={column type=|C{.75cm}, column name=$\beta_{GH}$},
    columns/GW/.style={column type=|C{.75cm}, column name=$\beta_{GW}$},
    columns/SL/.style={column type=|C{.75cm}|, column name=$\beta_{SL}$}
]{
Corr., CSA, GI, GV, GH, GW, SL
$\beta_{CSA}$, 1, 0.43, 0.47, -0.51, -0.46, -0.24
$\beta_{GI}$, 0.43, 1, 0.01, -0.37, -0.22, -0.04
$\beta_{GV}$, 0.47, 0.01, 1, -0.08, -0.33, -0.16
$\beta_{GH}$, -0.51, -0.37, -0.08, 1, 0.67, 0.71
$\beta_{GW}$, -0.46, -0.22, -0.33, 0.67, 1, 0.59
$\beta_{SL}$, -0.24, -0.04, -0.16, 0.71, 0.59, 1 
}
\label{tab:correlation_anatomical_parameters_scapula}
\end{subtable}
\label{tab:correlation_anatomical_parameters}
\end{table}

\begin{table}[t]
\centering
\caption{Pearson correlation coefficients between anatomical parameters and shape coefficients in synthetic populations. Only the first 15 shape coefficients are reported.}
\begin{subtable}{.5\textwidth}
\centering
\caption{Correlation in femoral synthetic population.}
\pgfplotstabletypeset[%
    color cells={min=-1,max=1,textcolor=black},
    /pgfplots/colormap={orangewhiteorange}{rgb255=(255,170,0) color=(white) rgb255=(255,170,0)},
    /pgf/number format/fixed,
    /pgf/number format/precision=3,
    col sep=comma,
    columns/Corr./.style={reset styles,string type, column type=|C{.75cm}},
    before row=\hline,every last row/.style={after row=\hline},
    columns/NSA/.style={column type=|C{.75cm}, column name=$\beta_{NSA}$},
    columns/FV/.style={column type=|C{.75cm}, column name=$\beta_{FV}$},
    columns/BW/.style={column type=|C{.75cm}, column name=$\beta_{BW}$},
    columns/HD/.style={column type=|C{.75cm}, column name=$\beta_{HD}$},
    columns/FL/.style={column type=|C{.75cm}|, column name=$\beta_{FL}$}
]{
Corr., NSA, FV, BW, HD, FL
$\alpha_{1}$, -0.18, -0.02, -0.8, -0.85, -0.99
$\alpha_{2}$, 0.58,	0.36, -0.14, -0.1, 0.08
$\alpha_{3}$, -0.23, -0.33, -0.15, -0.14, -0.01
$\alpha_{4}$, -0.15, -0.55, -0.05, -0.14, -0.02
$\alpha_{5}$, 0.33, 0.48, -0.28, -0.18, 0.03
$\alpha_{6}$, -0.15, 0.04, -0.34, -0.18, 0.02
$\alpha_{7}$, 0.21, -0.33, 0.1, 0, 0.03
$\alpha_{8}$, -0.09, 0.23, 0.13, 0.05, -0.02
$\alpha_{9}$, 0.24, -0.1, -0.14, -0.04, 0.02
$\alpha_{10}$, 0.24, 0.01, -0.12, -0.01, -0.02
$\alpha_{11}$, 0.07, 0.07, 0.1, 0.02, -0.04
$\alpha_{12}$, -0.03, -0.01, 0.02, 0.01, 0
$\alpha_{13}$, -0.04, -0.04, -0.06, 0.01, -0.03
$\alpha_{14}$, 0.04, -0.01, 0, 0.14, -0.01
$\alpha_{15}$, -0.05, -0.03, 0.01, 0.02, 0.03
}
\end{subtable}

\bigskip
\begin{subtable}{.5\textwidth}
\centering
\caption{Correlation in scapular synthetic population.}
\pgfplotstabletypeset[%
    color cells={min=-1,max=1,textcolor=black},
    /pgfplots/colormap={orangewhiteorange}{rgb255=(255,170,0) color=(white) rgb255=(255,170,0)},
    /pgf/number format/fixed,
    /pgf/number format/precision=3,
    col sep=comma,
    columns/Corr./.style={reset styles,string type, column type=|C{.75cm}},
    before row=\hline,every last row/.style={after row=\hline},
    columns/CSA/.style={column type=|C{.75cm}, column name=$\beta_{CSA}$},
    columns/GI/.style={column type=|C{.75cm}, column name=$\beta_{GI}$},
    columns/GV/.style={column type=|C{.75cm}, column name=$\beta_{GV}$},
    columns/GH/.style={column type=|C{.75cm}, column name=$\beta_{GH}$},
    columns/GW/.style={column type=|C{.75cm}, column name=$\beta_{GW}$},
    columns/SL/.style={column type=|C{.75cm}|, column name=$\beta_{SL}$}
]{
Corr., CSA, GI, GV, GH, GW, SL
$\alpha_{1}$, -0.27, -0.25, -0.17, 0.76, 0.72, 0.89
$\alpha_{2}$, -0.27, 0.26, -0.16, 0.27, 0.09, 0.41
$\alpha_{3}$, 0.18, 0.12, 0, -0.05, -0.15, -0.07
$\alpha_{4}$, -0.24, 0.28, -0.22, -0.13, -0.01, 0.04
$\alpha_{5}$, -0.02, 0.28, -0.35, -0.20, 0.29, 0.02
$\alpha_{6}$, -0.51, -0.17, -0.33, 0.26, 0.23, -0.13
$\alpha_{7}$, 0.17, 0.33, 0.13, -0.22, -0.10, 0.02
$\alpha_{8}$, -0.09, -0.10, -0.20, -0.01, 0.16, -0.03
$\alpha_{9}$, -0.08, 0.08, 0.33, 0.11, 0.03, -0.09
$\alpha_{10}$, 0.07, 0.28, 0.19, 0.08, 0.07, 0.01
$\alpha_{11}$, -0.07, -0.22, -0.03, 0.15, 0.07, -0.06
$\alpha_{12}$, 0.19, 0.16, 0.04, -0.19, -0.19, 0.04
$\alpha_{13}$, 0.3, 0.13, 0.3, 0.07, 0.01, 0.05
$\alpha_{14}$, 0.07, 0.05, -0.17, -0.16, -0.01, 0.01
$\alpha_{15}$, -0.04, 0.16, -0.22, 0.09, -0.14, 0.05
}
\end{subtable}
\label{tab:correlation_anatomical_parameters_coefficients}
\end{table}

\subsection{Assessment of synthetically generated population}
\label{sec:synthetic_population_characteristics}

Each anatomical parameter passed the test of normality with \textit{p}-value greater than the significance level of $0.01$ (Fig. \ref{fig:anatomical_parameters_histograms}). In both datasets, it was confirmed that the mean values of the anatomical parameters ($\mu_{NSA}, ..., \mu_{FL}$ and $\mu_{CSA}, ..., \mu_{SL}$) corresponded with the anatomical measurements obtained on the mean shape of the femoral and scapular BASE$_{\text{SSM}}$, while the variance ($\sigma_{NSA}, ..., \sigma_{FL}$ and $\sigma_{CSA}, ..., \sigma_{SL}$) represented the variability of the generative models (Fig. \ref{fig:anatomical_parameters_histograms}).

The Pearson correlation coefficients between anatomical parameters of femoral and scapular synthetic data ranged from $0.01$ to $0.89$ with both positive and negative correlations (Table \ref{tab:correlation_anatomical_parameters}). The highest correlation value ($\rho = 0.89$) was found between femoral BW-HD pairs while the lowest value ($\rho = 0.01$) was found between GI and GV scapular measures. In both anatomical structures, the correlation coefficients between size measurements were high and positive ($\rho \geq 0.59$), while angles measures were less correlated ($\rho \leq 0.47$). In the scapular population, all size measures were negatively correlated with angle measurements.

The Pearson correlation coefficients between shape coefficients from BASE$_{\text{SSM}}$ and anatomical parameters from generated populations expectedly showed that in each bone, each anatomical parameter was correlated with multiple shape coefficients (Table \ref{tab:correlation_anatomical_parameters_coefficients}). Furthermore, the statistical relationships were sparse as $79\%$ of the femoral and $76\%$ of the scapular correlation coefficients were close to zero ($|\rho| \leq 0.1$). In both anatomical structures, the first shape coefficient was highly correlated with size measurements ($|\rho| \geq 0.72$), while being marginally correlated with angles measures ($|\rho| \leq 0.27$).

\subsection{Assessment of the learned matrices}
\label{sec:validation_matrices}

The learned weights of the femoral and scapular matrices $Q$ and $K$ were similar to the Pearson correlation coefficients previously obtained (Table \ref{tab:correlation_anatomical_parameters_coefficients}), with a mean absolute difference lower than $0.05$. This provided an indirect validation that the learned matrices correctly integrated the statistical relationship between shape coefficients and anatomical parameters computed in the synthetic populations. Furthermore, the matrices were sparse with at least $68\%$ of the learned weights close to zero ($|.|\leq 0.1$). This characteristic was also noted for the Pearson correlation coefficients in Section \ref{sec:synthetic_population_characteristics}. Finally, it was confirmed that in both structure the matrix $Q$ was of rank $m$.

The anatomical parameter covariance matrices  $QQ^T$ and $KK^T$ were computed and compared with the Pearson correlation coefficient of the synthetic data (Table \ref{tab:correlation_anatomical_parameters}). The weights of the femoral and scapular covariance $QQ^T$ were close ($0.01$ and $0.02$ mean absolute difference) to the reported Pearson correlation coefficients indicating that the learned matrices correctly integrated the statistical relationship between anatomical parameters computed in both synthetic populations. Second, in each anatomical structure, we performed the sanity check to confirm that the introduction of orthogonality constraints led to an identity covariance matrix $KK^T$. Please refer to the supplementary material for the weights of the matrices.

\subsection{Assessment of ANAT$_{\text{SSM}}$ and OC-ANAT$_{\text{SSM}}$}
\label{sec:validation_anat_ssm}

Both ANAT$_{\text{SSM}}$ and OC-ANAT$_{\text{SSM}}$ achieved satisfactory predictive performance compared with manual measurements on each anatomical structure (Table \ref{tab:automatic_computation_error}). Firstly, both ANAT$_{\text{SSM}}$ and OC-ANAT$_{\text{SSM}}$ achieved lower predictive performance than BASE$_{\text{SSM}}$ for every anatomical measurement. This was expected as both anatomical models were built on synthetic data generated by BASE$_{\text{SSM}}$ which were used as proxy for real measurements. For this reason, we expected ANAT$_{\text{SSM}}$ and OC-ANAT$_{\text{SSM}}$ to perform equally or worse than BASE$_{\text{SSM}}$ on real external shapes. More specifically, we observed maximum error for extreme measures (e.g. SL $= 180$ mm), which were scarce in the synthetically generated population. Hence, ANAT$_{\text{SSM}}$ and ANAT$_{\text{SSM}}$ did not generalize well on these extreme shapes while the method based on BASE$_{\text{SSM}}$ and landmark tracking did not suffer from this limitation. Secondly, ANAT$_{\text{SSM}}$ outperformed OC-ANAT$_{\text{SSM}}$ for every anatomical measurements except for neck shaft angle ($2.1\degree$ and $2.0\degree$ error). For both SSMs, the maximum angle measurement error was reported for femoral version ($9.9\degree$ and $10.1\degree$) while the maximum size measurements error was reported for femur length ($0.3$ cm and $15.8$ cm). Hence, the additional constraints on OC-ANAT$_{\text{SSM}}$ reduced its predictive performance compared to the ANAT$_{\text{SSM}}$. This was expected as $Q$ corresponded to the best fitted solution. Please refer to the supplementary material for further evaluation of the predictive performance.

We then visually confirmed that each anatomical parameter variation induced the correct modification in femoral and scapular shapes (Fig. \ref{fig:visual_comparison_shape_variation}). With regards to the femoral parameters, the NSA altered the orientation of the femoral head with regards to the femoral shaft, while the FV changed its orientation with respect to the femoral condyles. The BW controlled the size of the condyles, whereas the HD modified the size of the femoral head and the FL changed the size of the femur. For the scapular shape, the CSA changed the orientation of the acromion with regards to the glenoid, while the GI and GV modified the glenoid orientation. The glenoid dimension was altered by both GH and GW anatomical parameters, whereas the SL changed the size of the scapula. We also visually validated the constraints added to the OC-ANAT$_{\text{SSM}}$ led to independent shape variation patterns. For instance, the variation of the GW less altered the size of the scapula in OC-ANAT$_{\text{SSM}}$ as compared to ANAT$_{\text{SSM}}$. Please refer to the supplementary material for a direct evaluation of orthogonality and video demonstrations of the shape variation patterns.

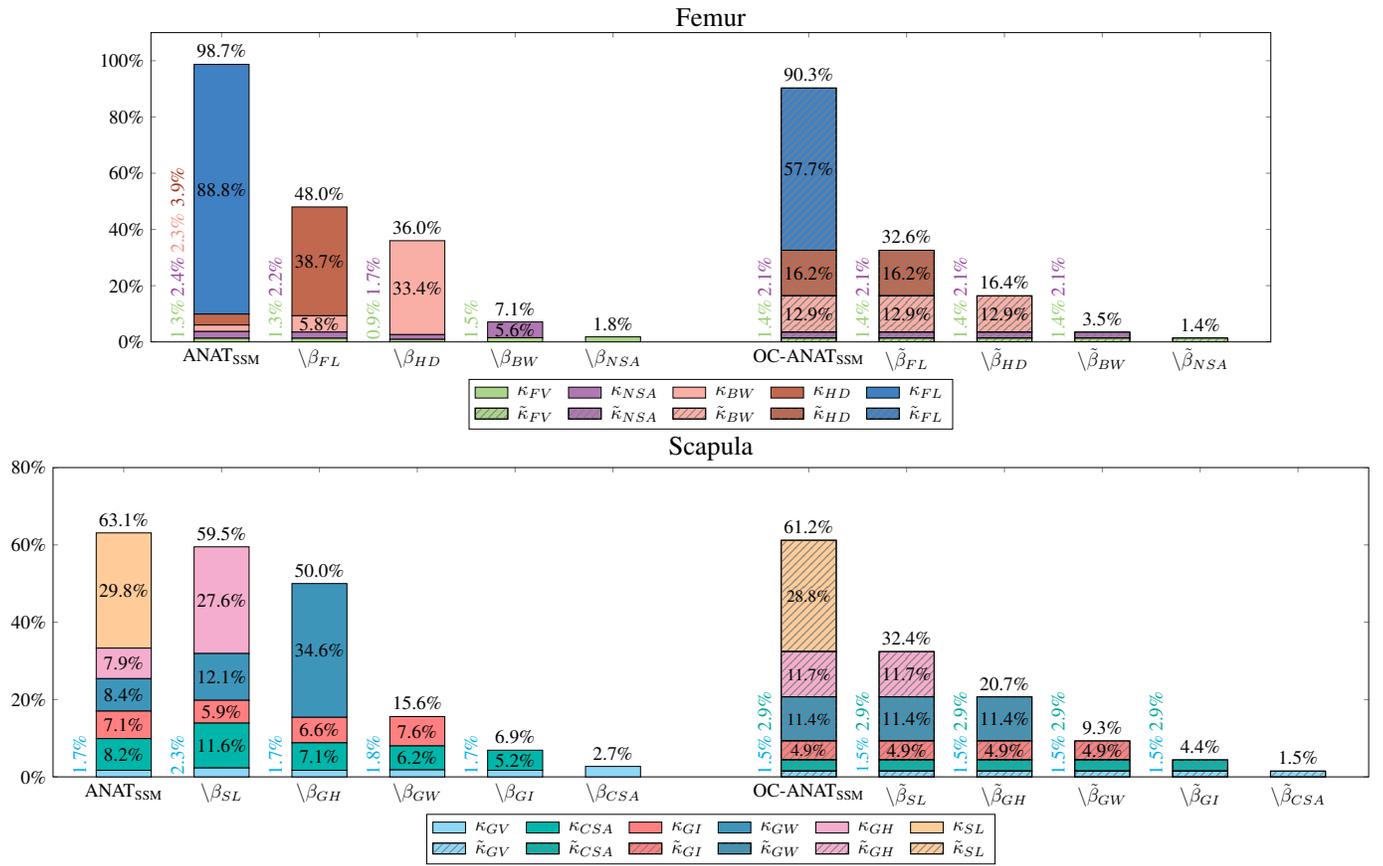
\begin{figure*}[t]
\centering
\begin{adjustbox}{width=\textwidth}
\begin{tikzpicture}
\begin{axis}[
    name=plot1,
    at={(0,0))},
    anchor=center,
    title={Femur},
    title style={font=\Large, yshift=-6},
    ybar stacked,
    ytick={0,20,40,60,80,100},
    yticklabels={0\%,20\%,40\%,60\%,80\%,100\%},
    yticklabel style={font=\normalsize},
    bar width=28.8pt,
    ymajorgrids = false,
    ymin=0,ymax=110,
    legend image code/.code={%
                    \draw[#1] (0cm,-0.1cm) rectangle (0.6cm,0.1cm);},
    legend style={at={(0.5,-0.12)},
      anchor=north, legend columns=5, font=\normalsize, column sep=.2cm, /tikz/every odd column/.append style={column sep=.1cm}},
    symbolic x coords={ANAT, ANAT$_{\backslash\text{FL}}$, ANAT$_{\backslash\text{HD}}$, ANAT$_{\backslash\text{BW}}$, ANAT$_{\backslash\text{NSA}}$, ,OC-ANAT, OC$_{\backslash\widetilde{\text{FL}}}$, OC$_{\backslash\widetilde{\text{HD}}}$, OC$_{\backslash\widetilde{\text{BW}}}$, OC$_{\backslash\widetilde{\text{NSA}}}$},
    xticklabel style={font=\normalsize},
    xticklabels={ANAT$_{\text{SSM}}$, $\setminus\beta_{FL}$, $\setminus\beta_{HD}$, $\setminus\beta_{BW}$, $\setminus\beta_{NSA}$, OC-ANAT$_{\text{SSM}}$, $\setminus\Tilde{\beta}_{FL}$, $\setminus\Tilde{\beta}_{HD}$, $\setminus\Tilde{\beta}_{BW}$, $\setminus\Tilde{\beta}_{NSA}$},
    x=1.8cm,
    enlarge x limits={abs=1.3cm},
    xtick=data,
    tickwidth=.9mm
    ]

\node[rotate=90] at (-4.5, 8) {\textcolor{YellowGreen}{1.3\%}};
\node[rotate=90] at (-4.5, 23.4) {\textcolor{Purple}{2.4\%}};
\node[rotate=90] at (-4.5, 38.8) {\textcolor{Salmon}{2.3\%}};
\node[rotate=90] at (-4.5, 55.2) {\textcolor{Mahogany}{3.9\%}};
\node at (0,54.3) {88.8\%};

\node[rotate=90] at (5.5, 8) {\textcolor{YellowGreen}{1.3\%}};
\node[rotate=90] at (5.5, 23.4) {\textcolor{Purple}{2.2\%}};
\node at (10, 6.4) {5.8\%};
\node at (10, 28.65) {38.7\%};

\node[rotate=90] at (15.5, 8) {\textcolor{YellowGreen}{0.9\%}};
\node[rotate=90] at (15.5, 23.4) {\textcolor{Purple}{1.7\%}};
\node at (20, 19.3) {33.4\%};

\node[rotate=90] at (25.5, 8) {\textcolor{YellowGreen}{1.5\%}};
\node at (30, 4.3) {5.6\%};

\node[rotate=90] at (55.5, 8) {\textcolor{YellowGreen}{1.4\%}};
\node[rotate=90] at (55.5, 23.4) {\textcolor{Purple}{2.1\%}};
\node at (60, 9.95) {12.9\%};
\node at (60, 24.5) {16.2\%};
\node at (60,61.45) {57.7\%};

\node[rotate=90] at (65.5, 8) {\textcolor{YellowGreen}{1.4\%}};
\node[rotate=90] at (65.5, 23.4) {\textcolor{Purple}{2.1\%}};
\node at (70, 9.95) {12.9\%};
\node at (70, 24.5) {16.2\%};

\node[rotate=90] at (75.5, 8) {\textcolor{YellowGreen}{1.4\%}};
\node[rotate=90] at (75.5, 23.4) {\textcolor{Purple}{2.1\%}};
\node at (80, 9.95) {12.9\%};

\node[rotate=90] at (85.5, 8) {\textcolor{YellowGreen}{1.4\%}};
\node[rotate=90] at (85.5, 23.4) {\textcolor{Purple}{2.1\%}};

\node[anchor=south] at (0, 98.7) {98.7\%};
\node[anchor=south] at (10, 48) {48.0\%};
\node[anchor=south] at (20, 36) {36.0\%};
\node[anchor=south] at (30, 7.1) {7.1\%};
\node[anchor=south] at (40, 1.8) {1.8\%};
\node[anchor=south] at (60, 90.3) {90.3\%};
\node[anchor=south] at (70, 32.6) {32.6\%};
\node[anchor=south] at (80, 16.4) {16.4\%};
\node[anchor=south] at (90, 3.5) {3.5\%};
\node[anchor=south] at (100, 1.4) {1.4\%};

\addplot[fill=YellowGreen!80] coordinates{(ANAT,1.3) (ANAT$_{\backslash\text{FL}}$,1.3) (ANAT$_{\backslash\text{HD}}$,0.9) (ANAT$_{\backslash\text{BW}}$,1.5) (ANAT$_{\backslash\text{NSA}}$,1.8) (OC-ANAT,0) (OC$_{\backslash\widetilde{\text{FL}}}$,0) (OC$_{\backslash\widetilde{\text{HD}}}$,0) (OC$_{\backslash\widetilde{\text{BW}}}$,0) (OC$_{\backslash\widetilde{\text{NSA}}}$,0)};
\addplot[fill=Purple!70] coordinates{(ANAT,2.4) (ANAT$_{\backslash\text{FL}}$,2.2) (ANAT$_{\backslash\text{HD}}$,1.7) (ANAT$_{\backslash\text{BW}}$,5.6) (ANAT$_{\backslash\text{NSA}}$,0) (OC-ANAT,0) (OC$_{\backslash\widetilde{\text{FL}}}$,0) (OC$_{\backslash\widetilde{\text{HD}}}$,0) (OC$_{\backslash\widetilde{\text{BW}}}$,0) (OC$_{\backslash\widetilde{\text{NSA}}}$,0)};
\addplot[fill=Salmon!70] coordinates{(ANAT,2.3) (ANAT$_{\backslash\text{FL}}$,5.8) (ANAT$_{\backslash\text{HD}}$,33.4) (ANAT$_{\backslash\text{BW}}$,0) (ANAT$_{\backslash\text{NSA}}$,0)  (OC-ANAT,0) (OC$_{\backslash\widetilde{\text{FL}}}$,0) (OC$_{\backslash\widetilde{\text{HD}}}$,0) (OC$_{\backslash\widetilde{\text{BW}}}$,0) (OC$_{\backslash\widetilde{\text{NSA}}}$,0)};
\addplot[fill=Mahogany!70] coordinates{(ANAT,3.9) (ANAT$_{\backslash\text{FL}}$,38.7) (ANAT$_{\backslash\text{HD}}$,0) (ANAT$_{\backslash\text{BW}}$,0) (ANAT$_{\backslash\text{NSA}}$,0) (OC-ANAT,0) (OC$_{\backslash\widetilde{\text{FL}}}$,0) (OC$_{\backslash\widetilde{\text{HD}}}$,0) (OC$_{\backslash\widetilde{\text{BW}}}$,0) (OC$_{\backslash\widetilde{\text{NSA}}}$,0)};
\addplot[fill=NavyBlue!80] coordinates{(ANAT,88.8) (ANAT$_{\backslash\text{FL}}$,0) (ANAT$_{\backslash\text{HD}}$,0) (ANAT$_{\backslash\text{BW}}$,0) (ANAT$_{\backslash\text{NSA}}$,0) (OC-ANAT,0) (OC$_{\backslash\widetilde{\text{FL}}}$,0) (OC$_{\backslash\widetilde{\text{HD}}}$,0) (OC$_{\backslash\widetilde{\text{BW}}}$,0) (OC$_{\backslash\widetilde{\text{NSA}}}$,0)};
\addplot[fill=YellowGreen!80, postaction={pattern=north east lines, pattern color=Gray}] coordinates{(ANAT,0) (ANAT$_{\backslash\text{FL}}$,0) (ANAT$_{\backslash\text{HD}}$,0) (ANAT$_{\backslash\text{BW}}$,0) (ANAT$_{\backslash\text{NSA}}$,0) (OC-ANAT,1.4) (OC$_{\backslash\widetilde{\text{FL}}}$,1.4) (OC$_{\backslash\widetilde{\text{HD}}}$,1.4) (OC$_{\backslash\widetilde{\text{BW}}}$,1.4) (OC$_{\backslash\widetilde{\text{NSA}}}$,1.4)};
\addplot[fill=Purple!70, postaction={pattern=north east lines, pattern color=Gray}] coordinates{(ANAT,0) (ANAT$_{\backslash\text{FL}}$,0) (ANAT$_{\backslash\text{HD}}$,0) (ANAT$_{\backslash\text{BW}}$,0) (ANAT$_{\backslash\text{NSA}}$,0) (OC-ANAT,2.1) (OC$_{\backslash\widetilde{\text{FL}}}$,2.1) (OC$_{\backslash\widetilde{\text{HD}}}$,2.1) (OC$_{\backslash\widetilde{\text{BW}}}$,2.1) (OC$_{\backslash\widetilde{\text{NSA}}}$,0)};
\addplot[fill=Salmon!70, postaction={pattern=north east lines, pattern color=Gray}] coordinates{(ANAT,0) (ANAT$_{\backslash\text{FL}}$,0) (ANAT$_{\backslash\text{HD}}$,0) (ANAT$_{\backslash\text{BW}}$,0) (ANAT$_{\backslash\text{NSA}}$,0) (OC-ANAT,12.9) (OC$_{\backslash\widetilde{\text{FL}}}$,12.9) (OC$_{\backslash\widetilde{\text{HD}}}$,12.9) (OC$_{\backslash\widetilde{\text{BW}}}$,0) (OC$_{\backslash\widetilde{\text{NSA}}}$,0)};
\addplot[fill=Mahogany!70, postaction={pattern=north east lines, pattern color=Gray}] coordinates{(ANAT,0) (ANAT$_{\backslash\text{FL}}$,0) (ANAT$_{\backslash\text{HD}}$,0) (ANAT$_{\backslash\text{BW}}$,0) (ANAT$_{\backslash\text{NSA}}$,0) (OC-ANAT,16.2) (OC$_{\backslash\widetilde{\text{FL}}}$,16.2) (OC$_{\backslash\widetilde{\text{HD}}}$,0) (OC$_{\backslash\widetilde{\text{BW}}}$,0) (OC$_{\backslash\widetilde{\text{NSA}}}$,0)};
\addplot[fill=NavyBlue!80, postaction={pattern=north east lines, pattern color=Gray}] coordinates{(ANAT,0) (ANAT$_{\backslash\text{FL}}$,0) (ANAT$_{\backslash\text{HD}}$,0) (ANAT$_{\backslash\text{BW}}$,0) (ANAT$_{\backslash\text{NSA}}$,0) (OC-ANAT,57.7) (OC$_{\backslash\widetilde{\text{FL}}}$,0) (OC$_{\backslash\widetilde{\text{HD}}}$,0) (OC$_{\backslash\widetilde{\text{BW}}}$,0) (OC$_{\backslash\widetilde{\text{NSA}}}$,0)};
\legend{$\kappa_{FV}$,$\kappa_{NSA}$,$\kappa_{BW}$,$\kappa_{HD}$,$\kappa_{FL}$,$\Tilde{\kappa}_{FV}$,$\Tilde{\kappa}_{NSA}$,$\Tilde{\kappa}_{BW}$,$\Tilde{\kappa}_{HD}$,$\Tilde{\kappa}_{FL}$}
\end{axis}

\begin{axis}[
    name=plot2,
    at={(0,-1125))},
    anchor=center,
    title={Scapula},
    title style={font=\Large, yshift=-6},
    ybar stacked,
    ytick={0,20,40,60,80},
    yticklabels={0\%,20\%,40\%,60\%,80\%},
    yticklabel style={font=\normalsize},
    bar width=28.8pt,
    ymajorgrids = false,
    ymin=0,ymax=80,
    legend image code/.code={%
                    \draw[#1] (0cm,-0.1cm) rectangle (0.6cm,0.1cm);},
    legend style={at={(0.5,-0.12)},
      anchor=north, legend columns=6, font=\normalsize, column sep=.2cm, /tikz/every odd column/.append style={column sep=.1cm},},
    symbolic x coords={ANAT, ANAT$_{\backslash\text{SL}}$, ANAT$_{\backslash\text{GH}}$, ANAT$_{\backslash\text{GW}}$, ANAT$_{\backslash\text{GI}}$, ANAT$_{\backslash\text{CSA}}$, ,OC-ANAT, OC$_{\backslash\widetilde{\text{SL}}}$, OC$_{\backslash\widetilde{\text{GH}}}$, OC$_{\backslash\widetilde{\text{GW}}}$, OC$_{\backslash\widetilde{\text{GI}}}$, OC$_{\backslash\widetilde{\text{CSA}}}$},
    xticklabel style={font=\normalsize},
    xticklabels={ANAT$_{\text{SSM}}$, $\setminus\beta_{SL}$, $\setminus\beta_{GH}$, $\setminus\beta_{GW}$, $\setminus\beta_{GI}$, $\setminus\beta_{CSA}$, OC-ANAT$_{\text{SSM}}$, $\setminus\Tilde{\beta}_{SL}$, $\setminus\Tilde{\beta}_{GH}$, $\setminus\Tilde{\beta}_{GW}$, $\setminus\Tilde{\beta}_{GI}$, $\setminus\Tilde{\beta}_{CSA}$},
    x=1.8cm,
    enlarge x limits={abs=1.3cm},
    xtick=data,
    tickwidth=.9mm
    ]
    
\node[rotate=90] at (-4.5, 58) {\textcolor{cyan}{1.7\%}};
\node at (0, 58) {8.2\%};
\node at (0, 134.5) {7.1\%};
\node at (0, 212) {8.4\%};
\node at (0, 293.5) {7.9\%};
\node at (0, 482) {29.8\%};

\node[rotate=90] at (5.5, 58) {\textcolor{cyan}{2.3\%}};
\node at (10, 81) {11.6\%};
\node at (10, 168.5) {5.9\%};
\node at (10, 258.5) {12.1\%};
\node at (10, 457) {27.6\%};

\node[rotate=90] at (15.5, 58) {\textcolor{cyan}{1.7\%}};
\node at (20, 52.5) {7.1\%};
\node at (20, 121) {6.6\%};
\node at (20, 327) {34.6\%};

\node[rotate=90] at (25.5, 58) {\textcolor{cyan}{1.8\%}};
\node at (30, 49) {6.2\%};
\node at (30, 118) {7.6\%};

\node[rotate=90] at (35.5, 58) {\textcolor{cyan}{1.7\%}};
\node at (40, 43) {5.2\%};

\node[rotate=90] at (65.5, 58) {\textcolor{cyan}{1.5\%}};
\node[rotate=90] at (65.5, 170) {\textcolor{JungleGreen}{2.9\%}};
\node at (70, 68.5) {\small{4.9\%}};
\node at (70, 150) {\small{11.4\%}};
\node at (70, 265.5) {\small{11.7\%}};
\node at (70, 468) {\small{28.8\%}};

\node[rotate=90] at (75.5, 58) {\small{\textcolor{cyan}{1.5\%}}};
\node[rotate=90] at (75.5, 170) {\textcolor{JungleGreen}{2.9\%}};
\node at (80, 68.5) {4.9\%};
\node at (80, 150) {11.4\%};
\node at (80, 265.5) {11.7\%};

\node[rotate=90] at (85.5, 58) {\textcolor{cyan}{1.5\%}};
\node[rotate=90] at (85.5, 170) {\textcolor{JungleGreen}{2.9\%}};
\node at (90, 68.5) {4.9\%};
\node at (90, 150) {11.4\%};

\node[rotate=90] at (95.5, 58) {\textcolor{cyan}{1.5\%}};
\node[rotate=90] at (95.5, 170) {\textcolor{JungleGreen}{2.9\%}};
\node at (100, 68.5) {4.9\%};

\node[rotate=90] at (105.5, 58) {\textcolor{cyan}{1.5\%}};
\node[rotate=90] at (105.5, 170) {\textcolor{JungleGreen}{2.9\%}};

\node[anchor=south] at (0, 631) {63.1\%};
\node[anchor=south] at (10, 595) {59.5\%};
\node[anchor=south] at (20, 500) {50.0\%};
\node[anchor=south] at (30, 156) {15.6\%};
\node[anchor=south] at (40, 69) {6.9\%};
\node[anchor=south] at (50, 27) {2.7\%};
\node[anchor=south] at (70, 612) {61.2\%};
\node[anchor=south] at (80, 324) {32.4\%};
\node[anchor=south] at (90, 207) {20.7\%};
\node[anchor=south] at (100, 93) {9.3\%};
\node[anchor=south] at (110, 44) {4.4\%};
\node[anchor=south] at (120, 15) {1.5\%};

\addplot[fill=cyan!40] coordinates{(ANAT,1.7) (ANAT$_{\backslash\text{SL}}$,2.3) (ANAT$_{\backslash\text{GH}}$,1.7) (ANAT$_{\backslash\text{GW}}$,1.8) (ANAT$_{\backslash\text{GI}}$,1.7) (ANAT$_{\backslash\text{CSA}}$,2.7) (OC-ANAT,0) (OC$_{\backslash\widetilde{\text{SL}}}$,0) (OC$_{\backslash\widetilde{\text{GH}}}$,0) (OC$_{\backslash\widetilde{\text{GW}}}$,0) (OC$_{\backslash\widetilde{\text{GI}}}$,0) (OC$_{\backslash\widetilde{\text{CSA}}}$,0)};
\addplot[fill=JungleGreen!80] coordinates{(ANAT,8.2) (ANAT$_{\backslash\text{SL}}$,11.6) (ANAT$_{\backslash\text{GH}}$,7.1) (ANAT$_{\backslash\text{GW}}$,6.2) (ANAT$_{\backslash\text{GI}}$,5.2) (ANAT$_{\backslash\text{CSA}}$,0) (OC-ANAT,0) (OC$_{\backslash\widetilde{\text{SL}}}$,0) (OC$_{\backslash\widetilde{\text{GH}}}$,0) (OC$_{\backslash\widetilde{\text{GW}}}$,0) (OC$_{\backslash\widetilde{\text{GI}}}$,0) (OC$_{\backslash\widetilde{\text{CSA}}}$,0)};
\addplot[fill=red!50] coordinates{(ANAT,7.1) (ANAT$_{\backslash\text{SL}}$,5.9) (ANAT$_{\backslash\text{GH}}$,6.6) (ANAT$_{\backslash\text{GW}}$,7.6) (ANAT$_{\backslash\text{GI}}$,0) (ANAT$_{\backslash\text{CSA}}$,0) (OC-ANAT,0) (OC$_{\backslash\widetilde{\text{SL}}}$,0) (OC$_{\backslash\widetilde{\text{GH}}}$,0) (OC$_{\backslash\widetilde{\text{GW}}}$,0) (OC$_{\backslash\widetilde{\text{GI}}}$,0) (OC$_{\backslash\widetilde{\text{CSA}}}$,0)};
\addplot[fill=MidnightBlue!60] coordinates{(ANAT,8.4) (ANAT$_{\backslash\text{SL}}$,12.1) (ANAT$_{\backslash\text{GH}}$,34.6) (ANAT$_{\backslash\text{GW}}$,0) (ANAT$_{\backslash\text{GI}}$,0) (ANAT$_{\backslash\text{CSA}}$,0) (OC-ANAT,0) (OC$_{\backslash\widetilde{\text{SL}}}$,0) (OC$_{\backslash\widetilde{\text{GH}}}$,0) (OC$_{\backslash\widetilde{\text{GW}}}$,0) (OC$_{\backslash\widetilde{\text{GI}}}$,0) (OC$_{\backslash\widetilde{\text{CSA}}}$,0)};
\addplot[fill=magenta!40] coordinates{(ANAT,7.9) (ANAT$_{\backslash\text{SL}}$,27.6) (ANAT$_{\backslash\text{GH}}$,0) (ANAT$_{\backslash\text{GW}}$,0) (ANAT$_{\backslash\text{GI}}$,0) (ANAT$_{\backslash\text{CSA}}$,0) (OC-ANAT,0) (OC$_{\backslash\widetilde{\text{SL}}}$,0) (OC$_{\backslash\widetilde{\text{GH}}}$,0) (OC$_{\backslash\widetilde{\text{GW}}}$,0) (OC$_{\backslash\widetilde{\text{GI}}}$,0) (OC$_{\backslash\widetilde{\text{CSA}}}$,0)};
\addplot[fill=orange!40] coordinates{(ANAT,29.8) (ANAT$_{\backslash\text{SL}}$,0) (ANAT$_{\backslash\text{GH}}$,0) (ANAT$_{\backslash\text{GW}}$,0) (ANAT$_{\backslash\text{GI}}$,0) (ANAT$_{\backslash\text{CSA}}$,0) (OC-ANAT,0) (OC$_{\backslash\widetilde{\text{SL}}}$,0) (OC$_{\backslash\widetilde{\text{GH}}}$,0) (OC$_{\backslash\widetilde{\text{GW}}}$,0) (OC$_{\backslash\widetilde{\text{GI}}}$,0) (OC$_{\backslash\widetilde{\text{CSA}}}$,0)};
\addplot[fill=cyan!40, postaction={pattern=north east lines, pattern color=Gray}] coordinates{(ANAT,0) (ANAT$_{\backslash\text{SL}}$,0) (ANAT$_{\backslash\text{GH}}$,0) (ANAT$_{\backslash\text{GW}}$,0) (ANAT$_{\backslash\text{GI}}$,0) (ANAT$_{\backslash\text{CSA}}$,0) (OC-ANAT,1.5) (OC$_{\backslash\widetilde{\text{SL}}}$,1.5) (OC$_{\backslash\widetilde{\text{GH}}}$,1.5) (OC$_{\backslash\widetilde{\text{GW}}}$,1.5) (OC$_{\backslash\widetilde{\text{GI}}}$,1.5) (OC$_{\backslash\widetilde{\text{CSA}}}$,1.5)};
\addplot[fill=JungleGreen!80, postaction={pattern=north east lines, pattern color=Gray}] coordinates{(ANAT,0) (ANAT$_{\backslash\text{SL}}$,0) (ANAT$_{\backslash\text{GH}}$,0) (ANAT$_{\backslash\text{GW}}$,0) (ANAT$_{\backslash\text{GI}}$,0) (ANAT$_{\backslash\text{CSA}}$,0) (OC-ANAT,2.9) (OC$_{\backslash\widetilde{\text{SL}}}$,2.9) (OC$_{\backslash\widetilde{\text{GH}}}$,2.9) (OC$_{\backslash\widetilde{\text{GW}}}$,2.9) (OC$_{\backslash\widetilde{\text{GI}}}$,2.9) (OC$_{\backslash\widetilde{\text{CSA}}}$,0)};
\addplot[fill=red!50, postaction={pattern=north east lines, pattern color=Gray}] coordinates{(ANAT,0) (ANAT$_{\backslash\text{SL}}$,0) (ANAT$_{\backslash\text{GH}}$,0) (ANAT$_{\backslash\text{GW}}$,0) (ANAT$_{\backslash\text{GI}}$,0) (ANAT$_{\backslash\text{CSA}}$,0) (OC-ANAT,4.9) (OC$_{\backslash\widetilde{\text{SL}}}$,4.9) (OC$_{\backslash\widetilde{\text{GH}}}$,4.9) (OC$_{\backslash\widetilde{\text{GW}}}$,4.9) (OC$_{\backslash\widetilde{\text{GI}}}$,0) (OC$_{\backslash\widetilde{\text{CSA}}}$,0)};
\addplot[fill=MidnightBlue!60, postaction={pattern=north east lines, pattern color=Gray}] coordinates{(ANAT,0) (ANAT$_{\backslash\text{SL}}$,0) (ANAT$_{\backslash\text{GH}}$,0) (ANAT$_{\backslash\text{GW}}$,0) (ANAT$_{\backslash\text{GI}}$,0) (ANAT$_{\backslash\text{CSA}}$,0) (OC-ANAT,11.4) (OC$_{\backslash\widetilde{\text{SL}}}$,11.4) (OC$_{\backslash\widetilde{\text{GH}}}$,11.4) (OC$_{\backslash\widetilde{\text{GW}}}$,0) (OC$_{\backslash\widetilde{\text{GI}}}$,0) (OC$_{\backslash\widetilde{\text{CSA}}}$,0)};
\addplot[fill=magenta!40, postaction={pattern=north east lines, pattern color=Gray}] coordinates{(ANAT,0) (ANAT$_{\backslash\text{SL}}$,0) (ANAT$_{\backslash\text{GH}}$,0) (ANAT$_{\backslash\text{GW}}$,0) (ANAT$_{\backslash\text{GI}}$,0) (ANAT$_{\backslash\text{CSA}}$,0) (OC-ANAT,11.7) (OC$_{\backslash\widetilde{\text{SL}}}$,11.7) (OC$_{\backslash\widetilde{\text{GH}}}$,0) (OC$_{\backslash\widetilde{\text{GW}}}$,0) (OC$_{\backslash\widetilde{\text{GI}}}$,0) (OC$_{\backslash\widetilde{\text{CSA}}}$,0)};
\addplot[fill=orange!40, postaction={pattern=north east lines, pattern color=Gray, pattern color=Gray}] coordinates{(ANAT,0) (ANAT$_{\backslash\text{SL}}$,0) (ANAT$_{\backslash\text{GH}}$,0) (ANAT$_{\backslash\text{GW}}$,0) (ANAT$_{\backslash\text{GI}}$,0) (ANAT$_{\backslash\text{CSA}}$,0) (OC-ANAT,28.8) (OC$_{\backslash\widetilde{\text{SL}}}$,0) (OC$_{\backslash\widetilde{\text{GH}}}$,0) (OC$_{\backslash\widetilde{\text{GW}}}$,0) (OC$_{\backslash\widetilde{\text{GI}}}$,0) (OC$_{\backslash\widetilde{\text{CSA}}}$,0)};
\legend{$\kappa_{GV}$,$\kappa_{CSA}$,$\kappa_{GI}$,$\kappa_{GW}$,$\kappa_{GH}$,$\kappa_{SL}$,$\Tilde{\kappa}_{GV}$,$\Tilde{\kappa}_{CSA}$,$\Tilde{\kappa}_{GI}$,$\Tilde{\kappa}_{GW}$,$\Tilde{\kappa}_{GH}$,$\Tilde{\kappa}_{SL}$}
\end{axis}
\end{tikzpicture}
\end{adjustbox}
\caption{Shape variability induced by each of the femoral and scapular anatomical parameters in ANAT$_{\text{SSM}}$, OC-ANAT$_{\text{SSM}}$ and their respective sub-models built by retaining the anatomical parameter with largest anatomical variability at each step. The shape variability values of the models were normalized by total BASE$_{\text{SSM}}$ shape variability.}
\label{fig:shape_variability_anatomical_parameters}
\end{figure*}

Finally, the obtained shape variation values revealed that the femoral ANAT$_{\text{SSM}}$ and OC-ANAT$_{\text{SSM}}$ incorporated $98.7\%$ and $90.3\%$ of the total variability present in the BASE$_{\text{SSM}}$, while the scapular models accounted for $63.1\%$ and $61.2\%$ of the total variability (Fig. \ref{fig:shape_variability_anatomical_parameters}). Hence, in both anatomical structures, OC-ANAT$_{\text{SSM}}$ integrated marginally less shape variability. In femoral models, the FL represented the most shape variation ($\kappa_{FL} = 88.8\%$ and $\Tilde{\kappa}_{FL} = 57.7\%$) while the FV corresponded to the least shape variation ($\kappa_{FV} = 1.3\%$ and $\Tilde{\kappa}_{FV} = 1.4\%$). Scapular results were similar with SL accounting for the most shape variation ($\kappa_{SL} = 29.8\%$ and  $\Tilde{\kappa}_{SL} = 28.8\%$) and GV representing the least variability ($\kappa_{GV} = 1.7\%$ and $\Tilde{\kappa}_{GV} = 1.5\%$). However, in both bones, the order of the anatomical parameters differed between the two models, the CSA was for instance third in ANAT$_{\text{SSM}}$ while being fifth in OC-ANAT$_{\text{SSM}}$. Furthermore, as expected, the shape variability exhibited by the sub-models was lower than original models, indicating that shape variability is an increasing function with respect to the number of anatomical parameters. The results obtained from OC-ANAT$_{\text{SSM}}$ provided further validation of the orthogonality constraints, as the shape variability induced by the anatomical parameters remained unchanged across sub-models, contrary to ANAT$_{\text{SSM}}$ sub-models in which $\kappa_{GH}$ increased from $7.9\%$ in ANAT$_{\text{SSM}}$ to $27.6\%$ in ANAT$_{\setminus \beta_{SL}}$.

\begin{figure*}[t]
\centering
\begin{adjustbox}{width=\textwidth}
\begin{tikzpicture}
\begin{scope}[spy using outlines=
      {circle, magnification=2.25, size=.85cm, connect spies, rounded corners}]

\node[inner sep=0pt] at (0,0.05)
    {\includegraphics[width=.022\textwidth]{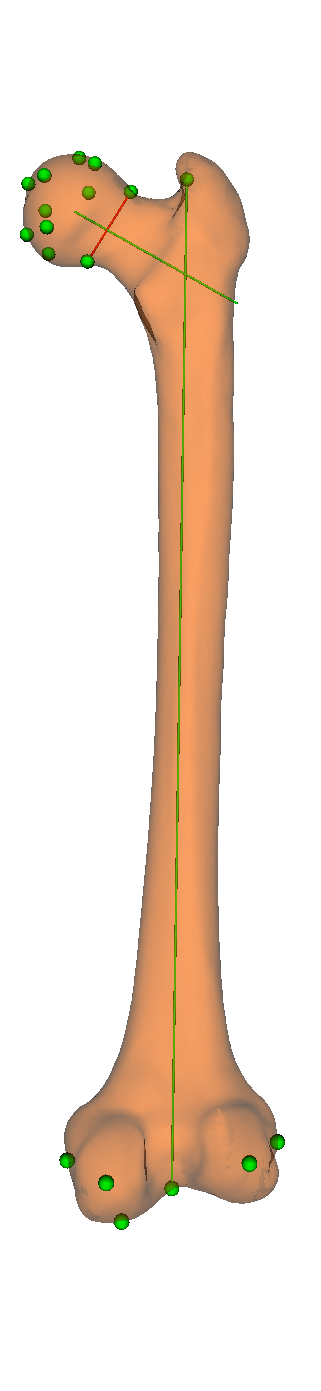}};
\node[inner sep=0pt] at (1.5,0.05)
    {\includegraphics[width=.022\textwidth]{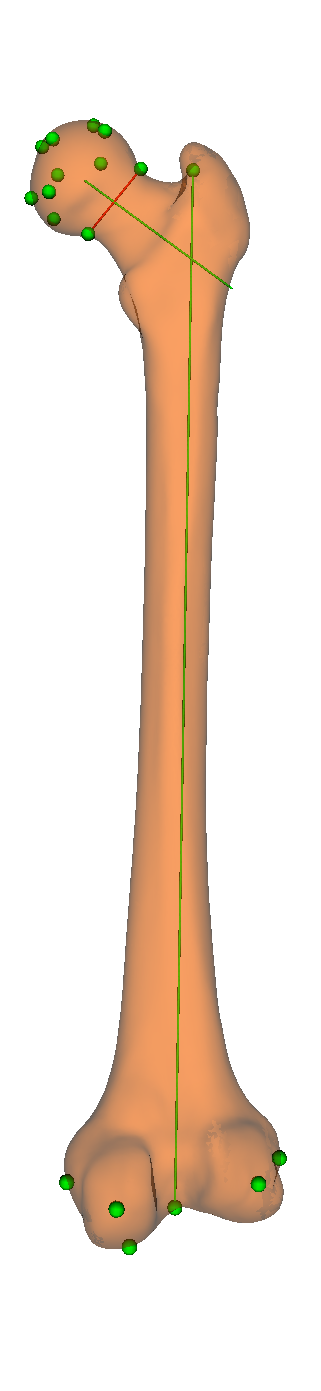}};
\node[inner sep=0pt] at (3,0.05)
    {\includegraphics[width=.022\textwidth]{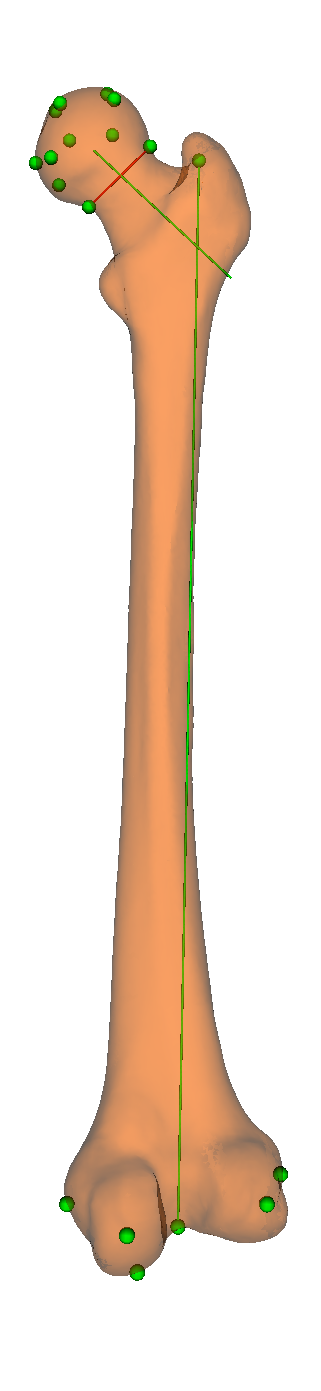}};
   
\node[inner sep=0pt] at (4.65,0.05)
    {\includegraphics[width=.022\textwidth]{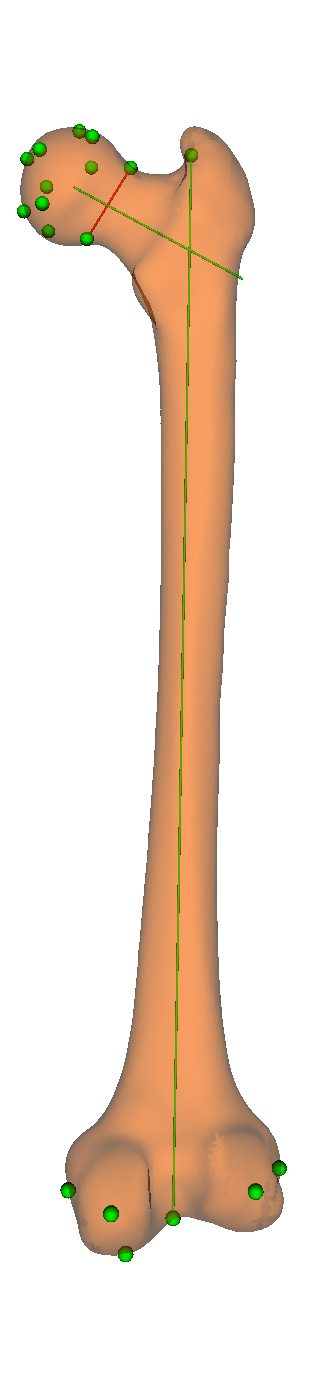}};
\node[inner sep=0pt] at (6.15,0.05)
    {\includegraphics[width=.022\textwidth]{NSA_mean.png}};
\node[inner sep=0pt] at (7.65,0.05)
    {\includegraphics[width=.022\textwidth]{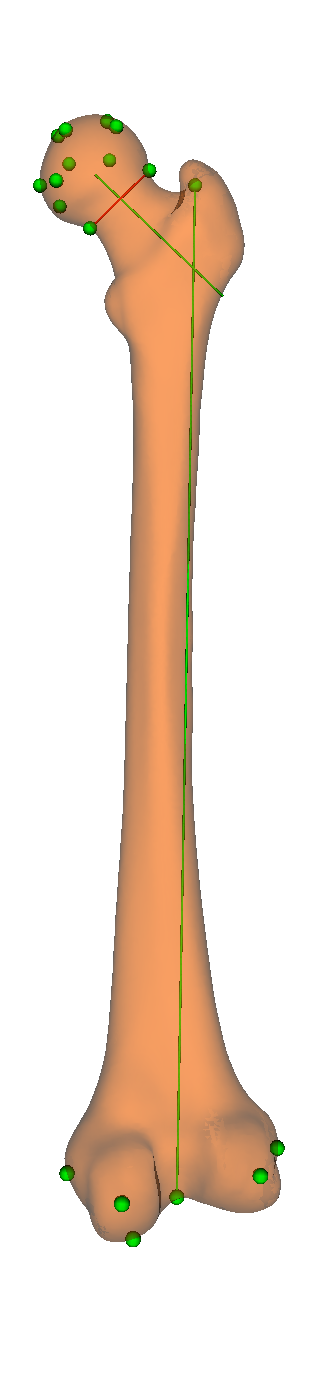}};
    
\node[inner sep=0pt] at (0,-1.9)
    {\includegraphics[width=.073\textwidth]{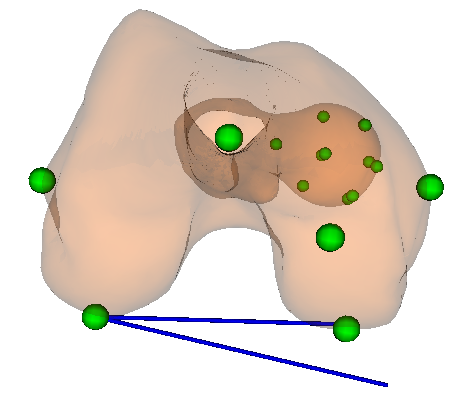}};
\node[inner sep=0pt] at (1.5,-1.9)
    {\includegraphics[width=.073\textwidth]{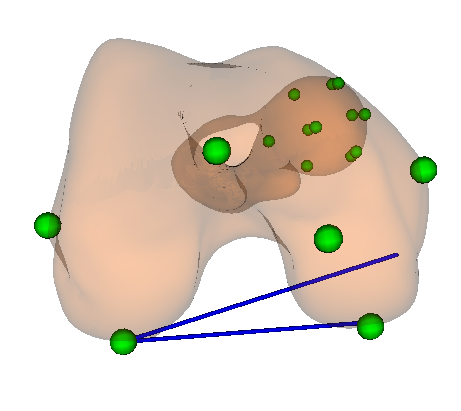}};
\node[inner sep=0pt] at (3,-1.9)
    {\includegraphics[width=.073\textwidth]{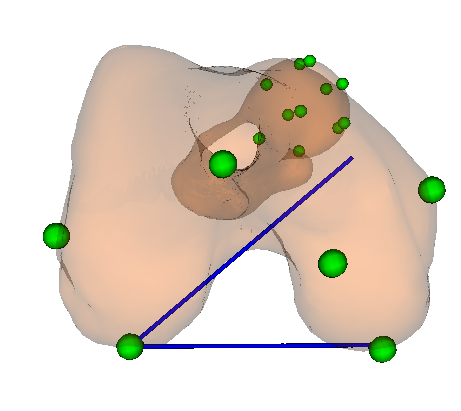}};
    
\node[inner sep=0pt] at (4.65,-1.9)
    {\includegraphics[width=.073\textwidth]{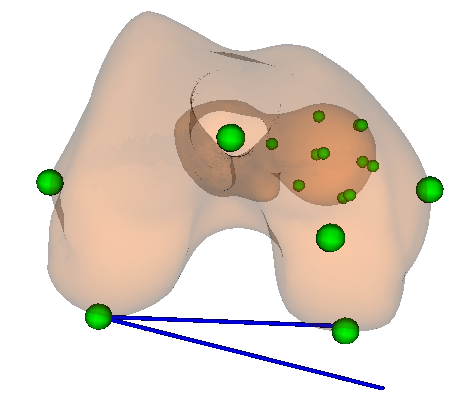}};
\node[inner sep=0pt] at (6.15,-1.9)
    {\includegraphics[width=.073\textwidth]{FV_mean.png}};
\node[inner sep=0pt] at (7.65,-1.9)
    {\includegraphics[width=.073\textwidth]{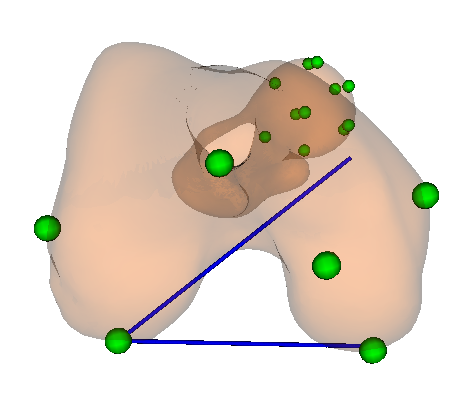}};
    
\node[inner sep=0pt] at (0,-3.75)
    {\includegraphics[width=.022\textwidth]{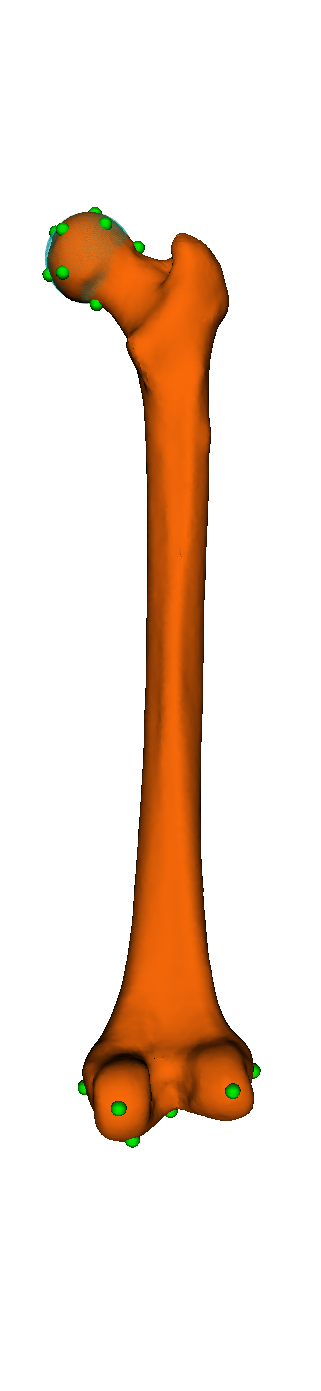}};
\node[inner sep=0pt] at (1.5,-3.75)
    {\includegraphics[width=.022\textwidth]{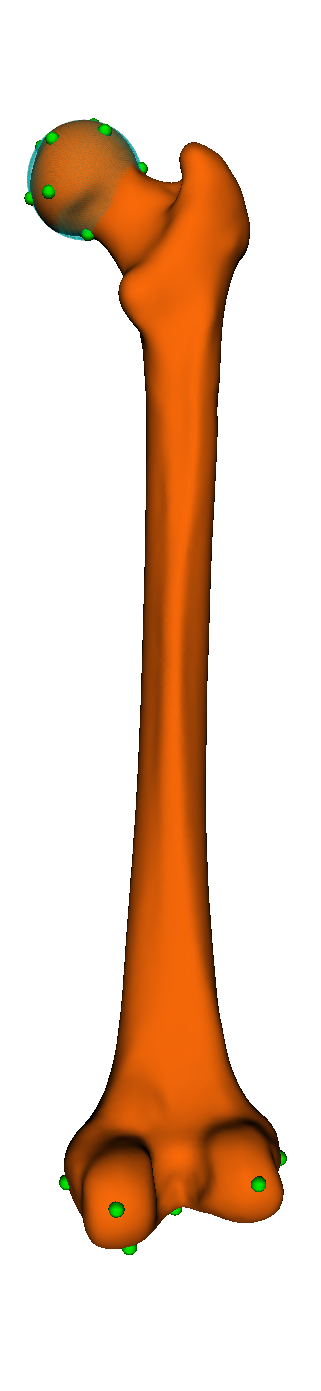}};
\node[inner sep=0pt] at (3,-3.75)
    {\includegraphics[width=.022\textwidth]{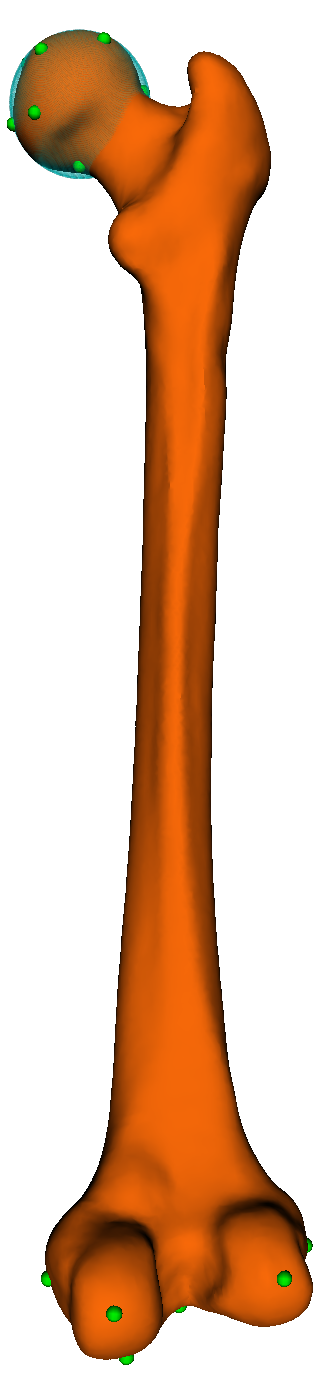}}; 
    
\node[inner sep=0pt] at (4.65,-3.75)
    {\includegraphics[width=.022\textwidth]{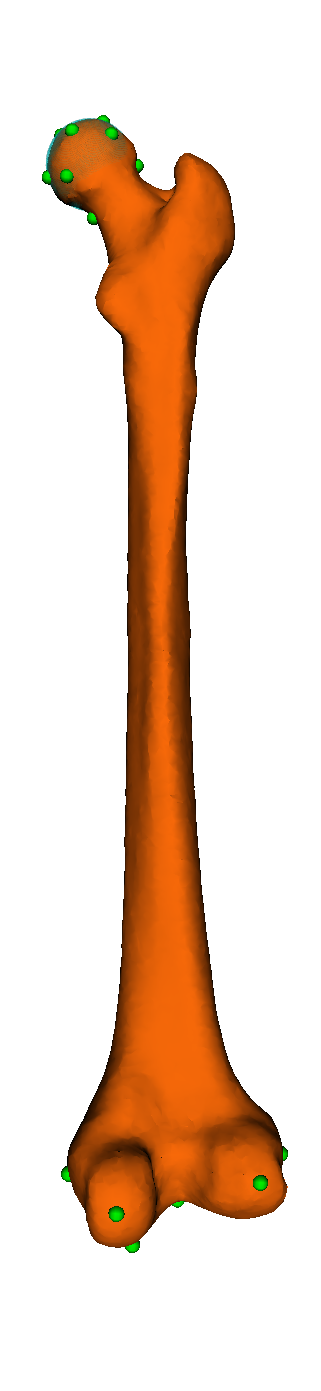}};
\node[inner sep=0pt] at (6.15,-3.75)
    {\includegraphics[width=.022\textwidth]{HD_mean.png}};
\node[inner sep=0pt] at (7.65,-3.75)
    {\includegraphics[width=.022\textwidth]{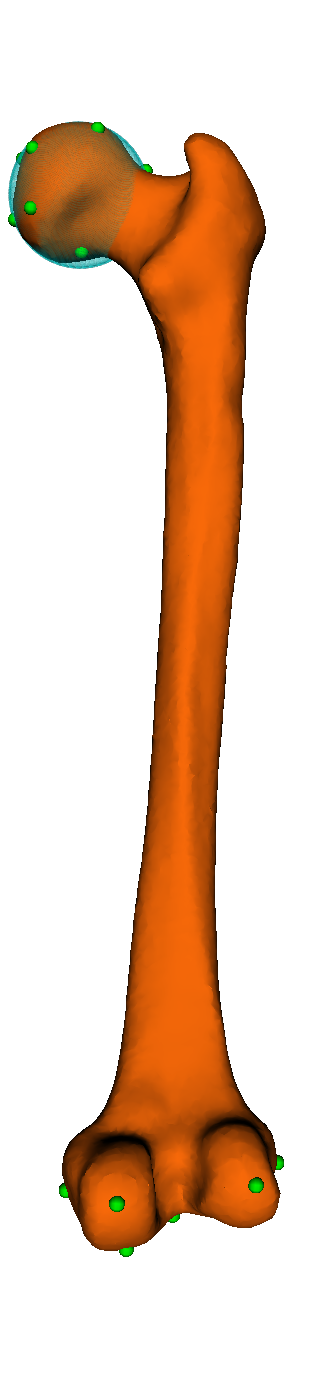}};
    
\node[inner sep=0pt] at (0,-5.65)
    {\includegraphics[width=.075\textwidth]{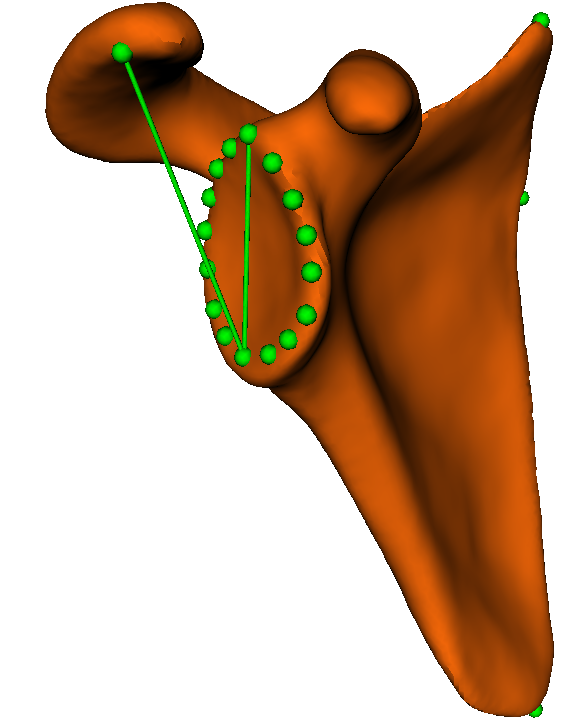}};
\node[inner sep=0pt] at (1.5,-5.65)
    {\includegraphics[width=.075\textwidth]{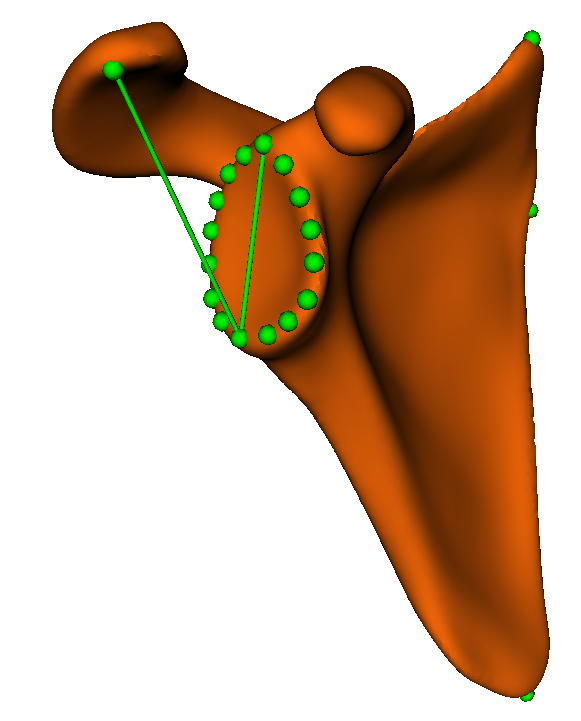}};
\node[inner sep=0pt] at (3,-5.65)
    {\includegraphics[width=.075\textwidth]{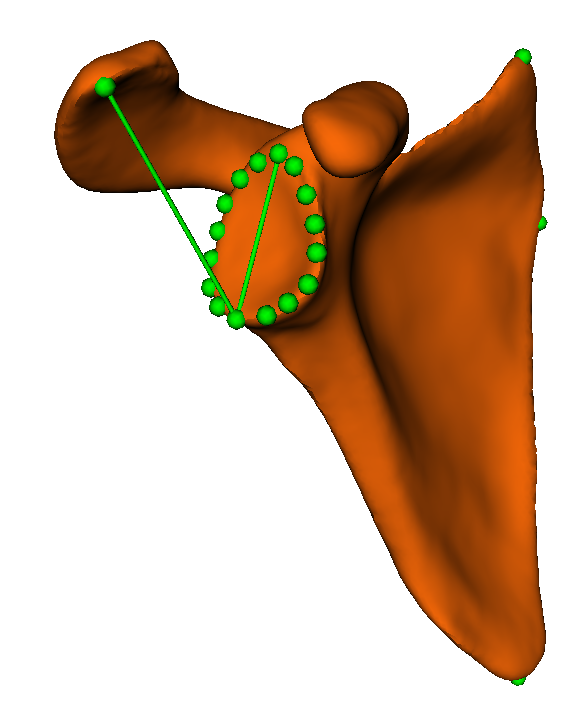}}; 
    
\node[inner sep=0pt] at (4.65,-5.65)
    {\includegraphics[width=.075\textwidth]{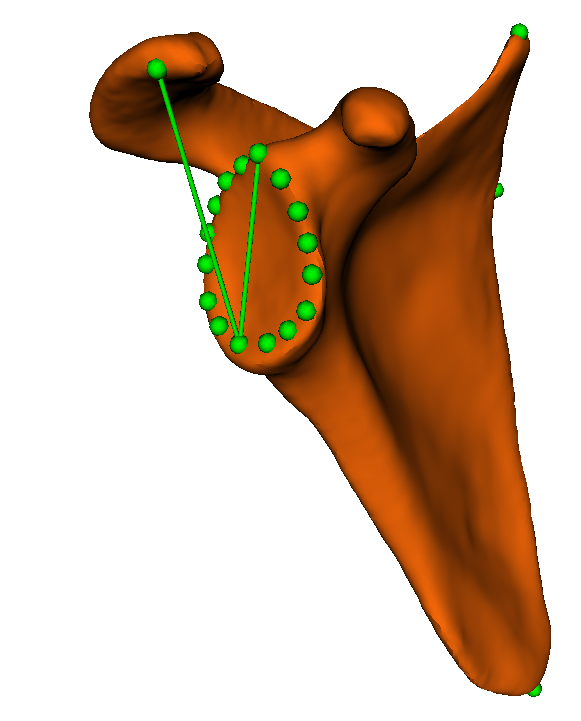}};
\node[inner sep=0pt] at (6.15,-5.65)
    {\includegraphics[width=.075\textwidth]{CSA_mean.png}};
\node[inner sep=0pt] at (7.65,-5.65)
    {\includegraphics[width=.075\textwidth]{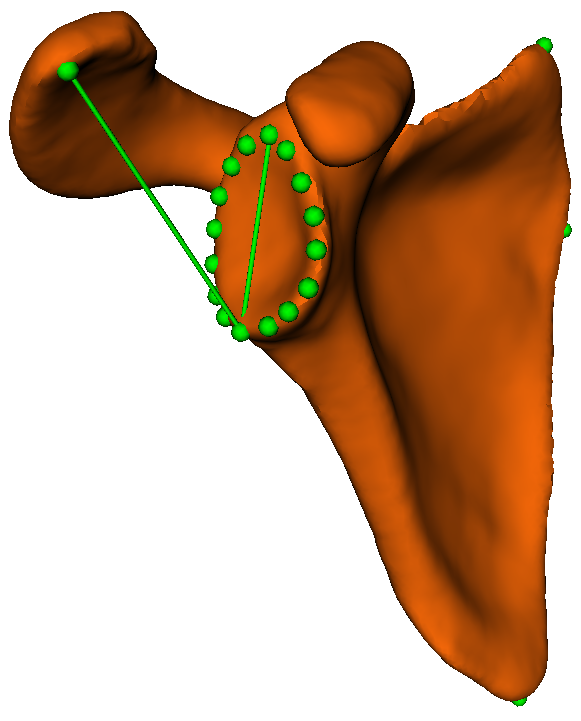}};
    
\node[inner sep=0pt] at (0,-7.55)
    {\includegraphics[width=.053\textwidth]{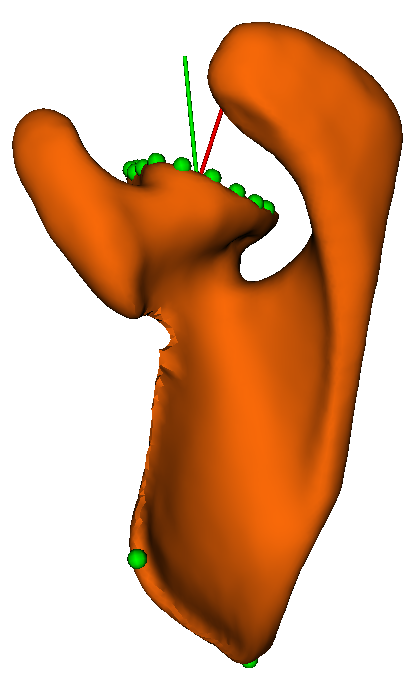}};
\node[inner sep=0pt] at (1.5,-7.55)
    {\includegraphics[width=.053\textwidth]{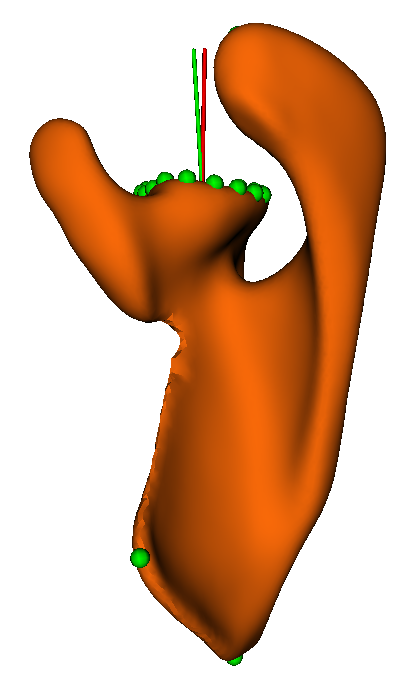}};
\node[inner sep=0pt] at (3,-7.55)
    {\includegraphics[width=.053\textwidth]{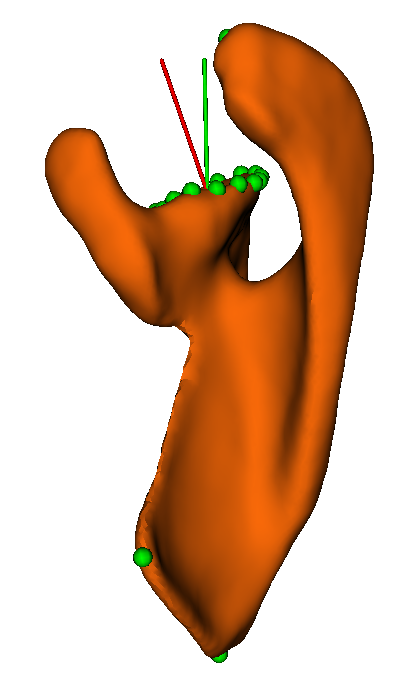}};
    
\node[inner sep=0pt] at (4.65,-7.55)
    {\includegraphics[width=.053\textwidth]{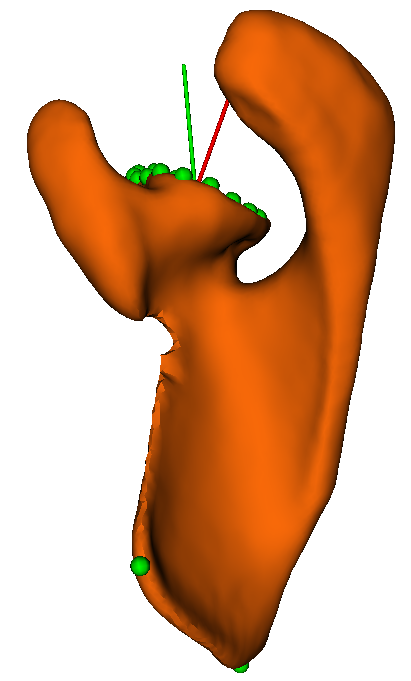}};
\node[inner sep=0pt] at (6.15,-7.55)
    {\includegraphics[width=.053\textwidth]{GV_mean.png}};
\node[inner sep=0pt] at (7.65,-7.55)
    {\includegraphics[width=.053\textwidth]{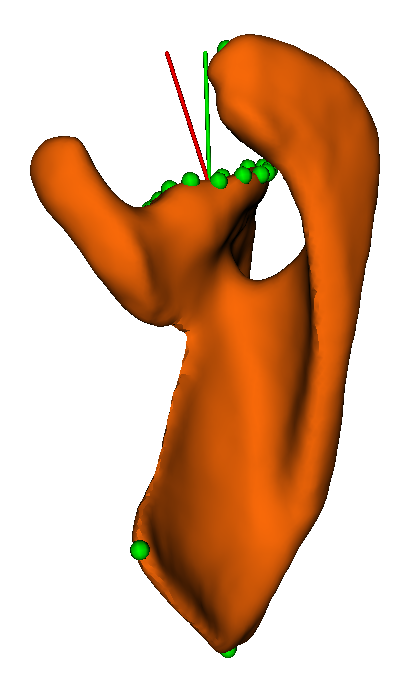}};
    
\node[inner sep=0pt] at (0,-9.5)
    {\adjincludegraphics[width=.058\textwidth]{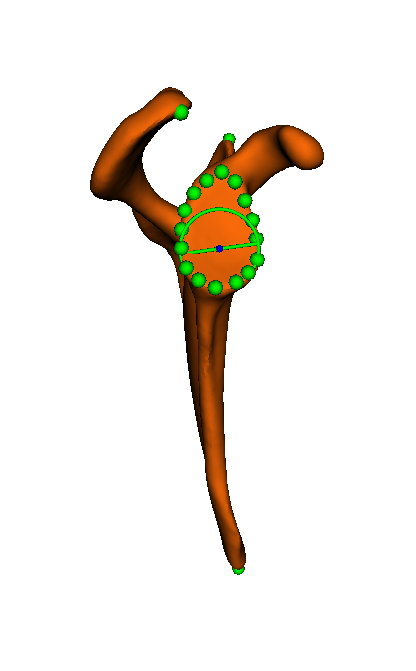}};
\node[inner sep=0pt] at (1.5,-9.5)
    {\includegraphics[width=.058\textwidth]{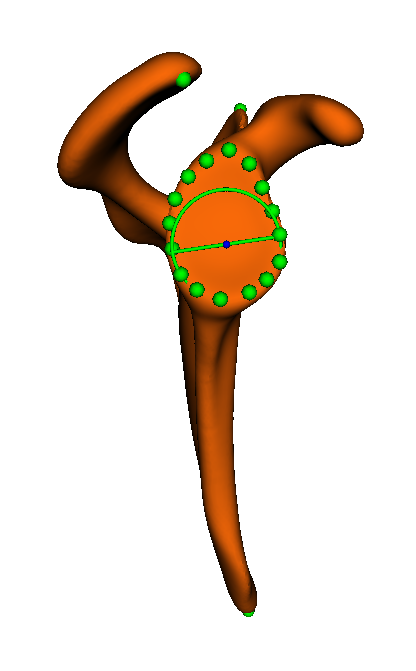}};
\node[inner sep=0pt] at (3,-9.5)
    {\includegraphics[width=.058\textwidth]{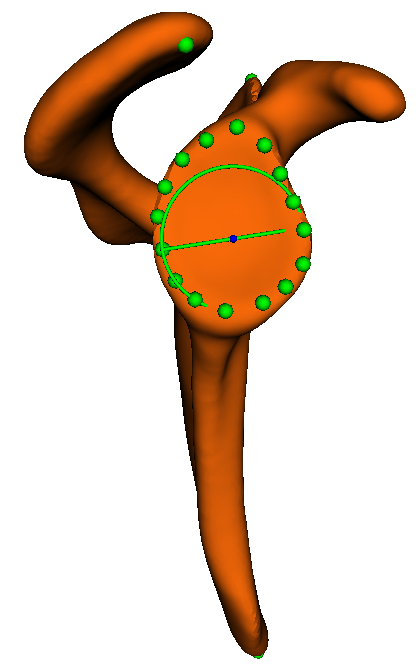}};
    
\node[inner sep=0pt] at (4.65,-9.5)
    {\includegraphics[width=.058\textwidth]{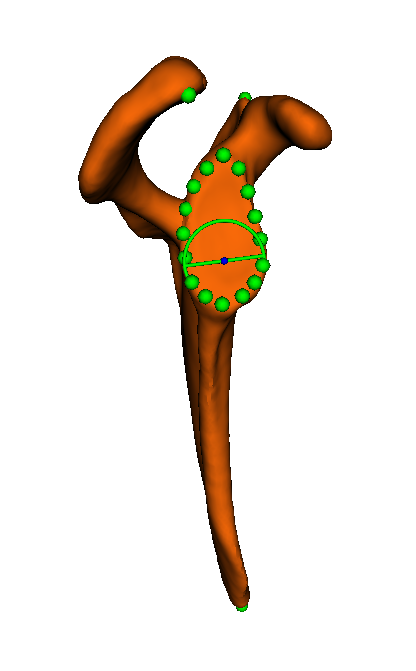}};
\node[inner sep=0pt] at (6.15,-9.5)
    {\includegraphics[width=.058\textwidth]{GW_mean.png}};
\node[inner sep=0pt] at (7.65,-9.5)
    {\includegraphics[width=.058\textwidth]{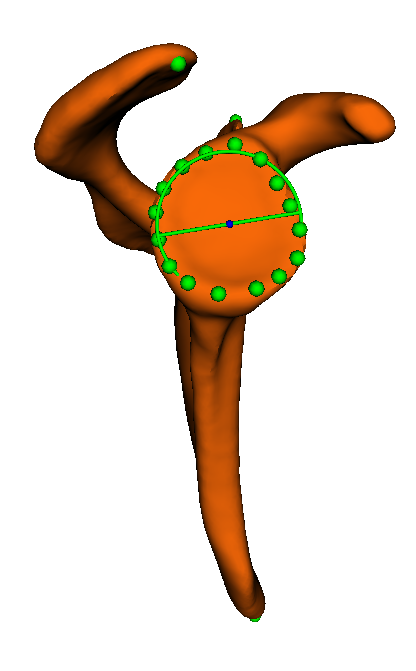}};

\node[anchor=north] at (1.5, 1.7) {\scalebox{.6}{\textbf{ANAT$_{\text{SSM}}$}}};
\draw[line width=.1mm, {Latex[length=4pt, width=4pt]}-{Latex[length=4pt, width=4pt]}] (-.25,1.3) -- (3.25,1.3);
\draw[line width=.1mm] (1.5,1.35) -- (1.5,1.25);
\draw[line width=.1mm] (0,1.35) -- (0,1.25);
\draw[line width=.1mm] (3,1.35) -- (3,1.25);
\node at (1.5, 1.1) {\scalebox{.6}{$\mu$}};
\node at (3, 1.1) {\scalebox{.6}{+3$\sigma_{c_j}$}};
\node at (0, 1.1) {\scalebox{.6}{-3$\sigma_{c_j}$}};

\node[anchor=north] at (6.25, 1.7) {\scalebox{.6}{\textbf{OC-ANAT$_{\text{SSM}}$}}};
\draw[line width=.1mm, {Latex[length=4pt, width=4pt]}-{Latex[length=4pt, width=4pt]}] (4.5,1.3) -- (8,1.3);
\draw[line width=.1mm] (4.75,1.35) -- (4.75,1.25);
\draw[line width=.1mm] (6.25,1.35) -- (6.25,1.25);
\draw[line width=.1mm] (7.75,1.35) -- (7.75,1.25);
\node at (6.25, 1.1) {\scalebox{.6}{$\mu$}};
\node at (7.75, 1.1) {\scalebox{.6}{+3$\sigma_{c_j}$}};
\node at (4.75, 1.1) {\scalebox{.6}{-3$\sigma_{c_j}$}};

\node[rotate=90] at (-1.35, -1.9) {\scalebox{.525}{Femur}};
\node[rotate=90] at (-1.35, -7.6) {\scalebox{.525}{Scapula}};

\node[rotate=90] at (-.95, 0) {\scalebox{.525}{Neck Shaft Angle}};
\node[rotate=90] at (-.95, -1.9) {\scalebox{.525}{Femoral Version}};
\node[rotate=90] at (-.95, -3.8) {\scalebox{.525}{Head Diameter}};
\node[rotate=90] at (-.95, -5.7) {\scalebox{.525}{Critical Shoulder Angle}};
\node[rotate=90] at (-.95, -7.6) {\scalebox{.525}{Glenoid Version}};
\node[rotate=90] at (-.95, -9.5) {\scalebox{.525}{Glenoid Width}};

\draw[line width=.2mm, color=black] (8.5,-0.95) -- (8.5,0.95) -- (-1.15,0.95) -- (-1.15, -0.95);
\draw[line width=.2mm, color=black] (8.5,-2.85) -- (8.5,-0.95) -- (-1.15,-0.95) -- (-1.15, -2.85);
\draw[line width=.2mm, color=black] (8.5,-4.75) -- (8.5,-2.85) -- (-1.15,-2.85) -- (-1.15, -4.75);
\draw[line width=.2mm, color=black] (8.5,-6.65) -- (8.5,-4.75) -- (-1.15,-4.75) -- (-1.15, -6.65);
\draw[line width=.2mm, color=black] (8.5,-8.55) -- (8.5,-6.65) -- (-1.15,-6.65) -- (-1.15, -8.55);
\draw[line width=.2mm, color=black] (8.5,-10.45) -- (8.5,-8.55) -- (-1.15,-8.55) -- (-1.15, -10.45) -- cycle;

\draw[line width=.2mm, color=black] (-.75,-10.45) -- (-.75,1.7) -- (3.875,1.7) -- (3.875,-10.45);
\draw[line width=.2mm, color=black] (3.875,1.7) -- (8.5,1.7) -- (8.5,0.95);
\draw[line width=.2mm, color=black] (-1.15,-10.45) -- (-1.55,-10.45) -- (-1.55,0.95) -- (-1.15,0.95);
\draw[line width=.2mm, color=black] (-1.15,-4.75) -- (-1.55,-4.75);

\node at (1.5, -.8) {\scalebox{.525}{$\beta_{NSA}$}};
\node at (1.5, -2.7) {\scalebox{.525}{$\beta_{FV}$}};
\node at (1.5, -4.6) {\scalebox{.525}{$\beta_{HD}$}};
\node at (1.5, -6.5) {\scalebox{.525}{$\beta_{CSA}$}};
\node at (1.5, -8.4) {\scalebox{.525}{$\beta_{GV}$}};
\node at (1.5, -10.3) {\scalebox{.525}{$\beta_{GW}$}};

\node at (6.15, -.8) {\scalebox{.525}{$\Tilde{\beta}_{NSA}$}};
\node at (6.15, -2.7) {\scalebox{.525}{$\Tilde{\beta}_{FV}$}};
\node at (6.15, -4.6) {\scalebox{.525}{$\Tilde{\beta}_{HD}$}};
\node at (6.15, -6.5) {\scalebox{.525}{$\Tilde{\beta}_{CSA}$}};
\node at (6.15, -8.4) {\scalebox{.525}{$\Tilde{\beta}_{GV}$}};
\node at (6.15, -10.3) {\scalebox{.525}{$\Tilde{\beta}_{GW}$}};

\spy [Dandelion] on (-.025,.6) in node [left] at (1.15,-.425);
\spy [Dandelion] on (1.475,.62) in node [left] at (1.15,.425);
\spy [Dandelion] on (2.985,.67) in node [left] at (2.65,.425);

\spy [Dandelion] on (4.625,.63) in node [left] at (5.80,-.425);
\spy [Dandelion] on (6.125,.62) in node [left] at (5.80,.425);
\spy [Dandelion] on (7.625,.62) in node [left] at (7.30,.425);

\spy [Dandelion] on (-.025,-3.28) in node [left] at (1.15,-4.225);
\spy [Dandelion] on (1.475,-3.15) in node [left] at (1.15,-3.375);
\spy [Dandelion] on (2.965,-3.055) in node [left] at (2.65,-3.375);

\spy [Dandelion] on (4.625,-3.15) in node [left] at (5.80,-4.225);
\spy [Dandelion] on (6.125,-3.15) in node [left] at (5.80,-3.375);
\spy [Dandelion] on (7.625,-3.15) in node [left] at (7.30,-3.375);

\end{scope}
\end{tikzpicture}
\end{adjustbox}
\caption{Visual comparison of shape variation patterns arising from anatomical parameters of femoral ANAT$_{\text{SSM}}$ ($\beta_{NSA}$, $\beta_{FV}$, $\beta_{HD}$) and OC-ANAT$_{\text{SSM}}$ ($\Tilde{\beta}_{NSA}$, $\Tilde{\beta}_{FV}$, $\Tilde{\beta}_{HD}$), as well as scapular ANAT$_{\text{SSM}}$ ($\beta_{CSA}$, $\beta_{GV}$, $\beta_{GW}$) and OC-ANAT$_{\text{SSM}}$ ($\Tilde{\beta}_{CSA}$, $\Tilde{\beta}_{GV}$, $\Tilde{\beta}_{GW}$). Each anatomical parameter $c_j$ is shown with the shape varied between three standard deviations ($\pm3\sigma_{c_j}$) from the mean shape ($\mu$). Please refer to the supplementary material for visualization of the remaining anatomical parameters.}
\label{fig:visual_comparison_shape_variation}
\end{figure*}

\section{Discussion}
\label{sec:discussion}

This study illustrated the development and evaluation of two novel SSMs, ANAT$_{\text{SSM}}$ and OC-ANAT$_{\text{SSM}}$, controlled by anatomical parameters derived from morphometric analysis. Experiments performed on the femoral and scapular bone shapes demonstrated that both SSMs integrated the statistical relationship between shape coefficients and anatomical parameters while preserving most of the BASE$_{\text{SSM}}$ shape variability. The exploration of the novel anatomical parameter representation further confirmed the feasibility and validity of the two proposed models. We finally validated that the added orthogonality constraints on OC-ANAT$_{\text{SSM}}$ resulted in independent shape variation patterns. To the best of our knowledge, the proposed modeling framework is the first illustration to build an SSM using anatomically relevant measures on a bone structure. The proposed models have deeper impact in multiple stages of any computer assisted surgery or even in intuitive understanding of morphometry and how multiple relevant anatomical measures are correlated.

\subsection{Synthetically generated population characteristics}
\label{sec:comparison_synthetic_real}

The proposed models were built on synthetically generated shapes from BASE$_{\text{SSM}}$. Although the descriptive statistics of synthetic data matched well with the original femoral and scapular datasets of real shapes, it was important to refer to the literature for larger acceptance. Thus, we compared the descriptive statistics of each of the anatomical parameters (Fig. \ref{fig:anatomical_parameters_histograms}) and their correlations (Table \ref{tab:correlation_anatomical_parameters}) with the values reported in the literature \cite{von_schroeder_osseous_2001, verhaegen_can_2018,hartel_determination_2016, terzidis_gender_2012, unnanuntana_evaluation_2010, verma_morphometry_2017, moor_is_2013, daggett_correlation_2015}. The means and variances computed on the synthetic population (Fig. \ref{fig:anatomical_parameters_histograms}) were comparable to the values reported from real femoral and scapular population: NSA ($121.8\pm3.9\degree$) \cite{hartel_determination_2016}, FV ($13.9\pm6.5\degree$) \cite{hartel_determination_2016}, BW ($83.9\pm6.3$ mm) \cite{terzidis_gender_2012}, HD ($52.1\pm4.4$ mm) \cite{unnanuntana_evaluation_2010}, FL ($42.8\pm2.9$ cm) \cite{verma_morphometry_2017},  CSA ($33.1\pm2.1\degree$) \cite{moor_is_2013}, GI ($11.0\pm4.0\degree$) \cite{verhaegen_can_2018}, GV ($-7.0\pm4.0\degree$) \cite{verhaegen_can_2018}, GH ($36.4\pm3.6$ mm) \cite{von_schroeder_osseous_2001}, GW ($28.6\pm3.3$ mm) \cite{von_schroeder_osseous_2001} and SL ($155.0\pm16.0$ mm) \cite{von_schroeder_osseous_2001}. Although few studies focused on analyzing the correlation between anatomical parameters, the scapular pair CSA and GI, and the femoral couple GW and FL were respectively reported to be positively correlated \cite{yazar_is_2012, daggett_correlation_2015}. Our observations on the synthetic data were similar (Table \ref{tab:correlation_anatomical_parameters}) and thus provided further validation of the population generated through BASE$_{\text{SSM}}$.

\subsection{Mapping assessment}
\label{sec:mapping_assessment}

We assumed that a linear mapping existed between the anatomical parameters and shape coefficients. While this could be a reasonable solution, the assumption may not be entirely valid. PCA linearizes the shape space during SSM building process, but whether the anatomical parameter representation also gets linearized is not yet understood. Further to this, we derived the inverse of this mapping to obtain ANAT$_{\text{SSM}}$ using the Moore-Penrose pseudo-inverse whose algebraic formula is valid only if the matrix is of full rank. Hence, the  anatomical parameters selected to build the models should respect this condition. Moreover, since the matrix $Q$ is a best fit solution, the approximated matrix $K$ derived after solving the orthogonal Procrustes problem needs to be evaluated to assure that it integrates the statistical relationship between shape coefficients and anatomical parameters (as in Section \ref{sec:validation_matrices}).  

\subsection{Benefits for clinical practice}
\label{sec:benefits_clinical_practice}

While previous studies already proposed to employ anatomical parameters for shape modeling, these works focused on predicting the complete shape from given partial observation \cite{albrecht_posterior_2013, blanc_statistical_2012}. Albrecht et al. \cite{albrecht_posterior_2013} proposed a posterior shape model to build a posterior distribution of the whole shape given the known parts, which one of the clinical applications consists in reconstructing a model of the premorbid shape as in \cite{salhi_statistical_2020, plessers_virtual_2018}. Additionally, the work of Blanc et al. \cite{blanc_statistical_2012} used anthropometric (patient age, weight, and height) and morphometric (bone height, width, and orientation) information to improve the prediction of the complete shape from sparse observation. To this end, known points information and meta-variables were concatenated into a single vector of predictors. Hence, our work differs from these studies as it aims at mapping two latent representations to build a new parameterization derived from anatomical parameters and therefore at controlling the whole shape variations through clinically relevant modes.

The integration of the two models in clinical workflow could provide significant benefits for the understanding of the relationship between bone shape and joint biomechanics, and the planning of patient-specific intervention. Specifically, as anatomical parameters can be used to characterize joint kinematics \cite{laxafoss_alignment_2013}, ANAT$_{\text{SSM}}$ could provide a formalized relationship between shape and biomechanics through controlled changes in anatomical parameters. This information is crucial to analyze morphological and biomechanical changes over time, and develop more accurate diagnostic tools. Furthermore, as ANAT$_{\text{SSM}}$ integrated correlation between anatomical parameters, this model could be used to analyze the difference in these correlations among different populations. Specifically, ANAT$_{\text{SSM}}$ could help understand the variation in bone morphometry across various gender and age groups. 

While OC-ANAT$_{\text{SSM}}$ is artificial due to the forced induction of orthogonal constraints, it may not reflect the true relationship between anatomical parameters. Hence, as opposed to ANAT$_\text{SSM}$, OC-ANAT$_\text{SSM}$ is not suitable to compare correlations among different groups, since the same identity covariance matrix would arise for different populations. However, this model could be effectively embedded into pre-surgery planning tools, as the orthogonality constraints would allow the clinician to modify one anatomical parameter at a time - as typically done during the surgery. For instance, in the context of bone loss in the glenoid region, the OC-ANAT$_{\text{SSM}}$ could be employed, first to predict the missing scapular bone \cite{plessers_virtual_2018, salhi_statistical_2020}, and next to modify the glenoid parameters independently without altering the other ones. This would allow the clinician to objectively calibrate the patient-specific model leading to improved accuracy and reliability of shoulder joint replacement procedures.

\subsection{Limitations}
\label{sec:limitations}

This study has certain limitations which are categorically listed in this section. First, our experiments were conducted on only two anatomical structures. Hence, the genericity of the method needs to be further evaluated. However, as formalized in the Sections \ref{sec:anat_ssm} and \ref{sec:ind_anat_ssm}, the proposed ANAT$_{\text{SSM}}$ and OC-ANAT$_{\text{SSM}}$ can be derived from any BASE$_{\text{SSM}}$ and for any anatomical structures. Furthermore, the number of anatomical parameters selected to develop ANAT$_{\text{SSM}}$ and OC-ANAT$_{\text{SSM}}$ is limited by $n - 1$, with $n$ the number of shapes in the training set which is typically greater than 50. Although we employed a limited number of anatomical parameters, the proposed framework allows us to study localized parameters (e.g. GV) and their relationship with global ones (e.g. SL).

Second, the automatic derivation of anatomical measurements was based on dense point-to-point correspondence and anatomical landmarks selected on the mean shape of the BASE$_{\text{SSM}}$. While the reliability of the selection of the majority of those landmarks (angulus inferior, angulus superior, trigonum spinae, inferior and superior points on the glenoid rim, and the most lateral point on the acromion) has been found to be excellent as reported in the literature \cite{borotikar_augmented_2017}, landmark transfer is still sensitive to the set of selected landmarks \cite{borotikar_augmented_2017}. This could affect the accuracy of the automatic derivation method and thus deteriorate the quality of the anatomical parameters used to build ANAT$_{\text{SSM}}$ and OC-ANAT$_{\text{SSM}}$. Hence, it could be beneficial to use automatic geometrical methods \cite{ghafurian_computerized_2016} that do not depend on landmarks and may further improve the reliability of the measurements.

Third, our experiments revealed that the six scapular parameters only accounted for $63.1\%$ of total shape variability (Fig. \ref{fig:shape_variability_anatomical_parameters}), indicating that more anatomical parameters are needed to attain the total BASE$_{\text{SSM}}$ shape variability. For instance, the coracoid morphology characterized by its length and thickness \cite{von_schroeder_osseous_2001} was not incorporated in the proposed models. While it is true that developing models with more anatomical parameters would increase their shape variability (as demonstrated by the comparison to sub-models) and provide a more complete description of the scapular shape, such a thorough representation is not needed in a clinical case. Hence, we made a deliberate attempt to limit the number of anatomical parameters and to select parameters relevant for morphometry around the glenoid region \cite{plessers_virtual_2018, salhi_statistical_2020}. Finally, we observed that the five femoral anatomical parameters represented $98.7\%$ of the total BASE$_{\text{SSM}}$ shape variability, illustrating the relative complexity to model the scapular morphology. Hence, the selection of the anatomical parameters should be determined according to the anatomical structure of interest and targeted applications. 

\section{Conclusion}
\label{sec:conclusion}


Understanding shape variations that are directly linked to anatomical measures of a biological structure is key in understanding the anatomo-physiological relationship between form and function. This paper introduced two data-driven generative models (ANAT$_{\text{SSM}}$ and OC-ANAT$_{\text{SSM}}$) controlled by anatomical parameters. These models were derived from the BASE$_{\text{SSM}}$ by integrating a mapping between shape coefficients and anatomical parameters. As opposed to traditional SSMs, the shape variation patterns of these models were tuned with selected anatomical parameters. Experiments conducted on the femoral and scapular bone shapes validated the predictive performance of the proposed models, and the shape variation patterns were found to agree with morphometrics. Such models will provide intuitive understanding to clinicians and surgeons on the selection of treatment strategies, and they will also improve the comprehension of joint pathomechanics in a population considered as risky based on anatomy. Future work will focus on extending the validation of the methodology to other anatomical structures. In addition, this approach will be integrated into biomechanical models to further investigate the relationship between bone shape and joint biomechanics.

\section*{Acknowledgment}
\noindent We would like to thank the Department of Anatomy at the Faculty of Medicine, CHRU Brest for making the scapular dry bones available and also the Department of Radiology for scanning the bones.

\bibliographystyle{ieeetr}
\bibliography{main}

\setcounter{table}{0}
\setcounter{section}{0}
\setcounter{figure}{0}
\setcounter{equation}{0}
\renewcommand{\thetable}{S\Roman{table}}
\renewcommand{\thesection}{S\Roman{section}}
\renewcommand{\thefigure}{S\arabic{figure}}
\renewcommand{\theequation}{S\arabic{equation}}

\section*{\Large{Supplementary Material}}

\vspace{1cm}

\section{BASE$_{\text{SSM}}$ robustness}

The robustness of the femoral and scapular BASE$_{\text{SSM}}$ was evaluated using the compactness $C(R)$, generality $G(R)$ and specificity $S(R)$ metrics, with $1\leq R \leq n - 1$ the number of retained principal components (Fig. \ref{fig:ssm_robustness}). The femoral and scapular BASE$_{\text{SSM}}$ were built using datasets of $n=50$ and $n=76$ shapes respectively. The metrics were defined as follows: 
\begin{align}
    C(R) &= \dfrac{\sum_{r=1}^{R} \lambda_{r}}{\sum_{i=1}^{n - 1} \lambda_{i}} \\
    G(R) &= \dfrac{1}{n - 1} \sum_{i=1}^{n - 1} \norm{s^{\prime}_i(R) - s_i}_2^2 \\
    S(R) &= \dfrac{1}{n - 1} \sum_{i=1}^{n - 1} \norm{s^{\prime\prime}_i(R) - s^{\prime}_i}_2^2
\end{align}
\noindent $s^{\prime}_i(R)$ was the best reconstruction of the instance $s_i$ from the model built excluding $s_i$ with $R$ principal components. Additionally, $s^{\prime\prime}_i(R)$ was a shape example randomly generated using the model with $R$ principal components and $s^{\prime}_i$ was the nearest instance of the training set to $s^{\prime\prime}_i(R)$. The specificity was computed using 200 randomly generated shapes.

\section{Automatic derivation of anatomical measurements from landmarks}

\noindent \textbf{Femoral landmarks.} The five femoral anatomical parameters (NSA, FV, BW, HD, and FL) were automatically computed using a set of 18 anatomical landmarks. Nine were equally distributed along the femoral head with one located at the most superior point of the femoral head (SFH). The nine remaining landmarks were positioned at anatomically relevant locations: most inferior, most medial and most posterior points of the medial condyle (IMC, MMC and PMC), most lateral and most posterior points of the lateral condyle (LLC and PLC), inferior and superior neck subcapital (ISN and SNS), facies patellaris saddle point (FP) and greater trochanter (GT) \cite{verma_morphometry_2017, hartel_determination_2016, terzidis_gender_2012, casciaro_towards_2014, unnanuntana_evaluation_2010, wei_approach_2020}.\\ 

\noindent \textbf{Femoral anatomical measurements.} Automatic extraction of the five femoral anatomical measures from the selected 18 landmarks comprised five steps. First, we fitted a sphere using least squares regression (LSR) on the nine points uniformly located on the femoral head. Center of this sphere was established as the femoral head center and its diameter defined the HD measure \cite{verma_morphometry_2017}. Second, we constructed the femoral neck axis as the line passing through the femoral head center and orthogonal to the ISN-SNS line \cite{wei_approach_2020}, and the femoral shaft axis as the FP-GT line \cite{hartel_determination_2016}. Third, the NSA measure was computed as the angle between the femoral neck axis and the femoral shaft axis \cite{hartel_determination_2016, casciaro_towards_2014, unnanuntana_evaluation_2010}. Fourth, the FV measure was calculated as the angle between the femoral neck axis and the PLC-PMC line projected onto the plane orthogonal to the femoral shaft axis \cite{hartel_determination_2016, casciaro_towards_2014}. Finally, the FL measure was determined from IMC to SFH \cite{verma_morphometry_2017}, and the BW measure was calculated from LLC to MMC \cite{terzidis_gender_2012}.\\

\noindent \textbf{Scapular landmarks.} The six scapular anatomical parameters (CSA, GI, GV, GH, GW, and SL) were automatically computed using a set of 20 anatomical landmarks. Sixteen landmarks were equally distributed along the glenoid rim, including glenoid superior (GS), glenoid inferior (GI), and eight inferior glenoid rim (IGR1-8) points, while four were placed at anatomically relevant locations: angulus inferior (AI), angulus superior (AS), trigonum spinae (TS), and the most lateral point on the acromion (LA) \cite{von_schroeder_osseous_2001, cherchi_critical_2016, plessers_virtual_2018, verhaegen_can_2018, burton_assessment_2019}.\\ 

\noindent \textbf{Scapular anatomical measurements.} Automatic extraction of the six scapular anatomical measures from the selected 20 landmarks comprised five steps. First, we defined the glenoid circle as a circle that best fitted the uniformly distributed points IGR1-8 by employing LSR \cite{plessers_virtual_2018}. Center of this circle was established as the glenoid center point and its diameter defined the GW measure \cite{burton_assessment_2019}. Second, we constructed a scapular plane using the glenoid center point, AI, and TS landmarks. An axial plane was constructed orthogonal to the scapular plane and parallel to trigonum spinae--glenoid center point (TS--GCP) axis. A glenoid plane was also established as the plane that best fitted the sixteen points on the glenoid rim. Third, the GV and GI measures were computed as the angle between the TS--GCP axis and the glenoid plane normal, projected to the axial plane and scapular plane respectively \cite{plessers_virtual_2018, verhaegen_can_2018}. Next, the CSA measure was computed as the angle formed by the line connecting GI and GS and a line drawn from GI to LA, projected onto the scapular plane \cite{cherchi_critical_2016}. Finally, the SL measure was determined from AI to AS landmarks, and the GH measure was calculated from GI to GS \cite{von_schroeder_osseous_2001}.

\section{Size of synthetic population}

\begin{figure*}[ht!]
\begin{adjustbox}{width=\textwidth}
\begin{tikzpicture}

\node[inner sep=0pt] at (0,0)
    {\includegraphics[width=.55\textwidth]{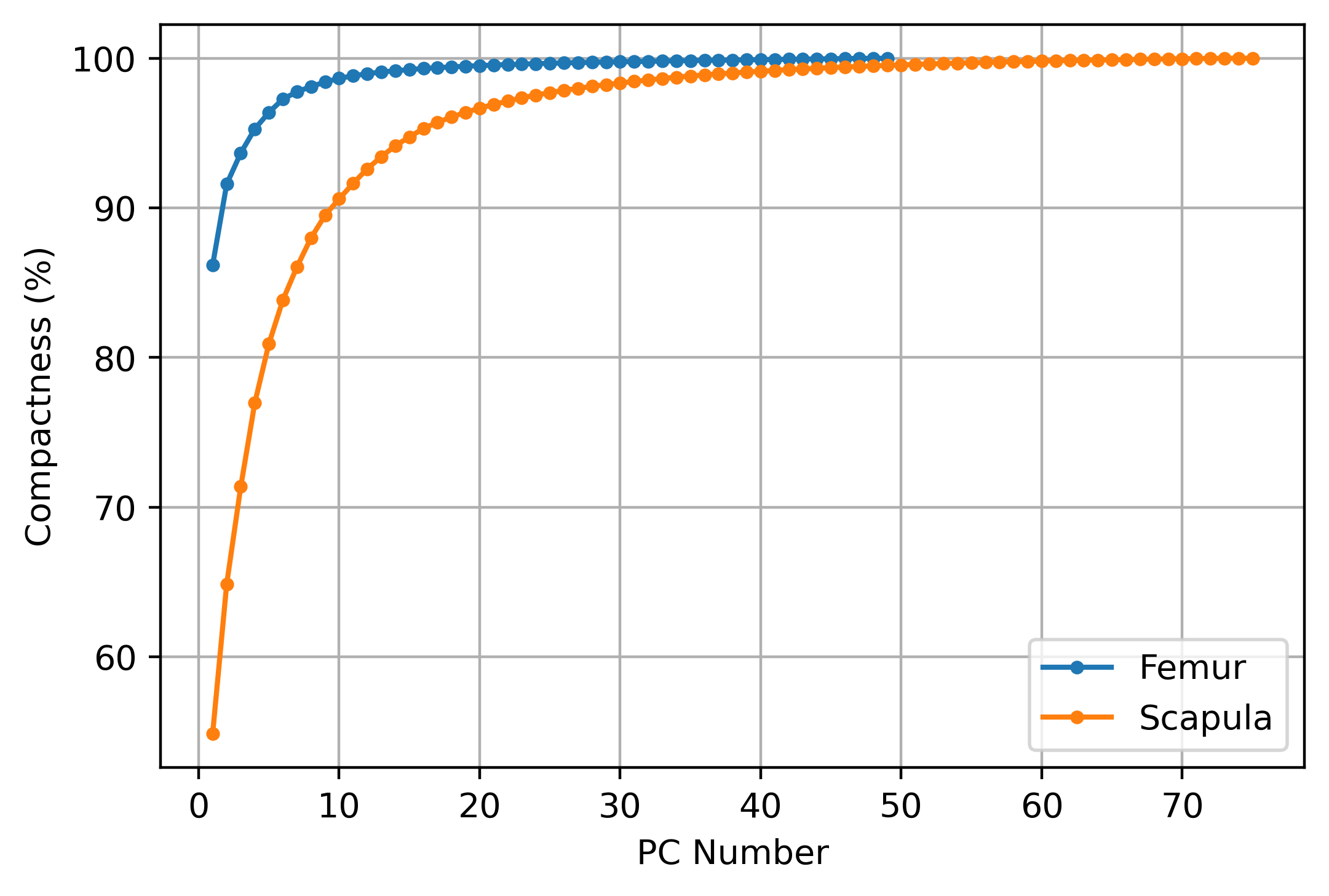}};
\node[inner sep=0pt] at (10,0)
    {\includegraphics[width=.55\textwidth]{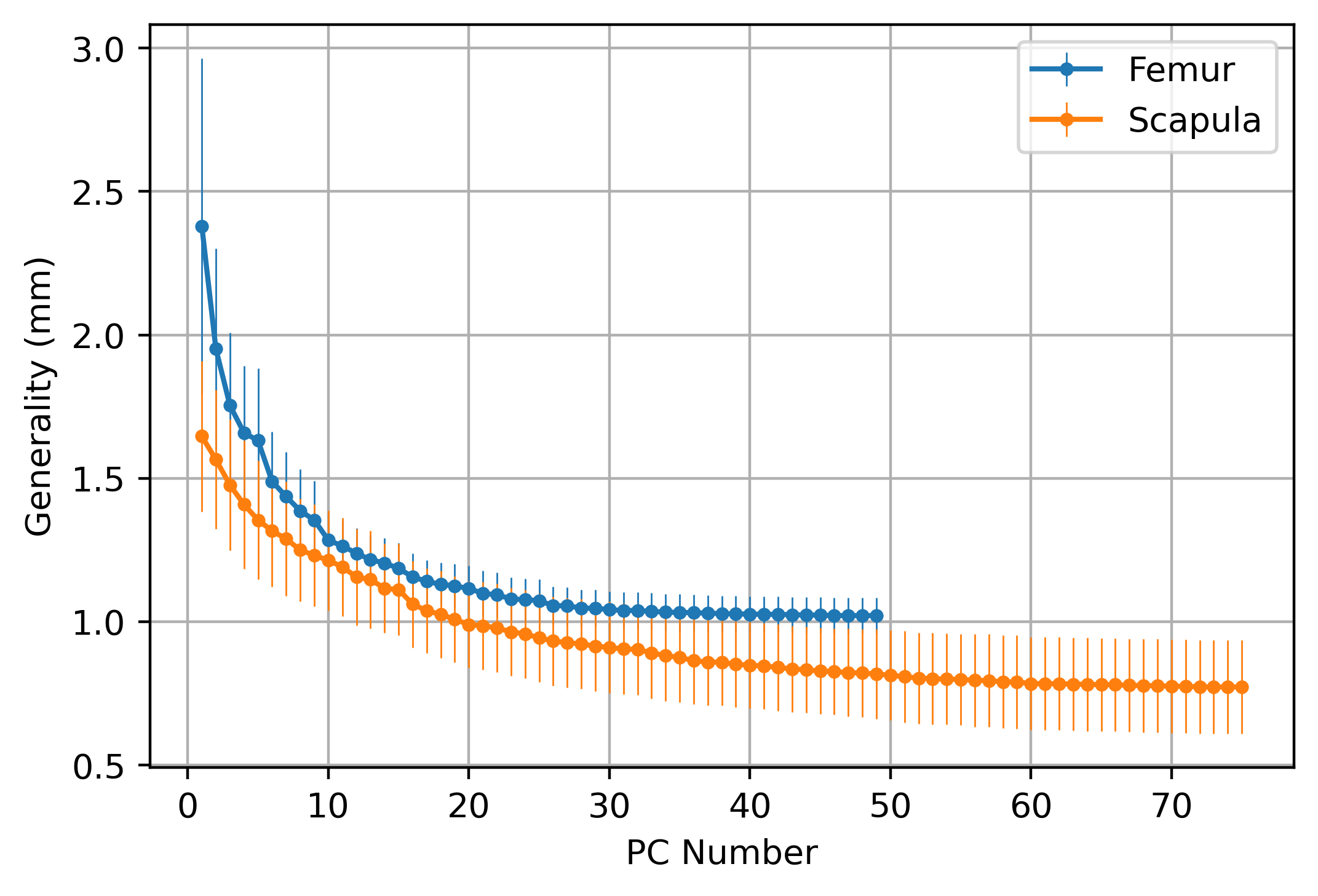}};
\node[inner sep=0pt] at (0,-7.5)
    {\includegraphics[width=.55\textwidth]{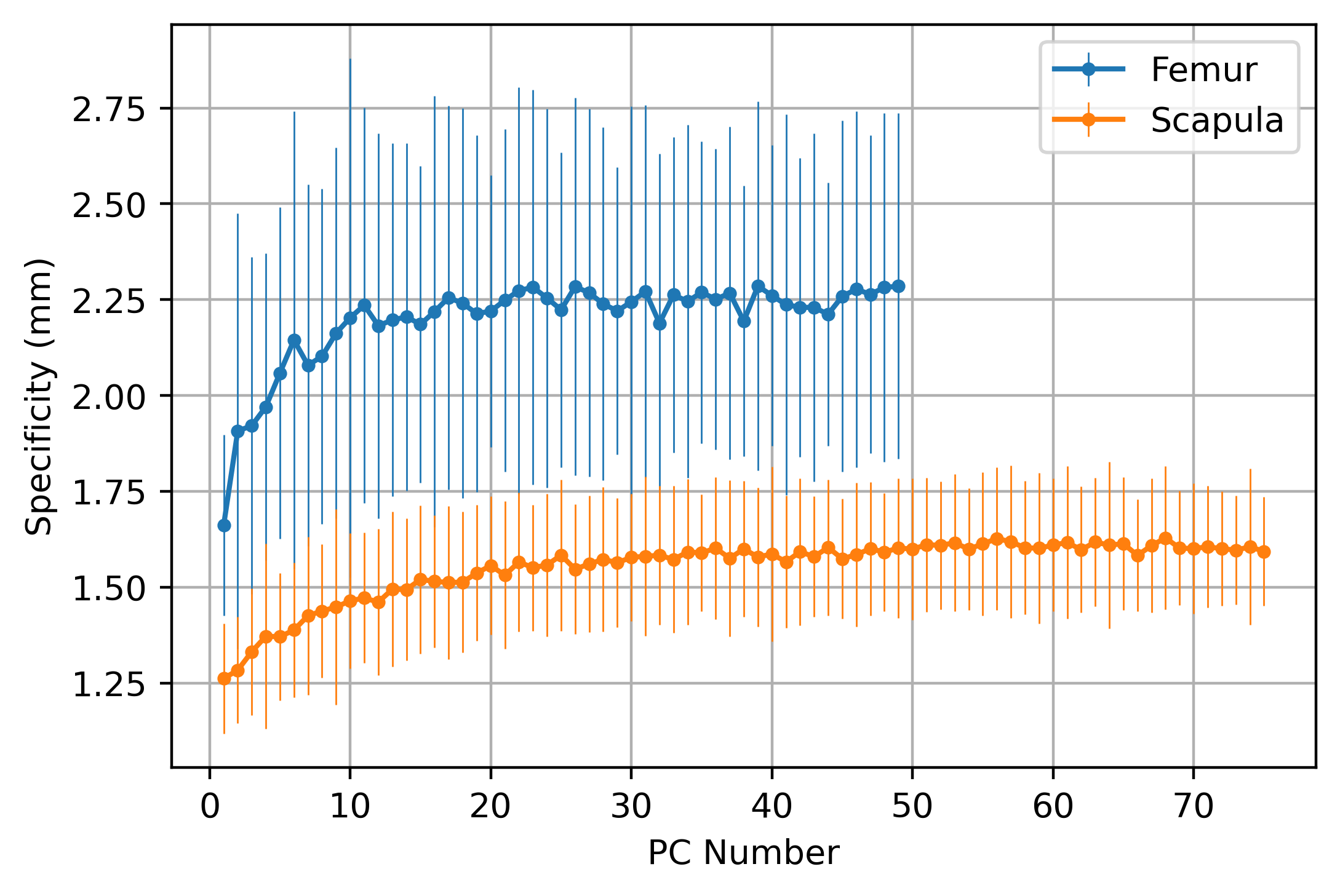}};
    
\node[anchor=west] at (6.35, -5.415) {\scalebox{1.25}{\textbf{Compactness:} Model’s ability to cover the}};
\node[anchor=west] at (6.35, -5.915) {\scalebox{1.25}{total variance.}};

\node[anchor=west] at (6.35, -6.915) {\scalebox{1.25}{\textbf{Generality:} Model’s ability to represent all}};
\node[anchor=west] at (6.35, -7.415) {\scalebox{1.25}{valid instances.}};

\node[anchor=west] at (6.35, -8.415) {\scalebox{1.25}{\textbf{Specificity:} Model’s ability to only represent}};
\node[anchor=west] at (6.35, -8.915) {\scalebox{1.25}{valid instances of the object.}};

\draw[line width=.3mm, color=black, rounded corners=10] (6.17, -4.95) -- (6.17,-9.38) -- (14.76,-9.38) -- (14.76,-4.95) -- cycle;

\end{tikzpicture}
\end{adjustbox}
\captionof{figure}{Compactness ($\%$), generality (mm), and specificity (mm) measures of femoral and scapular BASE$_{\text{SSM}}$. The first 4 principal component of the femoral BASE$_{\text{SSM}}$ represented 95$\%$ of total variance while its generality (respectively specificity) was found average, ranging from $1.0$ to $2.4$ mm (respectively from $1.7$ to $2.3$ mm). With respect to the scapular BASE$_{\text{SSM}}$, its first 15 principal component represented 95$\%$ of total variance and its generality (respectively specificity) was excellent, ranging from $0.8$ to $1.6$ mm (respectively from $1.3$ to $1.6$ mm).}
\label{fig:ssm_robustness}

\vspace*{\floatsep}
\centering
\begin{adjustbox}{width=.5\textwidth}
\begin{tikzpicture}

\node[inner sep=0pt] at (0,0)
    {\includegraphics[width=.5\textwidth]{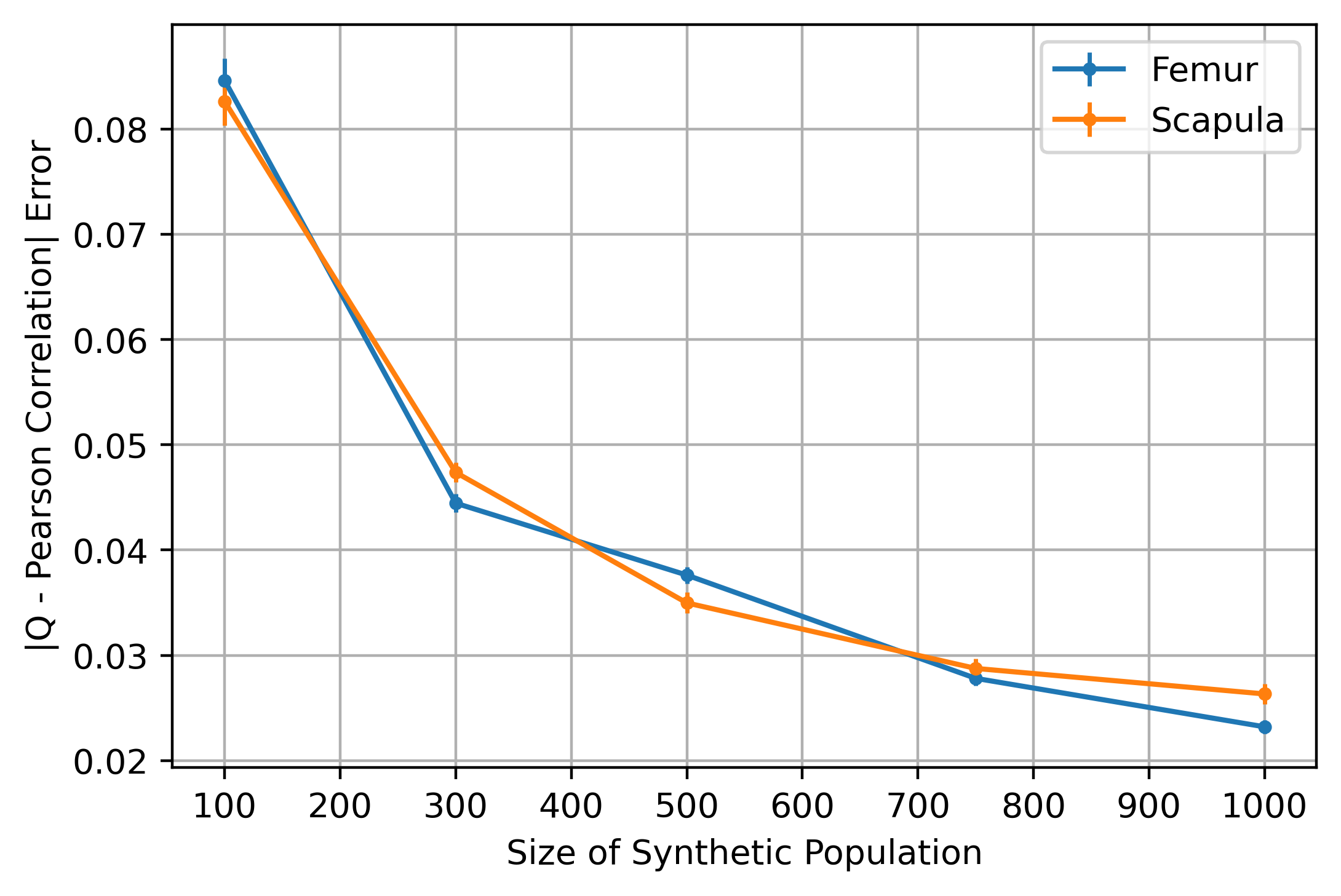}};

\end{tikzpicture}
\end{adjustbox}
\captionof{figure}{Absolute error between the learned matrix Q and the Pearson correlation coefficients computed between the shape coefficients and anatomical parameters. The absolute error was evaluated for different size of synthetic population (100, 300, 500, 750, 1000) and ranged from 0.08 to 0.03.}
\label{fig:population_size}
\end{figure*}

To assess the optimal size of the synthetic population generated to build the anatomical models, we computed the absolute error between the learned matrix $Q$ and the Pearson correlation coefficients for sizes ranging from 100 to 1000 (Fig. \ref{fig:population_size}). The error ranged from 0.08 to 0.03 with a steep decline between 100 and 300, and a slight decrease between 750 and 1000. Hence, the size of the synthetic population was optimally set to 1000 samples.

\section{Evaluation of orthogonality}

\begin{figure*}[ht!]
\begin{adjustbox}{width=\textwidth}
\begin{tikzpicture}

\node[inner sep=0pt] at (5,8)
    {\includegraphics[width=.55\textwidth]{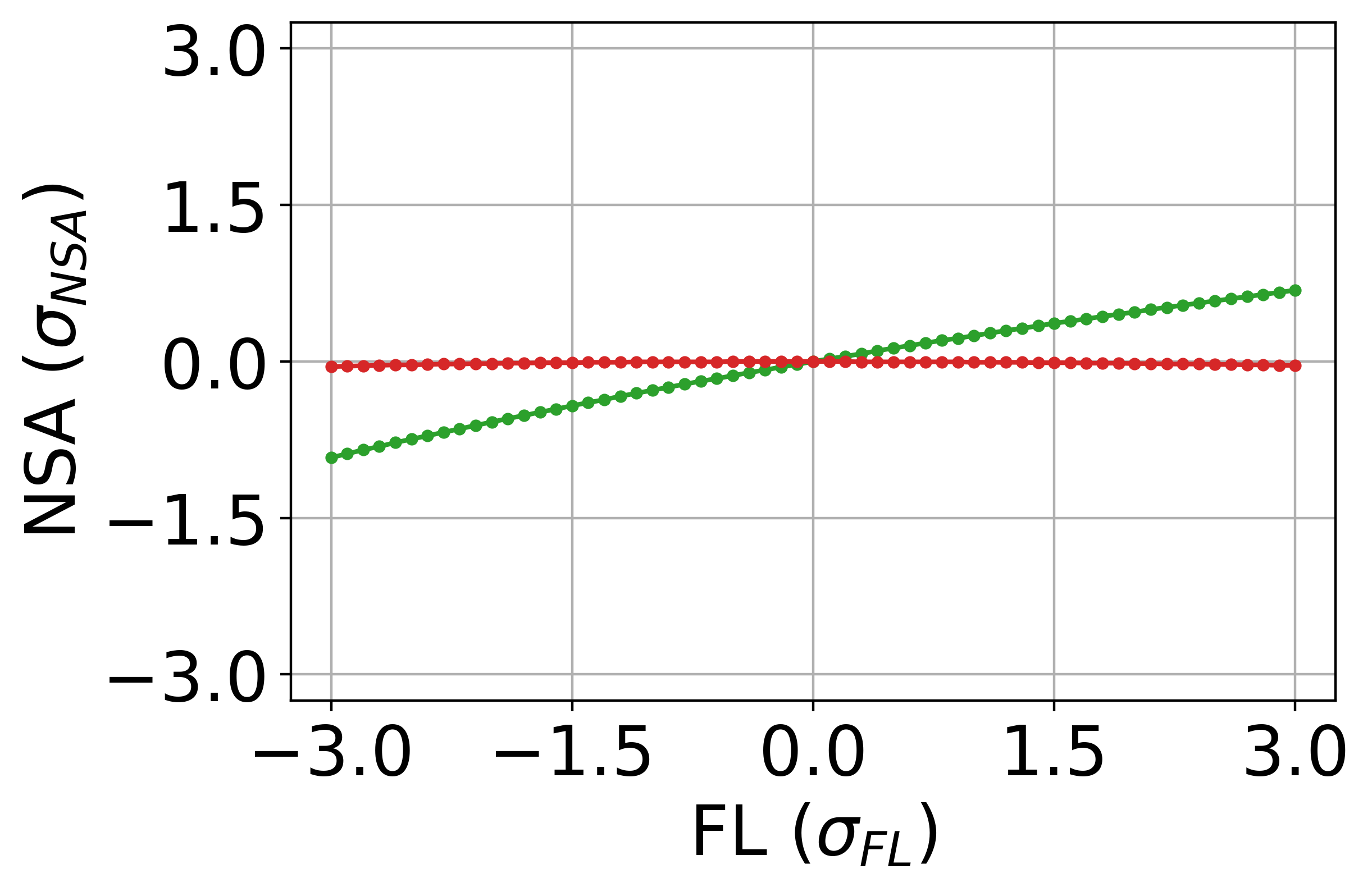}};
\node[inner sep=0pt] at (15,8)
    {\includegraphics[width=.55\textwidth]{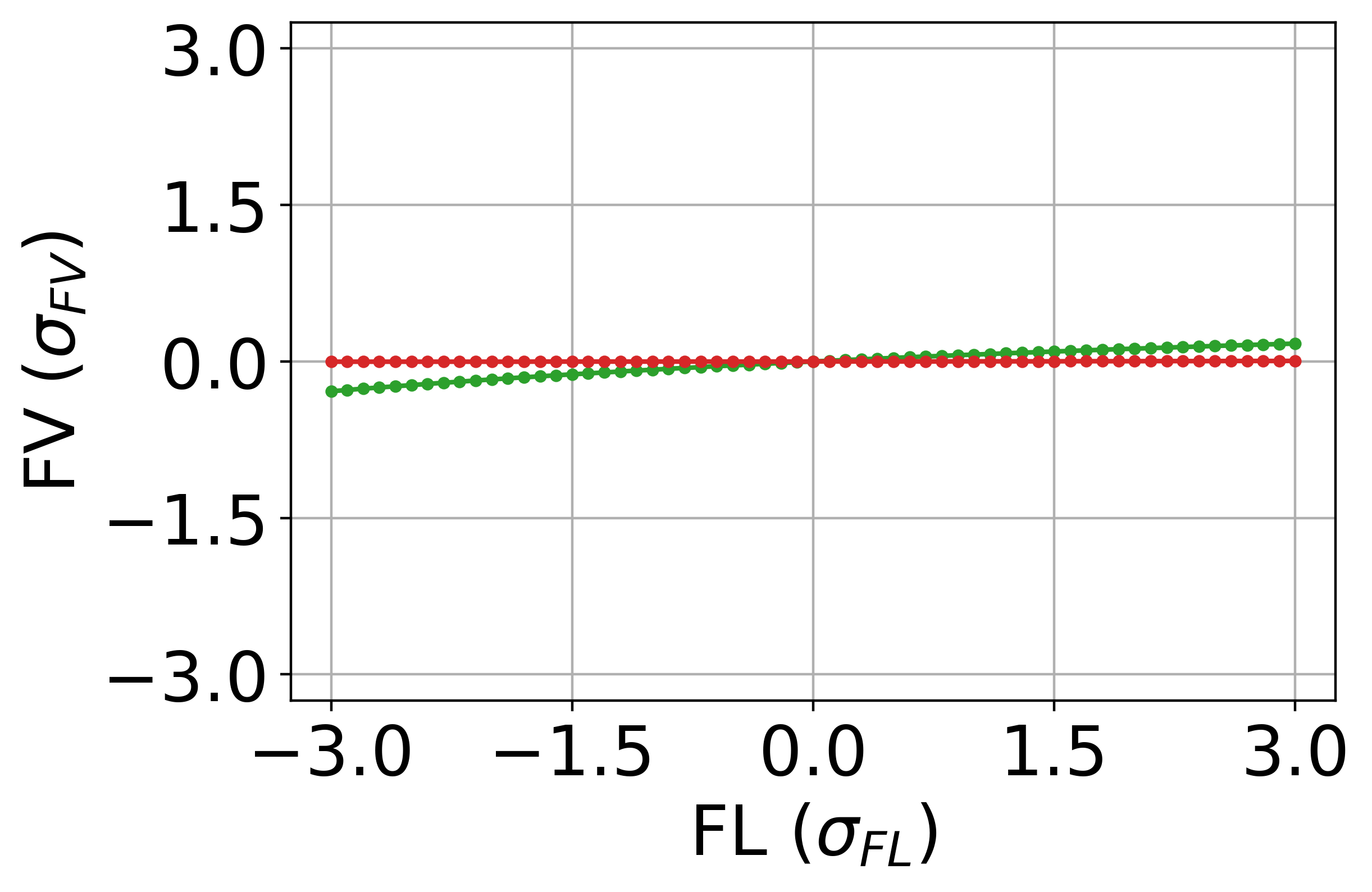}};
\node[inner sep=0pt] at (25,8)
    {\includegraphics[width=.55\textwidth]{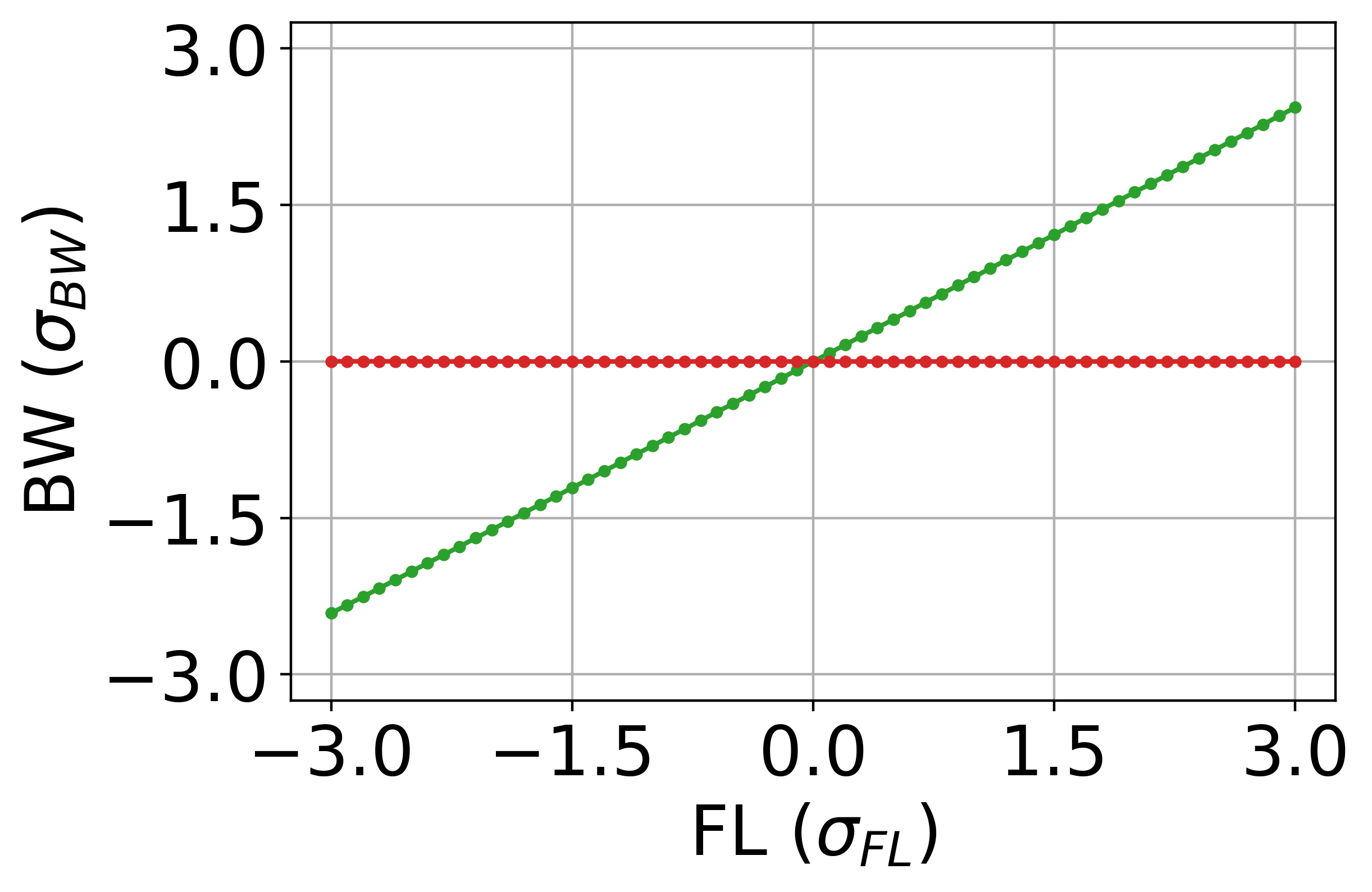}};
\node[inner sep=0pt] at (35,8)
    {\includegraphics[width=.55\textwidth]{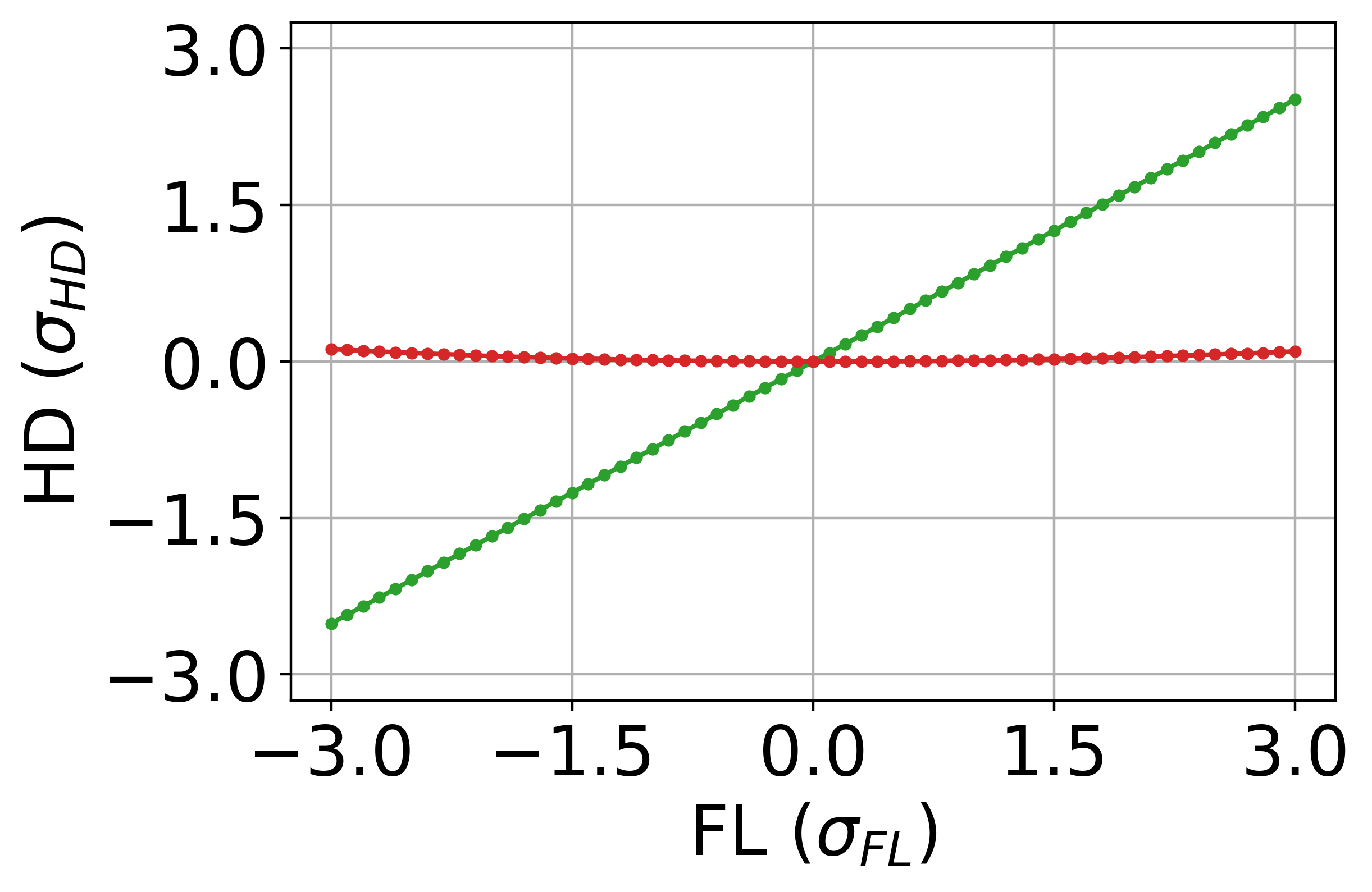}};
    
\node[inner sep=0pt] at (0,0)
    {\includegraphics[width=.55\textwidth]{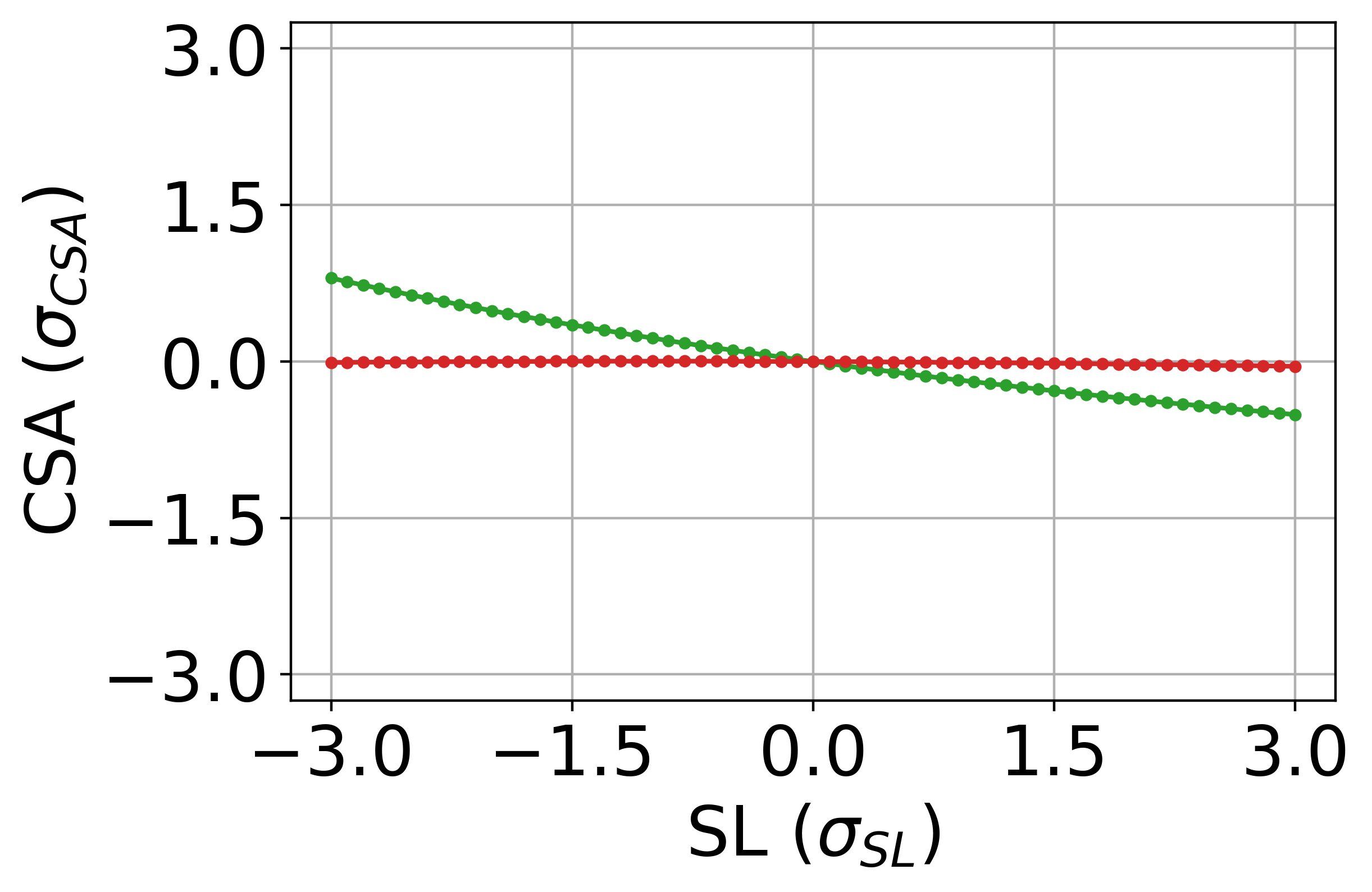}};
\node[inner sep=0pt] at (10,0)
    {\includegraphics[width=.55\textwidth]{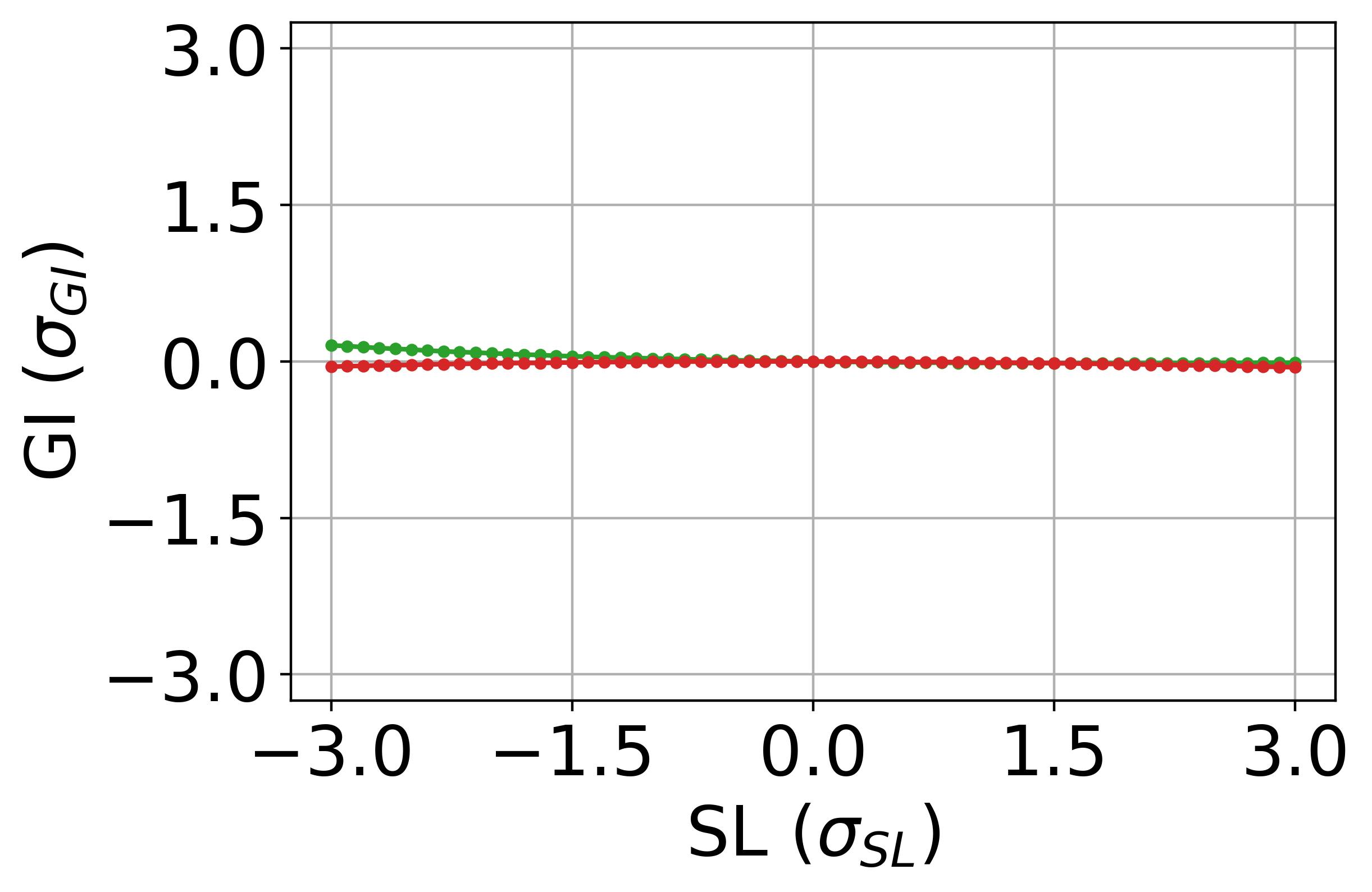}};
\node[inner sep=0pt] at (20,0)
    {\includegraphics[width=.55\textwidth]{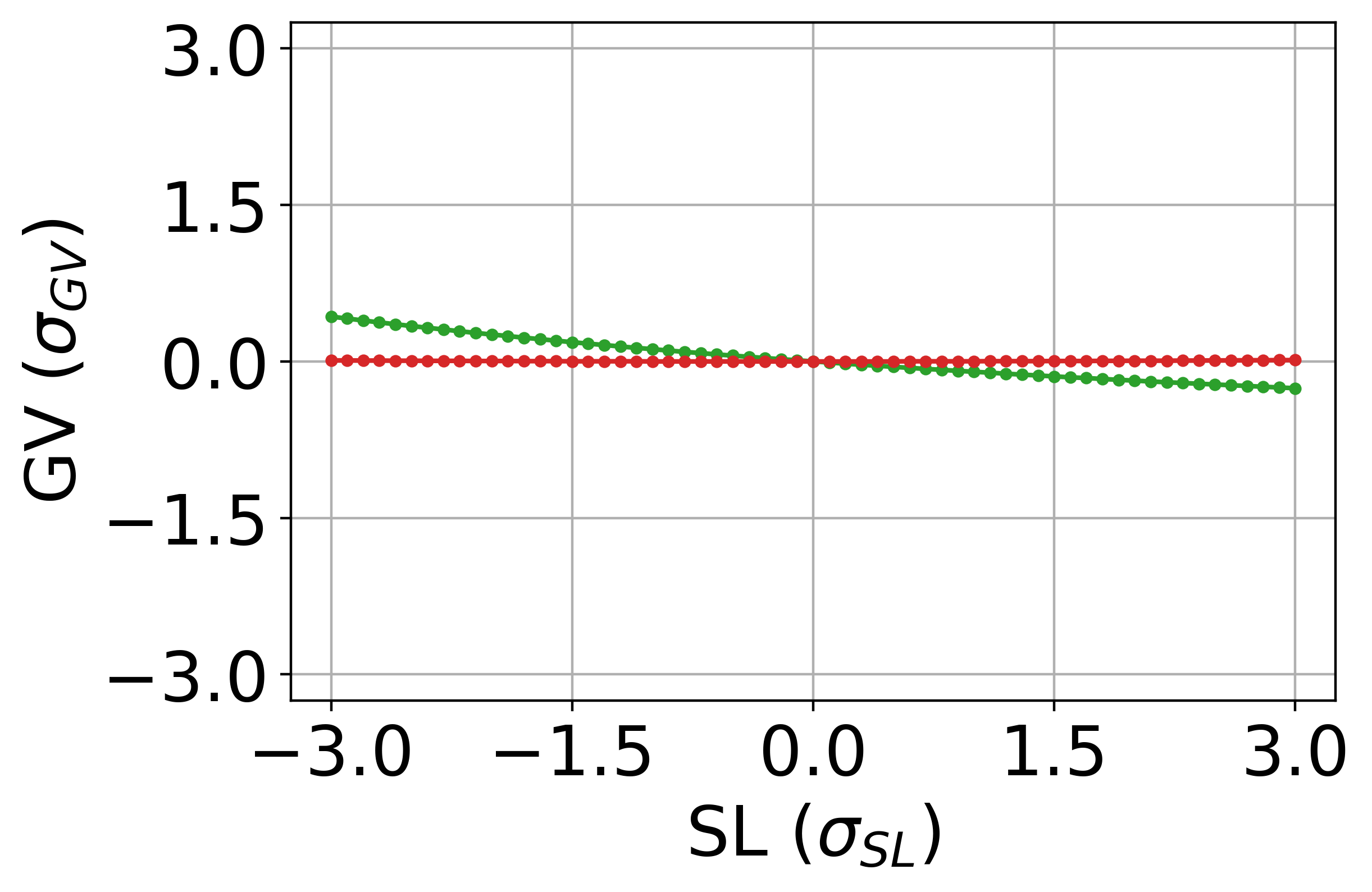}};
\node[inner sep=0pt] at (30,0)
    {\includegraphics[width=.55\textwidth]{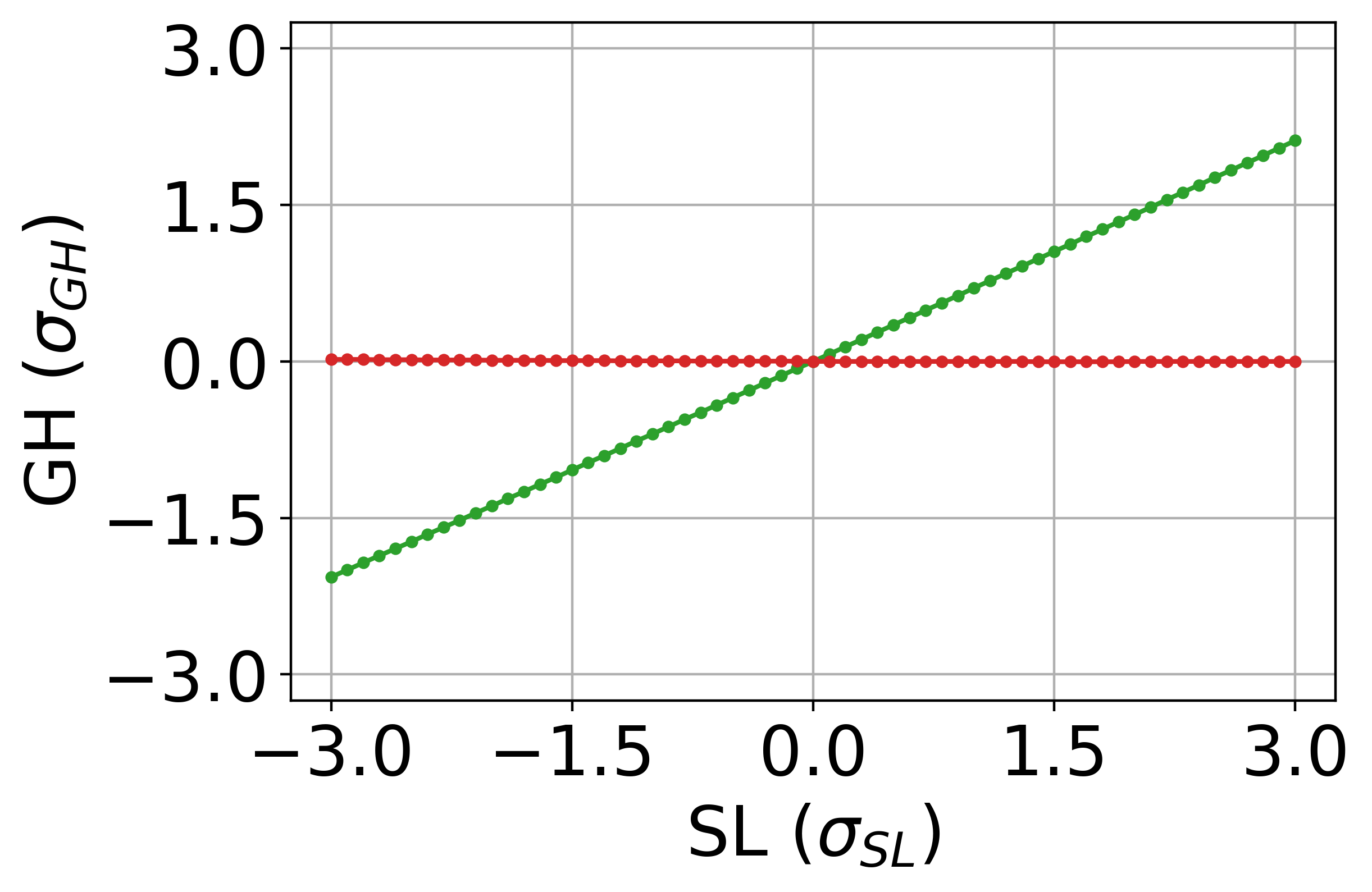}};
\node[inner sep=0pt] at (40,0)
    {\includegraphics[width=.55\textwidth]{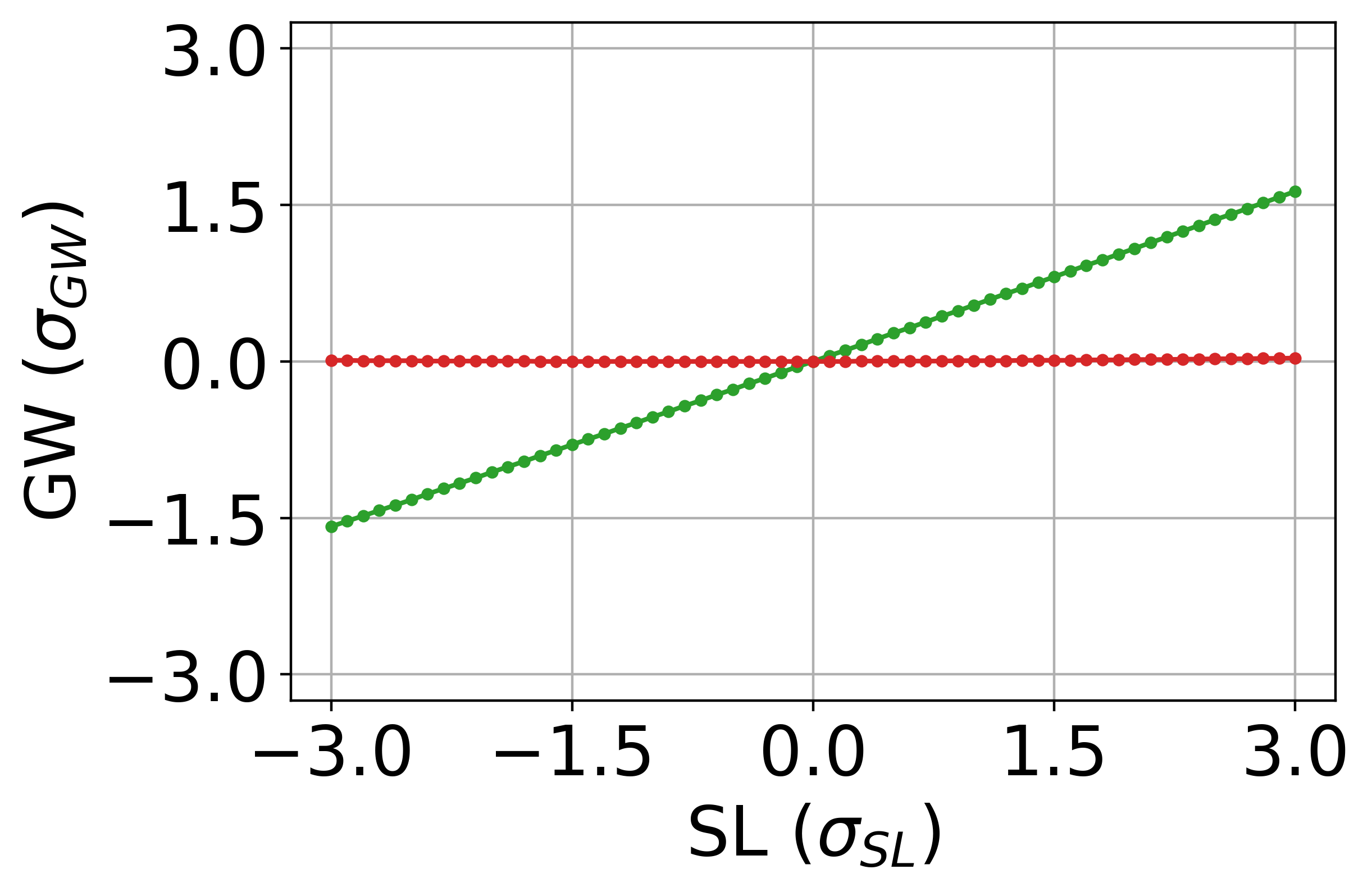}};
 
\node[inner sep=0pt] at (20,-5)
    {\includegraphics[width=.65\textwidth]{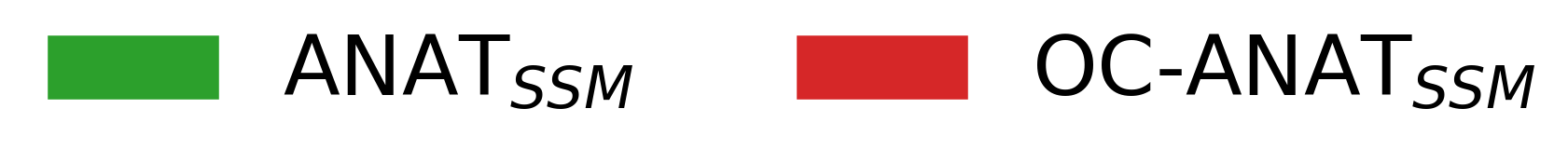}};
    
\node at (20,12.25) {\scalebox{2.75}{Femur}};
\node at (20,4.25) {\scalebox{2.75}{Scapula}};
    
\draw[line width=.1mm] (14, -4.25) -- (26, -4.25) -- (26,-5.5) -- (14,-5.5) -- cycle;

\end{tikzpicture}
\end{adjustbox}
\captionof{figure}{Evaluation of orthogonality by comparing the anatomical measurements of ANAT$_{\text{SSM}}$ and OC-ANAT$_{\text{SSM}}$ models with one varying anatomical parameter. The femoral varying parameter corresponded to the FL while the SL was the moving parameter in scapular models.}
\label{fig:orthogonality_measure}

\vspace*{\floatsep}

\captionof{table}{Weights of the learned linear mappings between the shape coefficients and the anatomical parameters. Only the first 15 shape coefficients are reported.}
\begin{subtable}{.5\textwidth}
\centering
\captionof{table}{Learned matrix $Q$ of femoral ANAT$_{\text{SSM}}$.}
\pgfplotstabletypeset[%
    color cells={min=-1,max=1,textcolor=black},
    /pgfplots/colormap={orangewhiteorange}{rgb255=(255,170,0) color=(white) rgb255=(255,170,0)},
    /pgf/number format/fixed,
    /pgf/number format/precision=3,
    col sep=comma,
    columns/$Q$/.style={reset styles,string type, column type=|C{.85cm}},
    before row=\hline,every last row/.style={after row=\hline},
    columns/NSA/.style={column type=|C{.75cm}, column name=$\beta_{NSA}$},
    columns/FV/.style={column type=|C{.75cm}, column name=$\beta_{FV}$},
    columns/BW/.style={column type=|C{.75cm}, column name=$\beta_{BW}$},
    columns/HD/.style={column type=|C{.75cm}, column name=$\beta_{HD}$},
    columns/FL/.style={column type=|C{.75cm}|, column name=$\beta_{FL}$}
]{
$Q$, NSA, FV, BW, HD, FL
$\alpha_{1}$, -0.19, -0.03, -0.82, -0.85, -0.99
$\alpha_{2}$, 0.57, 0.34, -0.11, -0.09, 0.08
$\alpha_{3}$, -0.25, -0.39, -0.13, -0.12, -0.01
$\alpha_{4}$, -0.15, -0.55, -0.08, -0.18, -0.06
$\alpha_{5}$, 0.28, 0.47, -0.27, -0.18, 0.01
$\alpha_{6}$, -0.17, 0.08, -0.39, -0.22, -0.01
$\alpha_{7}$, 0.24, -0.28, 0.14, 0.02, 0.05
$\alpha_{8}$, -0.1, 0.21, 0.11, 0.02, -0.03
$\alpha_{9}$, 0.22, -0.11, -0.16, -0.06, 0.02
$\alpha_{10}$, 0.2, -0.03, -0.11, 0, 0
$\alpha_{11}$, 0.07, 0.08, 0.11, 0.03, -0.02
$\alpha_{12}$, -0.07, -0.04, 0.03, 0.02, 0.02
$\alpha_{13}$, 0.04, 0.02, -0.07, 0.02, 0
$\alpha_{14}$, 0.02, -0.01, 0, 0.15, 0
$\alpha_{15}$, -0.09, -0.06, 0.03, 0.04, 0.01
}
\end{subtable}
\begin{subtable}{.5\textwidth}
\centering
\captionof{table}{Learned matrix $K$ of femoral OC-ANAT$_{\text{SSM}}$.}
\pgfplotstabletypeset[%
    color cells={min=-1,max=1,textcolor=black},
    /pgfplots/colormap={orangewhiteorange}{rgb255=(255,170,0) color=(white) rgb255=(255,170,0)},
    /pgf/number format/fixed,
    /pgf/number format/precision=3,
    col sep=comma,
    columns/$K$/.style={reset styles,string type, column type=|C{.85cm}},
    before row=\hline,every last row/.style={after row=\hline},
    columns/NSA/.style={column type=|C{.75cm}, column name=$\Tilde{\beta}_{NSA}$},
    columns/FV/.style={column type=|C{.75cm}, column name=$\Tilde{\beta}_{FV}$},
    columns/BW/.style={column type=|C{.75cm}, column name=$\Tilde{\beta}_{BW}$},
    columns/HD/.style={column type=|C{.75cm}, column name=$\Tilde{\beta}_{HD}$},
    columns/FL/.style={column type=|C{.75cm}|, column name=$\Tilde{\beta}_{SL}$}
]{
$K$, NSA, FV, BW, HD, FL
$\alpha_{1}$, -0.05, 0, -0.38, -0.43, -0.81
$\alpha_{2}$, 0.55, 0.24, -0.06, -0.24, 0.16
$\alpha_{3}$, -0.19, -0.38, -0.2, -0.09, 0.16
$\alpha_{4}$, -0.04, -0.55, -0.02, -0.26, 0.1
$\alpha_{5}$, 0.21, 0.43, -0.27, -0.21, 0.2
$\alpha_{6}$, -0.21, 0.09, -0.55, -0.09, 0.31
$\alpha_{7}$, 0.33, -0.34, 0.23, -0.15, 0
$\alpha_{8}$, -0.14, 0.25, 0.22, -0.04, -0.11
$\alpha_{9}$, 0.26, -0.18, -0.25, -0.02, 0.12
$\alpha_{10}$, 0.22, -0.09, -0.2, 0.08, 0.01
$\alpha_{11}$, 0.08, 0.09, 0.21, -0.04, -0.11
$\alpha_{12}$, -0.07, -0.03, 0.01, 0.02, 0.02
$\alpha_{13}$, 0.03, 0.01, -0.15, 0.11, 0.01
$\alpha_{14}$, 0.01, -0.02, -0.15, 0.34, -0.1
$\alpha_{15}$, -0.08, -0.05, 0.01, 0.07, -0.01
}
\end{subtable}

\bigskip
\begin{subtable}{.5\textwidth}
\centering
\captionof{table}{Learned matrix $Q$ of scapular ANAT$_{\text{SSM}}$.}
\pgfplotstabletypeset[%
    color cells={min=-1,max=1,textcolor=black},
    /pgfplots/colormap={orangewhiteorange}{rgb255=(255,170,0) color=(white) rgb255=(255,170,0)},
    /pgf/number format/fixed,
    /pgf/number format/precision=3,
    col sep=comma,
    columns/$Q$/.style={reset styles,string type, column type=|C{.85cm}},
    before row=\hline,every last row/.style={after row=\hline},
    columns/CSA/.style={column type=|C{.75cm}, column name=$\beta_{CSA}$},
    columns/GI/.style={column type=|C{.75cm}, column name=$\beta_{GI}$},
    columns/GV/.style={column type=|C{.75cm}, column name=$\beta_{GV}$},
    columns/GH/.style={column type=|C{.75cm}, column name=$\beta_{GH}$},
    columns/GW/.style={column type=|C{.75cm}, column name=$\beta_{GW}$},
    columns/SL/.style={column type=|C{.75cm}|, column name=$\beta_{SL}$}
]{
$Q$, CSA, GI, GV, GH, GW, SL
$\alpha_{1}$, -0.23, -0.25, -0.12, 0.78, 0.70, 0.89
$\alpha_{2}$, -0.25, 0.32, -0.17, 0.22, 0.07, 0.37
$\alpha_{3}$, 0.18, 0.11, 0.03, 0, -0.12, -0.01
$\alpha_{4}$, -0.27, 0.23, -0.16, -0.07, 0.01, -0.06
$\alpha_{5}$, -0.04, 0.27, -0.34, -0.23, 0.24, -0.03
$\alpha_{6}$, -0.50, -0.17, -0.36, 0.28, 0.27, -0.13
$\alpha_{7}$, 0.19, 0.31, 0.18, -0.22, -0.12, 0.03
$\alpha_{8}$, -0.08, -0.06, -0.21, -0.05, 0.14, -0.05
$\alpha_{9}$, -0.12, 0.09, 0.34, 0.11, 0.03, -0.09
$\alpha_{10}$, 0.02, 0.24, 0.19, 0.14, 0.15, 0.08
$\alpha_{11}$, -0.09, -0.24, -0.03, 0.16, 0.06, -0.08
$\alpha_{12}$, 0.19, 0.19, 0, -0.18, -0.18, 0.05
$\alpha_{13}$, 0.32, 0.14, 0.34, 0.08, -0.02, 0.06
$\alpha_{14}$, 0.06, 0.03, -0.16, -0.15, -0.01, 0.01
$\alpha_{15}$, -0.04, 0.15, -0.18, 0.08, -0.16, 0.02
}
\end{subtable}
\begin{subtable}{.5\textwidth}
\centering
\captionof{table}{Learned matrix $K$ of scapular OC-ANAT$_{\text{SSM}}$.}
\pgfplotstabletypeset[%
    color cells={min=-1,max=1,textcolor=black},
    /pgfplots/colormap={orangewhiteorange}{rgb255=(255,170,0) color=(white) rgb255=(255,170,0)},
    /pgf/number format/fixed,
    /pgf/number format/precision=3,
    col sep=comma,
    columns/$K$/.style={reset styles,string type, column type=|C{.85cm}},
    before row=\hline,every last row/.style={after row=\hline},
    columns/CSA/.style={column type=|C{.75cm}, column name=$\Tilde{\beta}_{CSA}$},
    columns/GI/.style={column type=|C{.75cm}, column name=$\Tilde{\beta}_{GI}$},
    columns/GV/.style={column type=|C{.75cm}, column name=$\Tilde{\beta}_{GV}$},
    columns/GH/.style={column type=|C{.75cm}, column name=$\Tilde{\beta}_{GH}$},
    columns/GW/.style={column type=|C{.75cm}, column name=$\Tilde{\beta}_{GW}$},
    columns/SL/.style={column type=|C{.75cm}|, column name=$\Tilde{\beta}_{SL}$}
]{
$K$, CSA, GI, GV, GH, GW, SL
$\alpha_{1}$, 0.02, -0.19, -0.05, 0.43, 0.43, 0.70
$\alpha_{2}$, -0.31, 0.41, -0.11, 0.16, -0.14, 0.36
$\alpha_{3}$, 0.19, 0.09, -0.04, 0.13, -0.14, -0.01
$\alpha_{4}$, -0.39, 0.29, -0.06, -0.17, -0.02, 0.11
$\alpha_{5}$, -0.05, 0.27, -0.28, -0.36, 0.39, -0.02
$\alpha_{6}$, -0.38, -0.02, -0.28, 0.33, 0.17, -0.36
$\alpha_{7}$, 0.04, 0.27, 0.19, -0.26, -0.03, 0.15
$\alpha_{8}$, -0.01, -0.07, -0.19, -0.11, -0.19, -0.07
$\alpha_{9}$, -0.24, 0.19, 0.41, 0.14, 0.08, -0.19
$\alpha_{10}$, -0.03, 0.31, 0.23, 0.16, 0.19, -0.03
$\alpha_{11}$, 0.01, -0.21, -0.05, 0.23, 0.02, -0.18
$\alpha_{12}$, 0.13, 0.12, -0.04, -0.17, -0.16, 0.17
$\alpha_{13}$, 0.31, 0.12, 0.29, 0.17, 0.04, 0
$\alpha_{14}$, 0.07, -0.03, -0.17, -0.21, 0.02, 0.09
$\alpha_{15}$, -0.02, 0.18, -0.24, 0.26, -0.31, -0.02
}
\end{subtable}
\label{tab:anat_ssms_weights}
\end{figure*}

To directly evaluate the orthogonality constraints imposed on OC-ANAT$_{\text{SSM}}$, we computed the measurements of shapes generated by varying the $j$-th anatomical parameter between $\pm3\sigma_{c_j}$ (Fig. \ref{fig:orthogonality_measure}). We compared OC-ANAT$_{\text{SSM}}$ with ANAT$_{\text{SSM}}$ and used our automatic method based on point-to-point correspondence and anatomical landmarks to derive the measurements. The graphs obtained by varying FL and SL illustrated an (approximately) linear relation between anatomical parameters with a slope corresponding to the covariance between parameters. Therefore, the orthogonality of the OC-ANAT$_{\text{SSM}}$ model was demonstrated as all its anatomical parameters remained unchanged (null slope).

\afterpage{\clearpage}

\setcounter{table}{1}
\begin{figure*}[ht!]

\captionof{table}{Weights of the covariance matrix $QQ^{T}$ associated with the distribution of the anatomical parameter.}
\begin{subtable}{.5\textwidth}
\centering
\captionof{table}{Covariance of femoral ANAT$_{\text{SSM}}$.}
\centering
\pgfplotstabletypeset[%
    color cells={min=-1,max=1,textcolor=black},
    /pgfplots/colormap={orangewhiteorange}{rgb255=(255,170,0) color=(white) rgb255=(255,170,0)},
    /pgf/number format/fixed,
    /pgf/number format/precision=3,
    col sep=comma,
    columns/$QQ^{T}$/.style={reset styles,string type, column type=|C{.75cm}},
    before row=\hline,every last row/.style={after row=\hline},
    columns/NSA/.style={column type=|C{.75cm}, column name=$\beta_{NSA}$},
    columns/FV/.style={column type=|C{.75cm}, column name=$\beta_{FV}$},
    columns/BW/.style={column type=|C{.75cm}, column name=$\beta_{BW}$},
    columns/HD/.style={column type=|C{.75cm}, column name=$\beta_{HD}$},
    columns/FL/.style={column type=|C{.75cm}|, column name=$\beta_{FL}$}
]{
$QQ^{T}$, NSA, FV, BW, HD, FL
$\beta_{NSA}$, 1, 0.36, 0.11, 0.24, 0.27
$\beta_{FV}$, 0.36, 1, -0.06, 0.03, 0.07
$\beta_{BW}$, 0.11, -0.06, 1, 0.89, 0.79
$\beta_{HD}$, 0.24, 0.03, 0.89, 1, 0.85
$\beta_{FL}$, 0.27, 0.07, 0.79, 0.85, 1
}
\end{subtable}
\begin{subtable}{.5\textwidth}
\centering
\captionof{table}{Covariance of scapular ANAT$_{\text{SSM}}$.}
\pgfplotstabletypeset[%
    color cells={min=-1,max=1,textcolor=black},
    /pgfplots/colormap={orangewhiteorange}{rgb255=(255,170,0) color=(white) rgb255=(255,170,0)},
    /pgf/number format/fixed,
    /pgf/number format/precision=3,
    col sep=comma,
    columns/$QQ^{T}$/.style={reset styles,string type, column type=|C{.75cm}},
    before row=\hline,every last row/.style={after row=\hline},
    columns/CSA/.style={column type=|C{.75cm}, column name=$\beta_{CSA}$},
    columns/GI/.style={column type=|C{.75cm}, column name=$\beta_{GI}$},
    columns/GV/.style={column type=|C{.75cm}, column name=$\beta_{GV}$},
    columns/GH/.style={column type=|C{.75cm}, column name=$\beta_{GH}$},
    columns/GW/.style={column type=|C{.75cm}, column name=$\beta_{GW}$},
    columns/SL/.style={column type=|C{.75cm}|, column name=$\beta_{SL}$}
]{
$QQ^{T}$, CSA, GI, GV, GH, GW, SL
$\beta_{CSA}$, 1, 0.42, 0.46, -0.49, -0.46, -0.21
$\beta_{GI}$, 0.42, 1, 0.03, -0.36, -0.22, -0.03
$\beta_{GV}$, 0.46, 0.03, 1, -0.06, -0.32, -0.11
$\beta_{GH}$, -0.49, -0.36, -0.06, 1, 0.69, 0.71
$\beta_{GW}$, -0.46, -0.22, -0.32, 0.69, 1, 0.59
$\beta_{SL}$, -0.21, -0.03, -0.11, 0.71, 0.59, 1 
}
\end{subtable}
\label{tab:anat_ssms_covariance}

\vspace*{\floatsep}

\centering
\begin{adjustbox}{width=\textwidth}
\begin{tikzpicture}
\begin{scope}[spy using outlines=
      {circle, magnification=2.25, size=.85cm, connect spies, rounded corners}]

\node[inner sep=0pt] at (0,0.05)
    {\includegraphics[width=.022\textwidth]{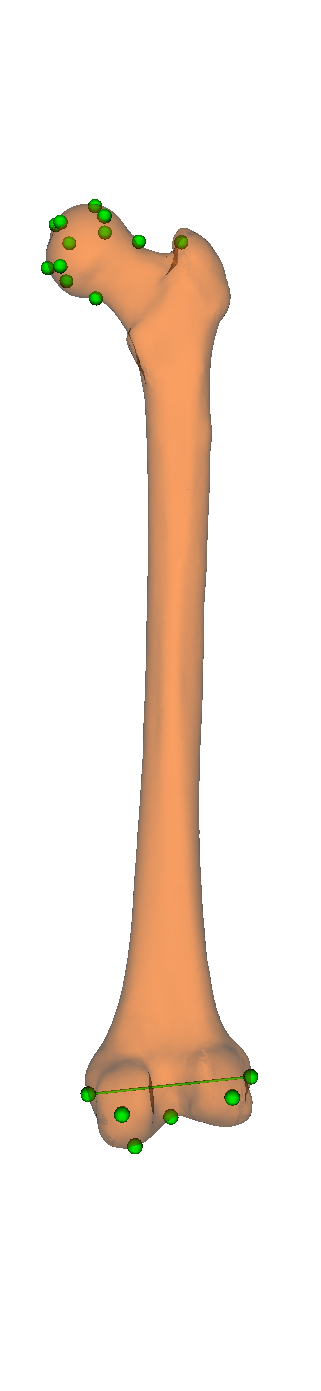}};
\node[inner sep=0pt] at (1.5,0.05)
    {\includegraphics[width=.022\textwidth]{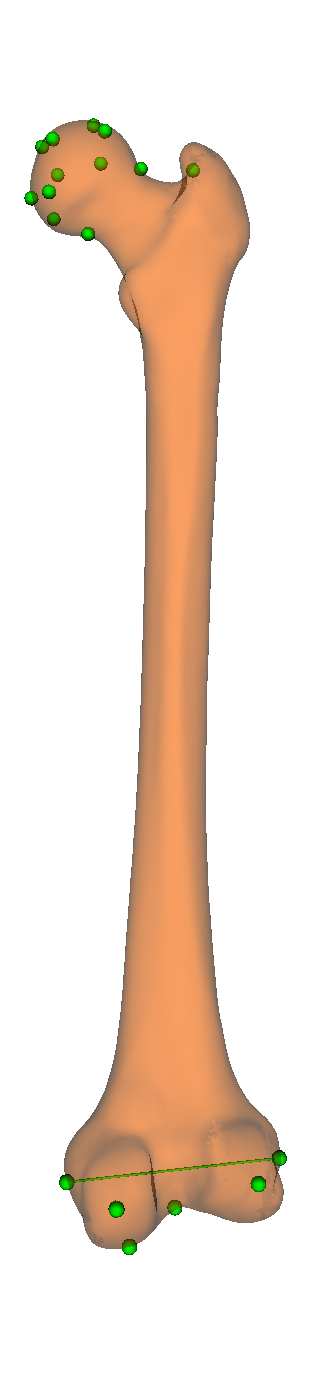}};
\node[inner sep=0pt] at (3,0.05)
    {\includegraphics[width=.022\textwidth]{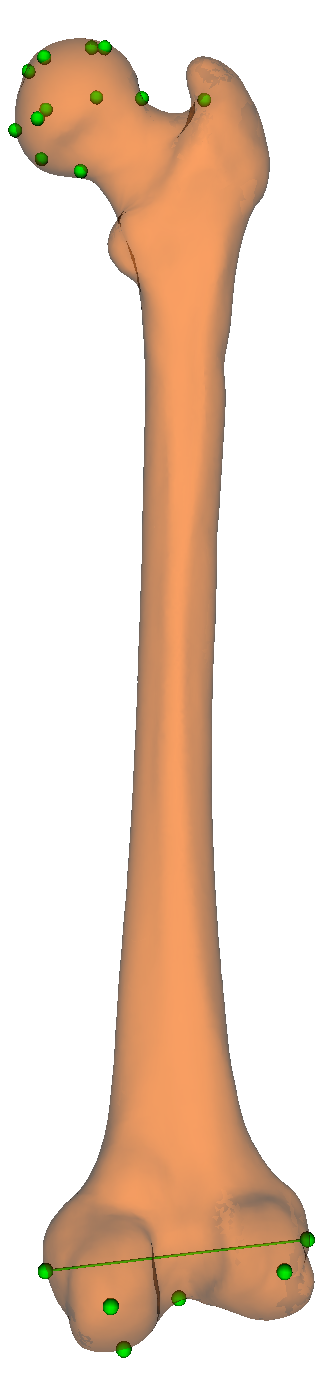}};
   
\node[inner sep=0pt] at (4.65,0.05)
    {\includegraphics[width=.022\textwidth]{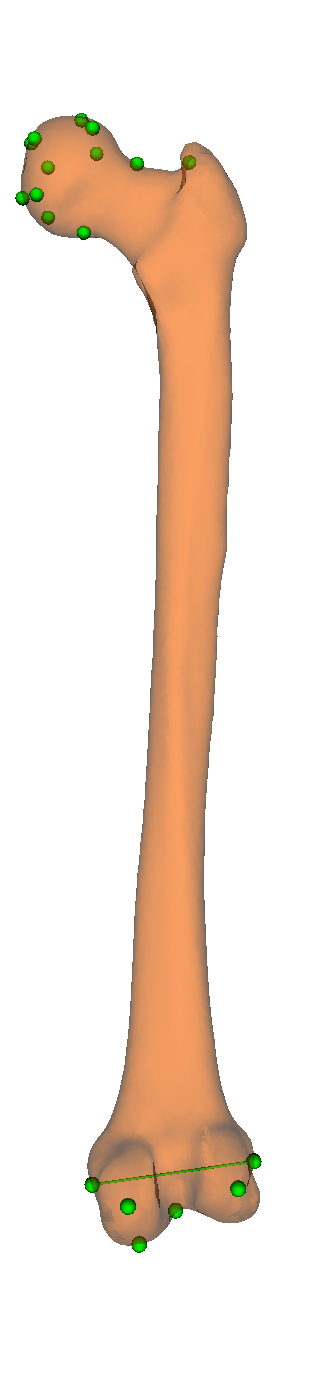}};
\node[inner sep=0pt] at (6.15,0.05)
    {\includegraphics[width=.022\textwidth]{BW_mean.png}};
\node[inner sep=0pt] at (7.65,0.05)
    {\includegraphics[width=.022\textwidth]{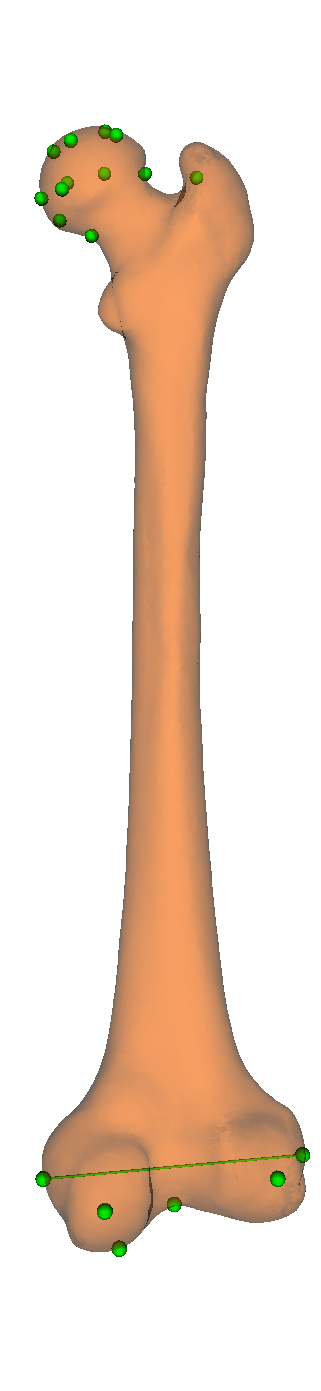}};
    
\node[inner sep=0pt] at (0,-1.85)
    {\includegraphics[width=.022\textwidth]{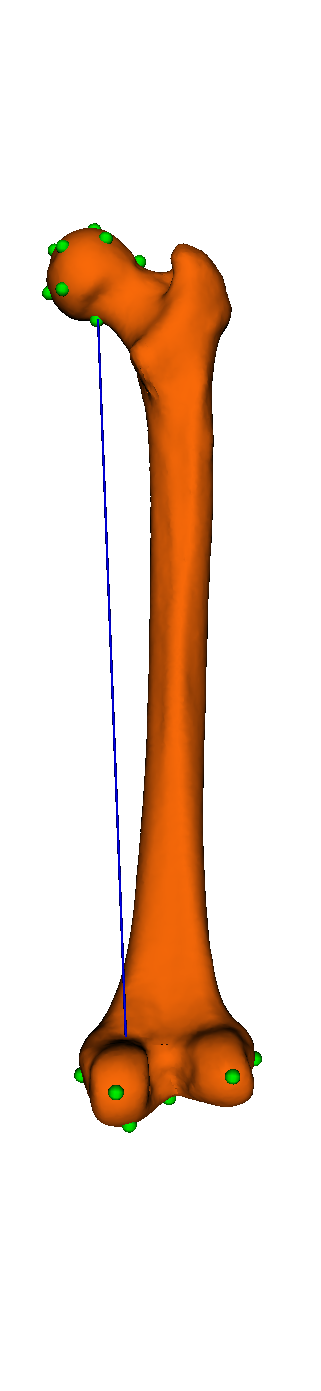}};
\node[inner sep=0pt] at (1.5,-1.85)
    {\includegraphics[width=.022\textwidth]{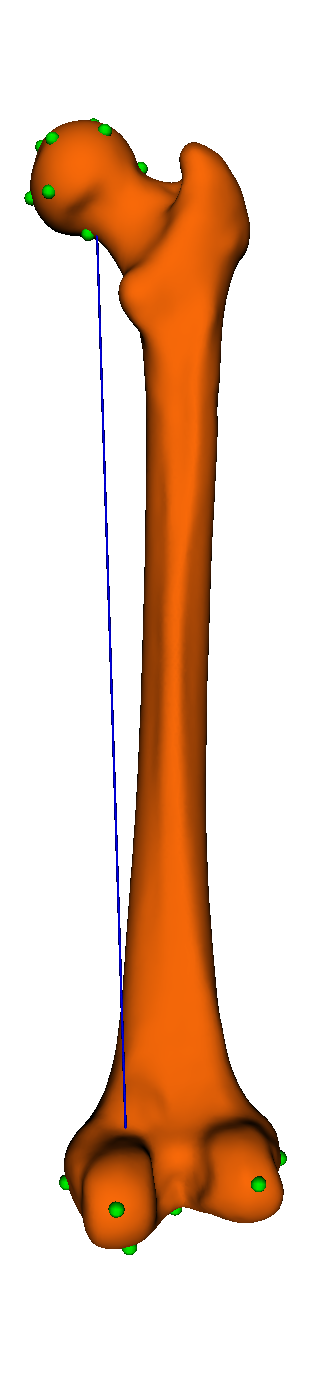}};
\node[inner sep=0pt] at (3,-1.85)
    {\includegraphics[width=.022\textwidth]{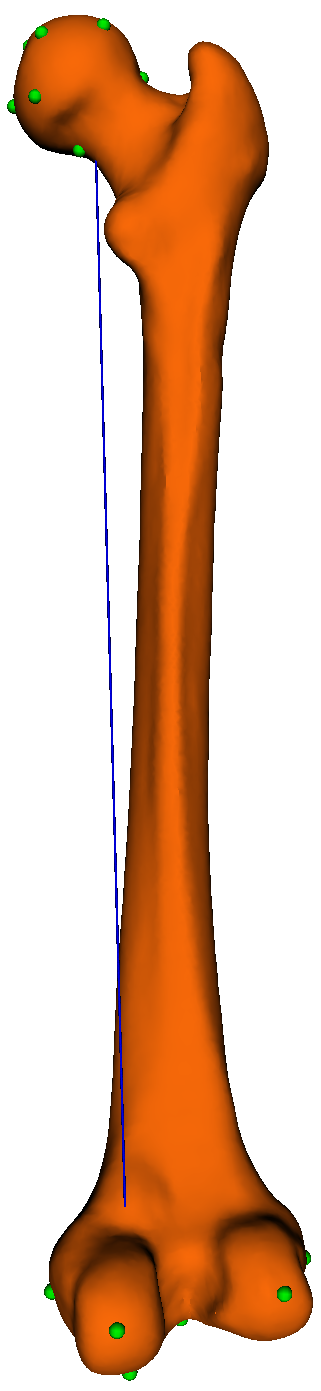}};
    
\node[inner sep=0pt] at (4.65,-1.85)
    {\includegraphics[width=.022\textwidth]{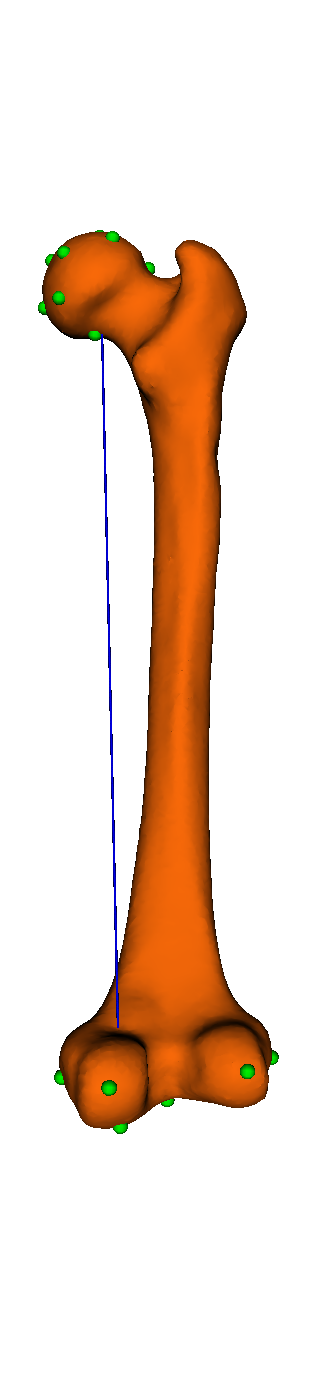}};
\node[inner sep=0pt] at (6.15,-1.85)
    {\includegraphics[width=.022\textwidth]{FL_mean.png}};
\node[inner sep=0pt] at (7.65,-1.85)
    {\includegraphics[width=.022\textwidth]{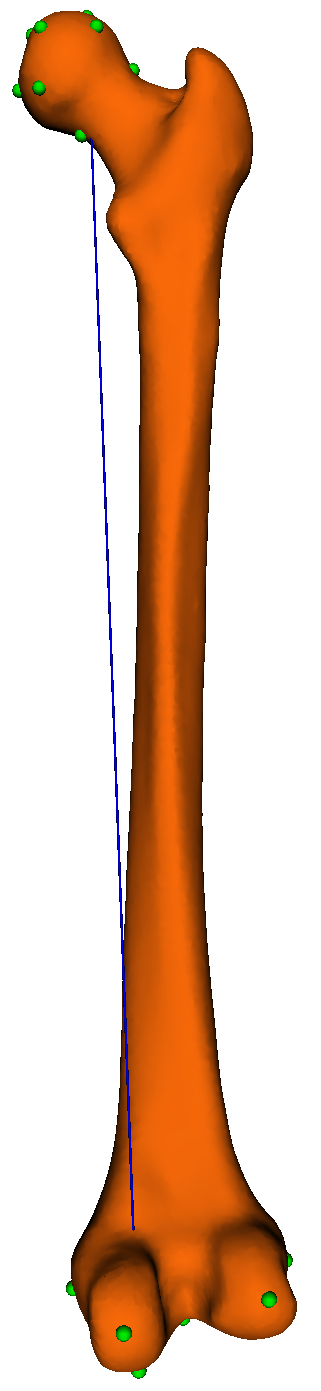}};
    
\node[inner sep=0pt] at (0,-3.75)
    {\includegraphics[width=.078\textwidth]{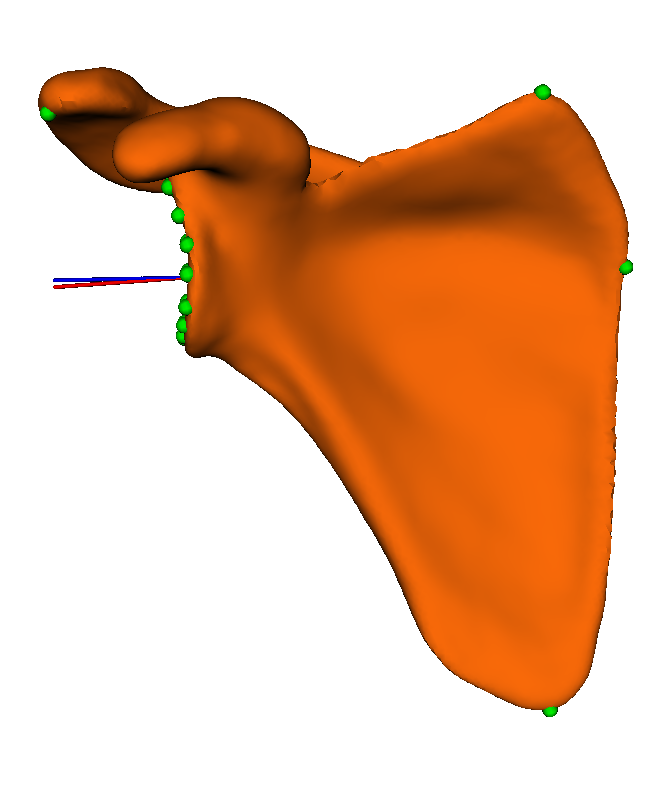}};
\node[inner sep=0pt] at (1.5,-3.75)
    {\includegraphics[width=.078\textwidth]{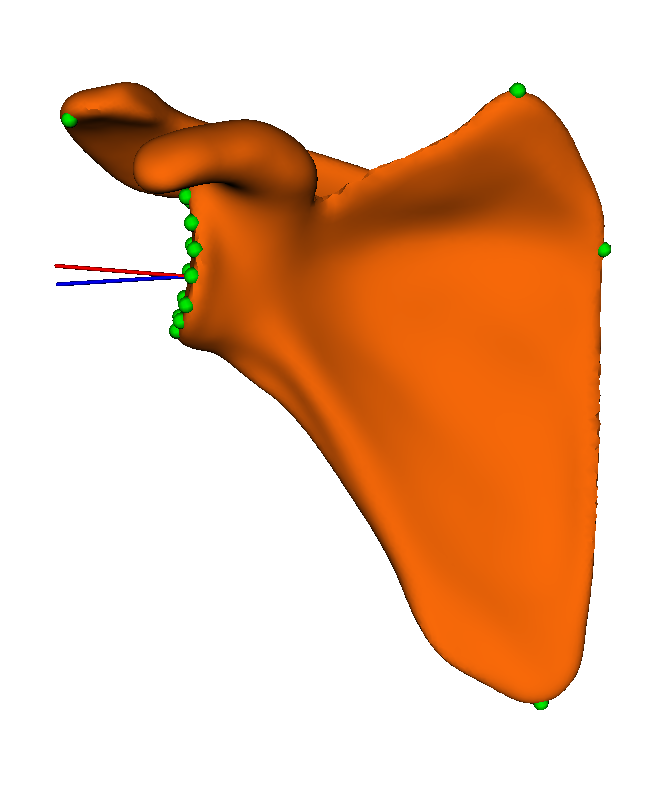}};
\node[inner sep=0pt] at (3,-3.75)
    {\includegraphics[width=.078\textwidth]{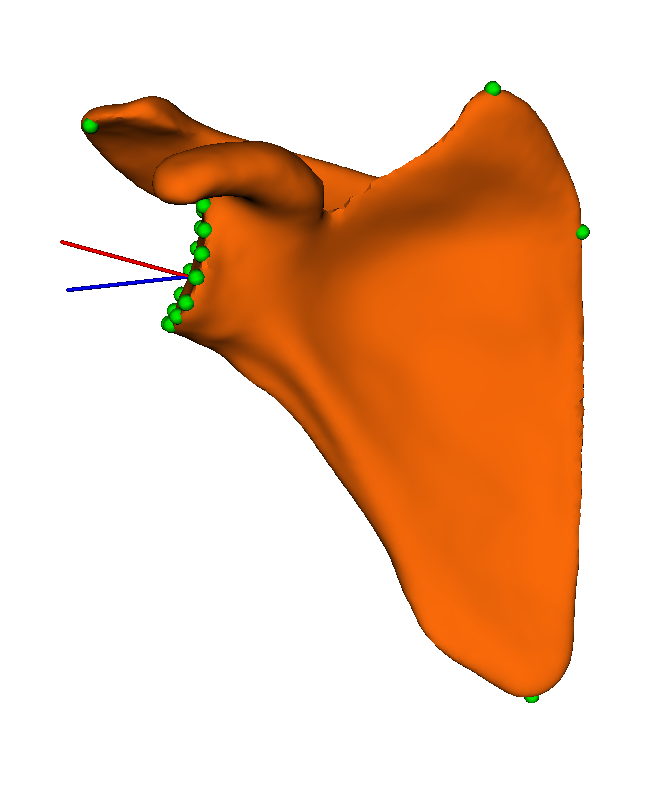}}; 
    
\node[inner sep=0pt] at (4.65,-3.75)
    {\includegraphics[width=.078\textwidth]{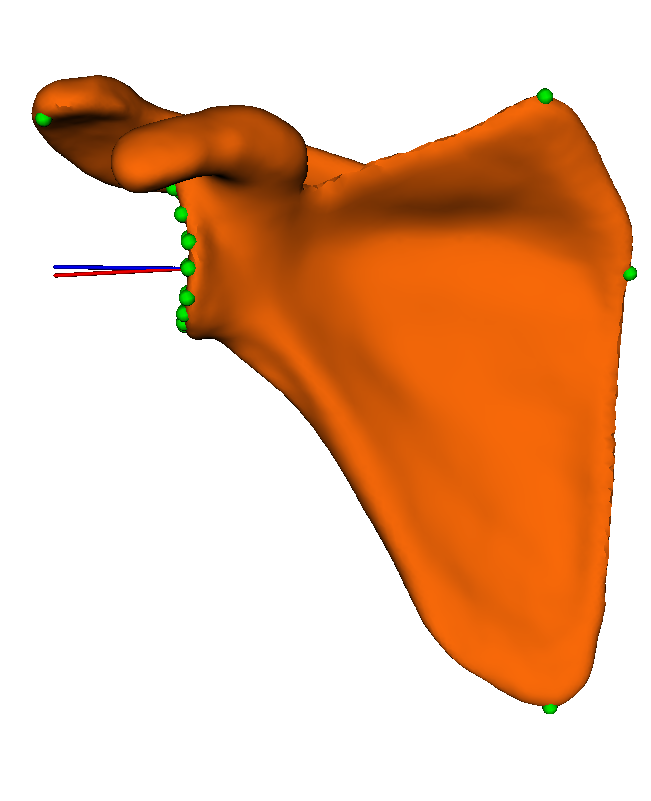}};
\node[inner sep=0pt] at (6.15,-3.75)
    {\includegraphics[width=.078\textwidth]{GI_mean.png}};
\node[inner sep=0pt] at (7.65,-3.75)
    {\includegraphics[width=.078\textwidth]{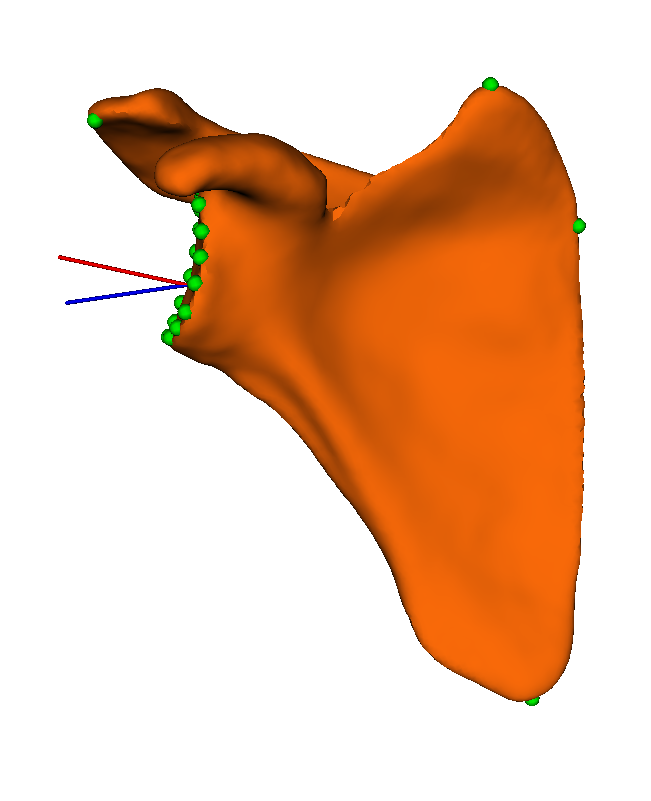}};
    
\node[inner sep=0pt] at (0,-5.65)
    {\includegraphics[width=.058\textwidth]{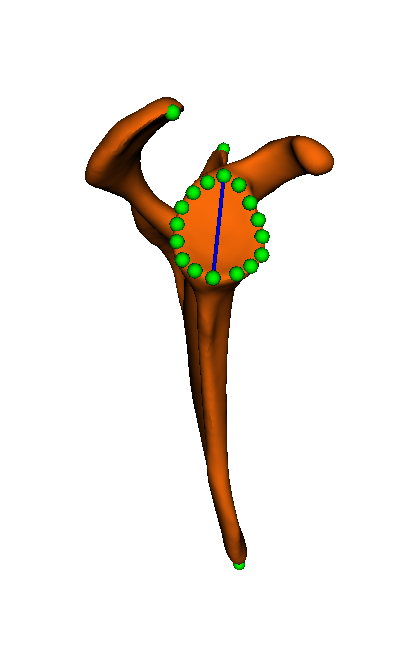}};
\node[inner sep=0pt] at (1.5,-5.65)
    {\includegraphics[width=.058\textwidth]{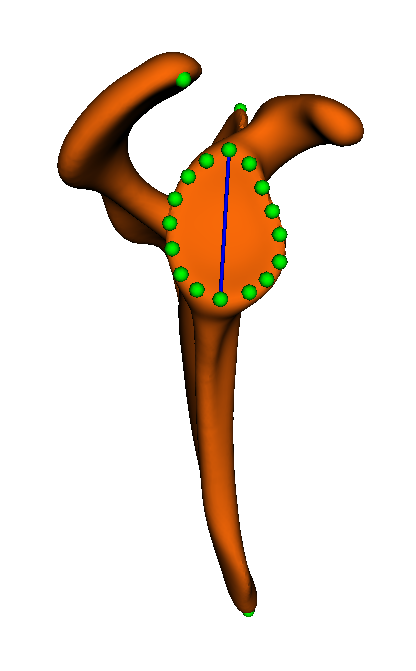}};
\node[inner sep=0pt] at (3,-5.65)
    {\includegraphics[width=.058\textwidth]{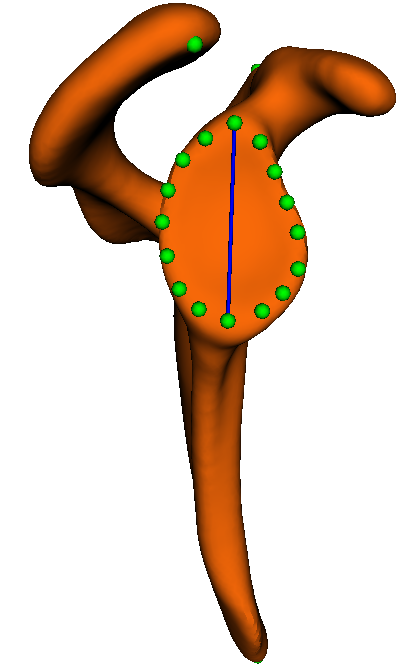}}; 
    
\node[inner sep=0pt] at (4.65,-5.65)
    {\includegraphics[width=.058\textwidth]{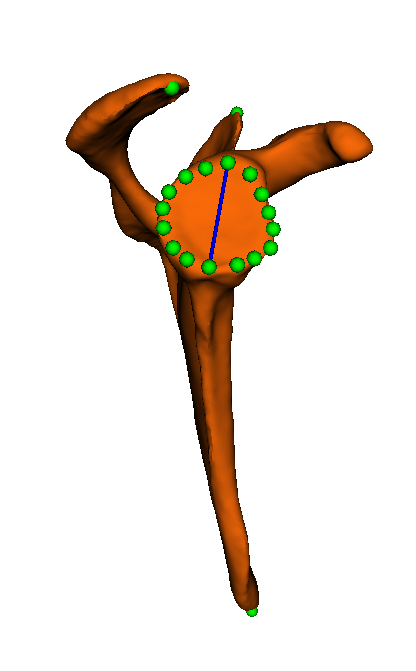}};
\node[inner sep=0pt] at (6.15,-5.65)
    {\includegraphics[width=.058\textwidth]{GH_mean.png}};
\node[inner sep=0pt] at (7.65,-5.65)
    {\includegraphics[width=.058\textwidth]{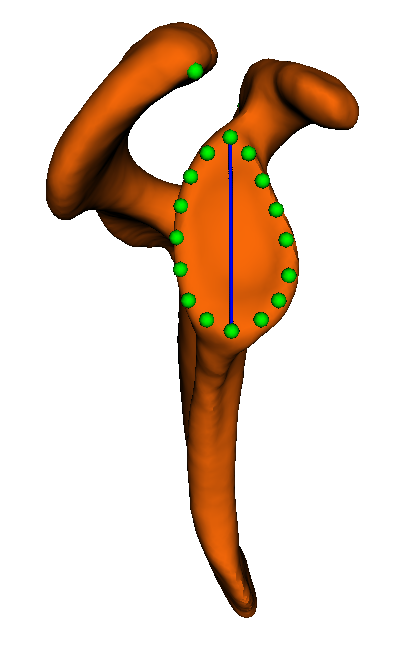}};
    
\node[inner sep=0pt] at (0,-7.55)
    {\includegraphics[width=.078\textwidth]{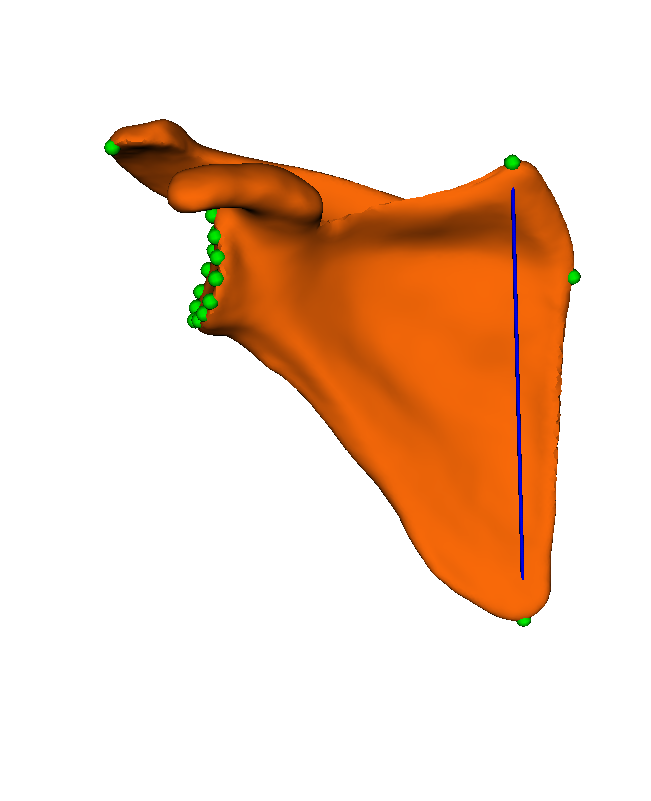}};
\node[inner sep=0pt] at (1.5,-7.55)
    {\includegraphics[width=.078\textwidth]{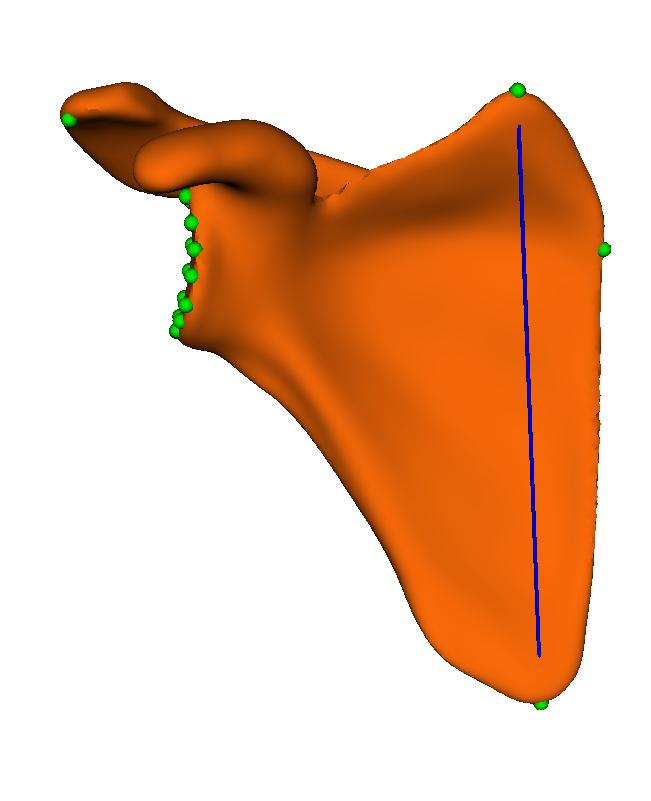}};
\node[inner sep=0pt] at (3,-7.55)
    {\includegraphics[width=.078\textwidth]{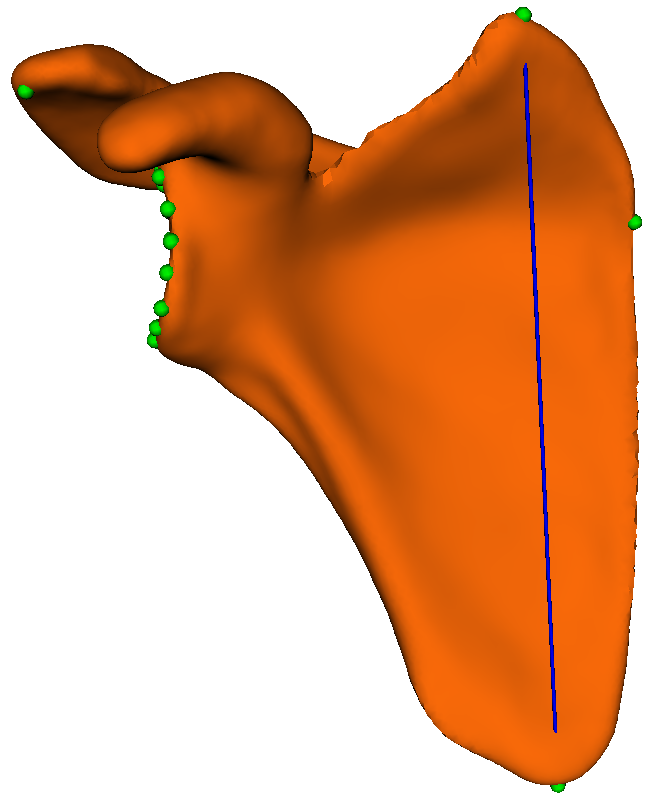}};
    
\node[inner sep=0pt] at (4.65,-7.55)
    {\includegraphics[width=.078\textwidth]{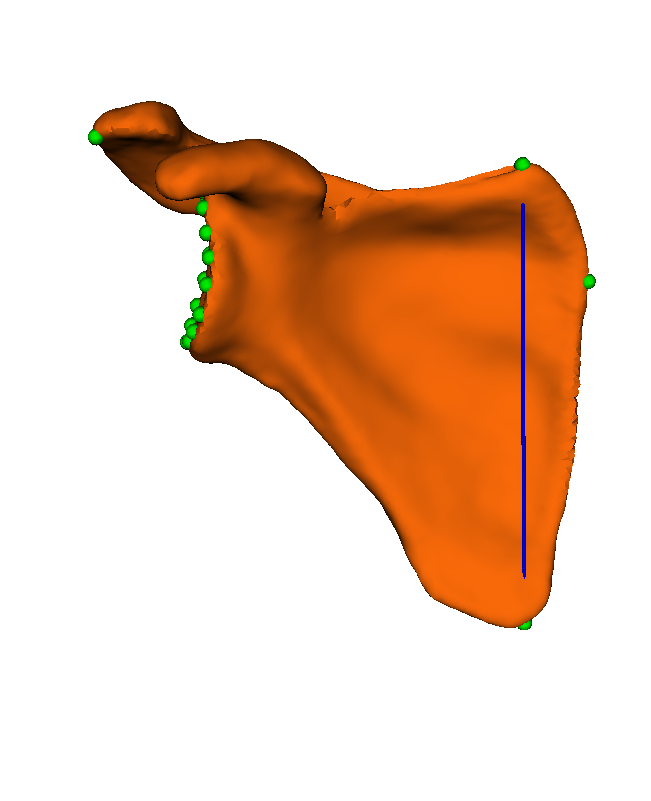}};
\node[inner sep=0pt] at (6.15,-7.55)
    {\includegraphics[width=.078\textwidth]{SL_mean.png}};
\node[inner sep=0pt] at (7.65,-7.55)
    {\includegraphics[width=.078\textwidth]{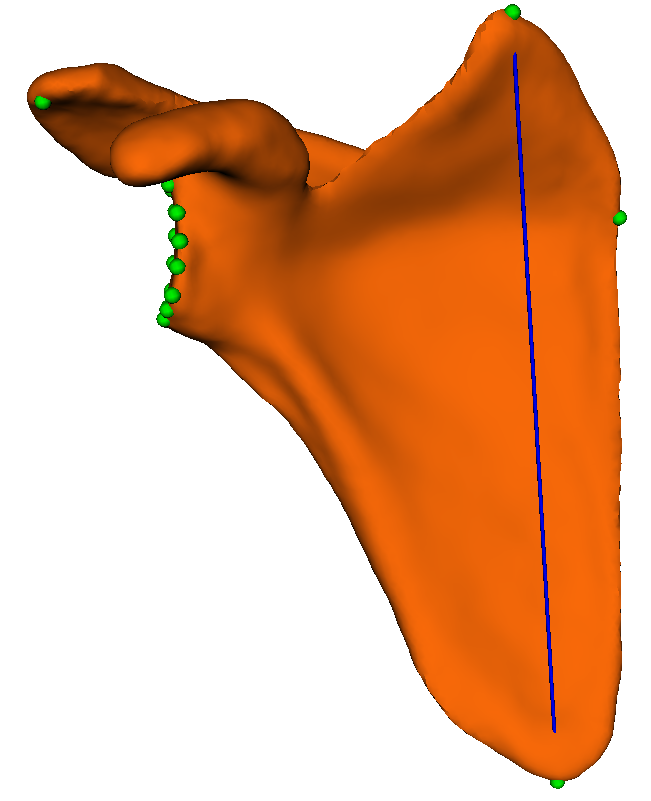}};

\node[anchor=north] at (1.5, 1.7) {\scalebox{.6}{\textbf{ANAT$_{\text{SSM}}$}}};
\draw[line width=.1mm, {Latex[length=4pt, width=4pt]}-{Latex[length=4pt, width=4pt]}] (-.25,1.3) -- (3.25,1.3);
\draw[line width=.1mm] (1.5,1.35) -- (1.5,1.25);
\draw[line width=.1mm] (0,1.35) -- (0,1.25);
\draw[line width=.1mm] (3,1.35) -- (3,1.25);
\node at (1.5, 1.1) {\scalebox{.6}{$\mu$}};
\node at (3, 1.1) {\scalebox{.6}{+3$\sigma_{c_j}$}};
\node at (0, 1.1) {\scalebox{.6}{-3$\sigma_{c_j}$}};

\node[anchor=north] at (6.25, 1.7) {\scalebox{.6}{\textbf{OC-ANAT$_{\text{SSM}}$}}};
\draw[line width=.1mm, {Latex[length=4pt, width=4pt]}-{Latex[length=4pt, width=4pt]}] (4.5,1.3) -- (8,1.3);
\draw[line width=.1mm] (4.75,1.35) -- (4.75,1.25);
\draw[line width=.1mm] (6.25,1.35) -- (6.25,1.25);
\draw[line width=.1mm] (7.75,1.35) -- (7.75,1.25);
\node at (6.25, 1.1) {\scalebox{.6}{$\mu$}};
\node at (7.75, 1.1) {\scalebox{.6}{+3$\sigma_{c_j}$}};
\node at (4.75, 1.1) {\scalebox{.6}{-3$\sigma_{c_j}$}};

\node[rotate=90] at (-1.35, -0.95) {\scalebox{.525}{Femur}};
\node[rotate=90] at (-1.35, -5.7) {\scalebox{.525}{Scapula}};

\node[rotate=90] at (-.95, 0) {\scalebox{.525}{Bicondylar Width}};
\node[rotate=90] at (-.95, -1.9) {\scalebox{.525}{Femur Length}};
\node[rotate=90] at (-.95, -3.8) {\scalebox{.525}{Glenoid Inclination}};
\node[rotate=90] at (-.95, -5.7) {\scalebox{.525}{Glenoid Height}};
\node[rotate=90] at (-.95, -7.6) {\scalebox{.525}{Scapula Length}};

\draw[line width=.2mm, color=black] (8.5,-0.95) -- (8.5,0.95) -- (-1.15,0.95) -- (-1.15, -0.95);
\draw[line width=.2mm, color=black] (8.5,-2.85) -- (8.5,-0.95) -- (-1.15,-0.95) -- (-1.15, -2.85);
\draw[line width=.2mm, color=black] (8.5,-4.75) -- (8.5,-2.85) -- (-1.15,-2.85) -- (-1.15, -4.75);
\draw[line width=.2mm, color=black] (8.5,-6.65) -- (8.5,-4.75) -- (-1.15,-4.75) -- (-1.15, -6.65);
\draw[line width=.2mm, color=black] (8.5,-8.55) -- (8.5,-6.65) -- (-1.15,-6.65) -- (-1.15, -8.55) -- cycle;

\draw[line width=.2mm, color=black] (-.75,-8.55) -- (-.75,1.7) -- (3.875,1.7) -- (3.875,-8.55);
\draw[line width=.2mm, color=black] (3.875,1.7) -- (8.5,1.7) -- (8.5,0.95);
\draw[line width=.2mm, color=black] (-1.15,-8.55) -- (-1.55,-8.55) -- (-1.55,0.95) -- (-1.15,0.95);
\draw[line width=.2mm, color=black] (-1.15,-2.85) -- (-1.55,-2.85);

\node at (1.5, -.8) {\scalebox{.525}{$\beta_{BW}$}};
\node at (1.5, -2.7) {\scalebox{.525}{$\beta_{FL}$}};
\node at (1.5, -4.6) {\scalebox{.525}{$\beta_{GI}$}};
\node at (1.5, -6.5) {\scalebox{.525}{$\beta_{GH}$}};
\node at (1.5, -8.4) {\scalebox{.525}{$\beta_{SL}$}};

\node at (6.15, -.8) {\scalebox{.525}{$\Tilde{\beta}_{BW}$}};
\node at (6.15, -2.7) {\scalebox{.525}{$\Tilde{\beta}_{FL}$}};
\node at (6.15, -4.6) {\scalebox{.525}{$\Tilde{\beta}_{GI}$}};
\node at (6.15, -6.5) {\scalebox{.525}{$\Tilde{\beta}_{GH}$}};
\node at (6.15, -8.4) {\scalebox{.525}{$\Tilde{\beta}_{SL}$}};

\spy [Dandelion] on (0.025,-.38) in node [left] at (1.15,.425);
\spy [Dandelion] on (1.525,-.51) in node [left] at (1.15,-.425);
\spy [Dandelion] on (3.025,-.65) in node [left] at (2.65,-.425);

\spy [Dandelion] on (4.675,-.51) in node [left] at (5.80,.425);
\spy [Dandelion] on (6.175,-.51) in node [left] at (5.80,-.425);
\spy [Dandelion] on (7.675,-.52) in node [left] at (7.30,-.425);

\end{scope}
\end{tikzpicture}
\end{adjustbox}
\captionof{figure}{Visual comparison of shape variation patterns arising from anatomical parameters of femoral ANAT$_{\text{SSM}}$ ($\beta_{BW}$, $\beta_{FL}$) and OC-ANAT$_{\text{SSM}}$ ($\Tilde{\beta}_{BW}$, $\Tilde{\beta}_{FL}$), as well as scapular ANAT$_{\text{SSM}}$ ($\beta_{GI}$, $\beta_{GH}$, $\beta_{SL}$) and OC-ANAT$_{\text{SSM}}$ ($\Tilde{\beta}_{GI}$, $\Tilde{\beta}_{GH}$, $\Tilde{\beta}_{SL}$). Each anatomical parameters $c_j$ is shown with the shape varied between three standard deviations ($\pm3\sigma_{c_j}$) from either side of the mean shape ($\mu$).}
\label{fig:visual_comparison_shape_variation}
\end{figure*}

\section{Learned matrices and additional visualization of shape variation patterns}

For both anatomical structures, the weights of the learned matrices $Q$, $K$, and $QQ^{T}$ are reported in Tables \ref{tab:anat_ssms_weights} and \ref{tab:anat_ssms_covariance}. Visualization of the shape variation patterns arising from the femoral (BW, FL) and scapular (GI, GH, SL) anatomical parameters are provided in Fig. \ref{fig:visual_comparison_shape_variation}.

\section{Predictive performance}

\begin{table}[t]
\caption{Leave-one-out assessment of the absolute error between OC-ANAT$_{\text{SSM}}$ predictions and manually derived anatomical measurements. Models were learned sequentially by retaining the anatomical parameter of $\Tilde{\beta}$ with largest anatomical variability at each step. Mean and standard deviation are reported.}
\begin{subtable}{.5\textwidth}
\centering
\captionof{table}{Prediction error of femoral OC-ANAT$_{\text{SSM}}$.}
    \begin{tabular}{|C{1.2cm}||C{.75cm}|C{.75cm}|C{.75cm}|C{.75cm}|C{.75cm}|} 
    \hline
    \multirow{2}{*}{\shortstack{Absolute\\Error}} & \multicolumn{5}{c|}{OC-ANAT$_{\text{SSM}}$} \\\cline{2-6}
    & \--- & $\setminus \Tilde{\beta}_{FL}$ & $\setminus \Tilde{\beta}_{HD}$ & $\setminus \Tilde{\beta}_{BW}$ & \!\!$\setminus \Tilde{\beta}_{NSA}$  \\ \hline\hline
    FV ($\degree$) & \!\!3.1$\pm$2.6 & \!\!3.1$\pm$2.6 & \!\!3.1$\pm$2.6 & \!\!3.1$\pm$2.6 & \!\!2.7$\pm$2.2 \\\hline
    NSA ($\degree$) & \!\!2.0$\pm$1.6 & \!\!2.0$\pm$1.5 & \!\!2.0$\pm$1.5 & \!\!2.0$\pm$1.5 & \--- \\\hline
    BW (mm) & \!\!3.4$\pm$2.5 & \!\!2.9$\pm$1.9 & \!\!1.1$\pm$0.9 & \--- & \--- \\\hline
    HD (mm) & \!\!2.2$\pm$1.6 & \!\!1.9$\pm$1.6 & \--- & \--- & \--- \\\hline
    FL (cm) & \!\!1.4$\pm$1.0 & \--- & \--- & \--- & \--- \\\hline
    \end{tabular}
\end{subtable}

\bigskip
\begin{subtable}{.5\textwidth}
\centering
\captionof{table}{Prediction error of scapular OC-ANAT$_{\text{SSM}}$.}
    \centering
    \begin{tabular}{|C{1.2cm}||C{.75cm}|C{.75cm}|C{.75cm}|C{.75cm}|C{.75cm}|C{.75cm}|} 
    \hline
    \multirow{2}{*}{\shortstack{Absolute\\Error}} & \multicolumn{6}{c|}{OC-ANAT$_{\text{SSM}}$} \\\cline{2-7}
    & \--- & $\setminus \Tilde{\beta}_{SL}$ & $\setminus \Tilde{\beta}_{GH}$ & $\setminus \Tilde{\beta}_{GW}$ & $\setminus \Tilde{\beta}_{GI}$ & \!\!$\setminus \Tilde{\beta}_{CSA}$\\
    \hline \hline
    GV ($\degree$) & \!\!2.6$\pm$2.1 & \!\!2.6$\pm$2.1 & \!\!2.6$\pm$2.0 & \!\!2.5$\pm$1.9 & \!\!2.4$\pm$1.9 & \!\!2.1$\pm$1.8 \\\hline
    CSA ($\degree$) & \!\!2.4$\pm$2.0 & \!\!2.4$\pm$2.0 & \!\!2.2$\pm$2.0 & \!\!2.2$\pm$1.9 & \!\!2.1$\pm$1.7 & \--- \\\hline
    GI ($\degree$) & \!\!2.6$\pm$2.3 & \!\!2.6$\pm$2.2 & \!\!2.6$\pm$2.2 & \!\!2.6$\pm$2.2 & \--- & \--- \\\hline
    GW (mm) & \!\!1.8$\pm$1.6 & \!\!1.8$\pm$1.6 & \!\!1.6$\pm$1.5 & \--- & \--- & \--- \\\hline
    GH (mm) & \!\!2.4$\pm$1.7 & \!\!2.2$\pm$1.7 & \--- & \--- & \--- & \--- \\\hline
    SL (mm) & \!\!5.1$\pm$3.7 & \--- & \--- & \--- & \--- & \--- \\\hline
    \end{tabular}
\end{subtable}
\label{tab:predictive_performance}
\end{table}

While the predictive performances of the BASE$_{\text{SSM}}$ and ANAT$_{\text{SSM}}$ models were not affected by the number of anatomical parameters, the orthogonality constraints and thus the predictive performance of OC-ANAT$_{\text{SSM}}$ were determined by the number of anatomical parameters employed. Hence, to assess the impact of the number of anatomical parameters on the predictive performance of OC-ANAT$_{\text{SSM}}$, we retained the anatomical parameter of $\Tilde{\beta}$ with largest anatomical variability sequentially. As expected, the mean absolute error of each remaining anatomical parameter decreased at each step due to reduced orthogonality constraints (Table \ref{tab:predictive_performance}), with largest decrease for highly correlated parameters (e.g. FL and BW). Most importantly, OC-ANAT$_{\text{SSM}}$ models parameterized by a unique anatomical parameter performed identically to ANAT$_{\text{SSM}}$ models with complete parameterization (FV = 2.7$\pm$2.2$\degree$, GV = 2.1$\pm$1.8$\degree$, Table I). Designing OC-ANAT$_{\text{SSM}}$ models with a limited number of anatomical parameters is therefore crucial to reach satisfactory predictive performance.

\section{Video demonstrations}

To compare the BASE$_{\text{SSM}}$ to the proposed models (ANAT$_{\text{SSM}}$ and OC-ANAT$_{\text{SSM}}$) of each bone, we provide video demonstrations \path{demo_bone_base_ssm.avi} and \path{demo_bone_anat_ssm.avi} (with \path{bone} = \path{femur} or \path{scapula}) which present the shape variation patterns of each model.

\end{document}